\documentclass[sort,preprint,12pt]{elsarticle}
\usepackage{natbib}
\usepackage{hyperref}
\usepackage{mathrsfs}
\usepackage{amsthm,amsmath,amssymb}
\usepackage{enumerate}
\usepackage{pifont}
\usepackage{booktabs}
\usepackage{caption}
\usepackage{graphicx}
\usepackage{float}
\usepackage{subcaption}
\usepackage{multirow}
\usepackage{cleveref}
\usepackage{stfloats}
\usepackage{color}
\usepackage{ulem}
\usepackage{makecell}  
\usepackage{fancyhdr}

\journal{journal}
\newcommand{\tabincell}[2]{\begin{tabular}{@{}#1@{}}#2\end{tabular}}
\newtheorem*{Lemma}{\textbf{Lemma}}
\begin{document}

\begin{frontmatter}

\title{Solving Partial Differential Equations in Different Domains by Operator Learning method Based on Boundary Integral Equations}

\author[a]{Bin Meng}
\author[a]{Yutong Lu}
\author[a,b]{Ying Jiang$^*$}

\affiliation[a]{organization={School of Computer Science and
Engineering, Sun Yat-sen University},
            city={Guangzhou},
            postcode={510006},
            country={People’s Republic of China}}
\affiliation[b]{organization={ Guangdong Province Key Laboratory of Computational Science},
            city={Guangzhou},
            postcode={510275},
            country={People’s Republic of China}}
\begin{abstract}
	This article explores operator learning models that can deduce solutions to partial differential equations (PDEs) on arbitrary domains without requiring retraining. We introduce two innovative models rooted in boundary integral equations (BIEs): the Boundary Integral Type Deep Operator Network (BI-DeepONet) and the Boundary Integral Trigonometric Deep Operator Neural Network (BI-TDONet), which are crafted to address PDEs across diverse domains. Once fully trained, these BIE-based models adeptly predict the solutions of PDEs in any domain without the need for additional training. BI-TDONet notably enhances its performance by employing the singular value decomposition (SVD) of bounded linear operators, allowing for the efficient distribution of input functions across its modules. Furthermore, to tackle the issue of function sampling values that do not effectively capture oscillatory and impulse signal characteristics, trigonometric coefficients are utilized as both inputs and outputs in BI-TDONet. Our numerical experiments robustly support and confirm the efficacy of this theoretical framework.
\end{abstract}




\end{frontmatter}

\section{Introduction}
In recent years, the integration of machine learning and data analytics has led to groundbreaking advancements across various scientific fields such as image recognition \cite{KeysersDeselaers2007} and natural language processing \cite{Nadkarni2011}. Specifically, deep neural networks have attracted considerable attention due to their ability to tackle complex challenges \cite{lecun2015deep,jan2021machine}. These networks are highly valued in mathematics for their capability to approximate virtually any function, a property confirmed by foundational theories \cite{hornik1990universal,cybenko1989approximation}. This functional approximation capability has significantly influenced both theoretical and practical aspects of various disciplines, facilitating a broad spectrum of applications, from simulating theoretical models to solving real-world problems in computational science.

Partial Differential Equations (PDEs) are fundamental in many scientific and engineering disciplines \cite{courant2008methods}. Typically, obtaining analytical solutions to PDEs is challenging, making numerical methods essential. These methods, including finite element \cite{hughes2012finite}, finite difference \cite{godunov1959finite}, and spectral methods \cite{feit1982solution}, have significantly evolved. However, the computational demands of these methods increase considerably with finer discretization grids, and the direct discretization of high-dimensional parametric spaces becomes computationally intensive.

Neural network-based approaches, in contrast, do not face such drastic increases in computational complexity with more extensive sampling within the solution domain. This advantage has positioned neural networks as a vibrant area of research for solving PDEs. Lagaris et al. \cite{Lagaris1998} introduced an Artificial Neural Network (ANN) strategy that constructs loss functions that adhere to boundary and initial conditions for solving Ordinary Differential Equations (ODEs) and PDEs. Raissi et al. \cite{raissi2019physics} developed physics-informed neural networks (PINNs), which utilize known physics laws to guide the learning process in solving forward and inverse problems related to nonlinear PDEs. Further developments include Ppinn \cite{Meng2020} and fpinns \cite{Pang2019}, which build on the PINN framework. Other innovative methods combine traditional numerical approaches with neural networks, such as the Deep Collocation Method (DCM) \cite{raissi2019physics}, the Deep Ritz Method (DRM) \cite{yu2018deep}, and the Deep Galerkin Method (DGM) \cite{sirignano2018dgm}. These neural network-based techniques have shown promising results, indicating their potential to revolutionize computational methods for solving PDEs across various fields.

Recently, the innovative concept of neural operators has emerged, where neural networks are employed to learn operators directly from data, without requiring any prior knowledge of the underlying physics such as PDEs \cite{li2020,luludeeponet,li2021fno}. Neural operators represent a significant advancement over traditional neural networks by possessing the ability to learn mappings between function spaces. This new capability allows them to tackle a broad spectrum of complex and abstract problems, which were previously difficult or impossible to address using conventional neural network methods. By learning the mapping from functionally dependent parameters to their solutions, neural operators are adept at handling intricate challenges often found in fields like PDEs. This ability not only enhances their versatility but also establishes neural operators as a potent tool for modeling and solving sophisticated mathematical problems. Their impact is expected to extend across various fields, including science, engineering, and technology, revolutionizing how complex systems are understood and analyzed.

Lu Lu et al. \cite{luludeeponet}, drawing inspiration from the universal approximation theorem \cite{chen1995universal} for nonlinear operators, introduced the neural operator known as DeepONet. This model has demonstrated significant success in learning and approximating nonlinear operators. Building on this, Jin et al. \cite{jin2022} further developed the concept, extending DeepONet into a more versatile nonlinear operator model capable of learning from multiple inputs, broadening its applicability.

In a parallel advancement,  Li et al. \cite{li2021fno} developed the Fourier Neural Operator (FNO), an architecture that operates directly in Fourier space. This method leverages the Fourier transform to shift input functions from their physical space to the frequency domain. Here, a series of learnable linear transformations coupled with nonlinear activation functions effectively capture the diverse frequency characteristics of the input functions. The transformations enable the FNO to discern and react to the various features of the inputs, subsequently generating appropriate output functions. These are then converted back to the physical space using the inverse Fourier transform.

Neural operators are highly regarded for their capacity to be trained on large datasets without needing a deep comprehension of the underlying physics, along with the swift inference capabilities of the trained models \cite{Kamyar2024}. However, one major challenge is that changes in the solution domain for a specific class of PDEs often require the retraining of neural operators. This retraining can be time-consuming and may exceed the time needed for traditional numerical methods, thereby not fully harnessing the potential of neural operators and adding extra overhead.

While transfer learning \cite{goswami2022} has somewhat reduced the time required for retraining operators, it still necessitates the transfer training of network parameters within the specified tasks. Despite researchers utilizing various techniques that enable operator learning methods such as DeepONet to train and infer solutions to PDEs in similar geometries \cite{he2024}, this approach is often inadequate. This methodology requires substantial training data encompassing the internal coordinates of geometries and necessitates repeated sampling in overlapping regions of different geometries, which invariably increases the training burden on the network. An alternative approach involves constructing body-fitted grids \cite{thompson1985numerical} for the geometries, which are then transformed via coordinate transformations to a fixed domain, somewhat addressing the issues related to geometric changes necessitating network retraining. However, the design and computation of body-fitted grids pose considerable challenges, especially when dealing with complex geometric shapes.


To maximize the benefits of neural operators and address the challenges of domain variability, this article introduces an operator learning framework based on Boundary Integral Equations (BIEs). This method optimizes the use of neural operators by providing a versatile and efficient solution framework that eliminates the need for retraining with changes in the domain. This significant enhancement greatly improves the practical utility and effectiveness of neural operator models in solving complex PDEs, enhancing their adaptability and efficiency across various computational environments.

Solutions to PDEs relevant to diverse fields such as fluid mechanics, elastostatics, and acoustic scattering can be represented through layer potential integrals of the layer potential that satisfy BIEs \cite{brebbia2012boundary, Aliabadi2003, atkinson_1997, Kress2014, colton1998inverse,Ludvig2020,Jiang2012}. The extensive practicality of BIE-based methods provides a solid foundation for developing operator learning frameworks that leverage these equations. Utilizing BIEs to solve PDEs offers several advantages. First, it shifts the focus from the domain's interior to its boundary, effectively reducing the problem's dimensionality by one. This reduction simplifies many complex PDE problems, especially those involving intricate domain geometries. Second, BIEs remove the need for domain discretization into a mesh, thereby avoiding issues associated with mesh generation in domains with complex geometries. Moreover, significant advances in fast solution algorithms for BIEs over the past decade \cite{jiang2010,Jiang2014,jiang2018} have made this approach particularly potent. However, the BIE method involves challenges such as handling kernel functions that often exhibit singularities or require the computation of oscillatory integrals, necessitating meticulous and robust algorithmic strategies. A crucial advantage of the BIE approach is that once the boundary and functions defined on it are parameterized, the solution domain remains fixed, irrespective of boundary changes. This feature allows neural operators to effectively solve PDEs across varying domains without the need for retraining.

Building on this advantage, we introduce the Boundary Integral Type Deep Operator Network (BI-DeepONet), which merges the capabilities of DeepONet with BIEs. Additionally, we have developed a novel operator learning model, the Boundary Integral Trigonometric Deep Operator Neural Network (BI-TDONet), which is predicated on the Singular Value Decomposition (SVD) of bounded linear operators. By meticulously aligning each segment of BI-TDONet with the respective components of the SVD, we underscore the structural soundness of BI-TDONet. To further tackle the challenges posed by oscillatory or impulsive signals, we model parameterized equations as trigonometric series, utilizing trigonometric coefficients as both inputs and outputs for BI-TDONet. This method not only maintains the periodicity of the functions but also enhances the capture of oscillatory or impulsive signal features, thus enabling BI-TDONet to achieve quicker convergence and heightened precision. Our numerical experiments confirm that both BI-DeepONet and BI-TDONet are capable of efficiently solving PDEs across diverse domains, with BI-TDONet demonstrating enhanced convergence speed and accuracy. Moreover, within our proposed framework, the average inference time for BIEs is notably swift, ranging from approximately \(10^{-5}\) to \(10^{-6}\) seconds. Post solving the BIE, the process to compute the PDE solution involves simply calculating potential integrals, substantially lowering the computational demands. Furthermore, the precision of potential integral calculations has seen significant advancements with the development of sophisticated and accurate fast algorithms such as Quadrature by Expansion (QBX), Quadrature by Two Expansion (QB2X), and Fast Multipole Methods (FMM) \cite{Andreas2013,Ding2021,chen2023}.

For partial differential equations (PDEs) that include time components or nonlinearities, it is feasible to linearize such equations using methods like explicit-implicit methods or the Rothe's method \cite{Mary2011,Andrew1978,Ascher1995,ascher1997implicit}. Consequently, this article exclusively investigates time-independent linear elliptic PDEs. The structure of the article is as follows: Section 2 revisits and summarizes potential representations of solutions to elliptic PDEs and redefines the Boundary Integral Operator (BIO). Section 3 provides an in-depth introduction to BI-DeepONet and explores the operator learning model BI-TDONet, which leverages the Singular Value Expansion (SVE) from the inverse operators derived from second-kind BIEs. This section also comprehensively describes the methods used to generate data necessary for model training and testing. Section 4 delivers several numerical examples, including scenarios related to Laplace Boundary Value Problems (LBVP), potential flow, elastostatics, and acoustic scattering. Finally, Section 5 concludes with a summary and future research directions.
\section{Preliminary}

Integral equations are extensively researched and used to solve boundary value problems related to elliptic PDEs. It is a recognized fact that solutions to boundary value problems for elliptic PDEs can be expressed as linear combinations of single-layer and double-layer potentials. For example, given a simply connected bounded domain $D\in\mathbb{R}^2$ with a twice continuously differentiable boundary $\partial D\in C^2$, the solution of Laplace boundary value problem (LBVP) with interior Dirichlet boundary condition
\begin{equation}\label{IDP}
	\left\{
	\begin{aligned}
		\Delta u(\boldsymbol{x})=0\qquad \boldsymbol{x}&\in D,\\
		u(\boldsymbol{x})=f(\boldsymbol{x}) \qquad \boldsymbol{x}&\in\partial  D,
	\end{aligned}
	\right.
\end{equation}
can be expressed as the layer potential integral
\begin{equation*}
	u(\boldsymbol{x})=\int_{\partial D}\varphi(\boldsymbol{y})\frac{\partial \Phi(\boldsymbol{x},\boldsymbol{y})}{\partial \nu(\boldsymbol{y})}ds(\boldsymbol{y}),\qquad \boldsymbol{x}\in \mathbb{R}^2\backslash \partial  D,
\end{equation*}
where $\Phi$ is fundamental solution,  $\partial \nu(\boldsymbol{y})$ denotes the outward unit normal vector at point $\boldsymbol{y}$, and $\varphi$, called the layer potential, is obtained by the BIE of the second kind
\begin{equation}\label{BIE}
	\varphi(\boldsymbol{x})-\int_{\partial D}\varphi(\boldsymbol{y})\frac{\partial \Phi(\boldsymbol{x},\boldsymbol{y})}{\partial \nu(\boldsymbol{y})}ds(\boldsymbol{y})=-2f(\boldsymbol{x}),\qquad \boldsymbol{x}\in \partial D.
\end{equation}

In fact, as we will review next, the solutions to both the interior and exterior boundary value problems of the Laplace equation can be expressed as layer potential integrals \cite{atkinson_1997, Kress2014}. For simplicity, we denote both Dirichlet boundary condition (DBC) and Neumann boundary condition (NBC) using the function $f$. In other words, under DBC, we have
$$
    u(\boldsymbol{x})=f(\boldsymbol{x}) \quad \boldsymbol{x}\in \partial D,
$$
and for NBC, it is formulated as
$$
	\frac{\partial u(\boldsymbol{x})}{\partial \nu(\boldsymbol{x})}=f(\boldsymbol{x}) \quad \boldsymbol{x}\in \partial D.
$$
We define the integral operators  $\mathcal{S},\mathcal{D}:L^2(\partial D)\to L^2(\partial D)$ as follows
\begin{equation}\label{single-layer potential}
	\mathcal{S}(\varphi)(\boldsymbol{x}):=\int_{\partial D}\varphi(\boldsymbol{y}) \Phi(\boldsymbol{x},\boldsymbol{y})ds(\boldsymbol{y}),\qquad \boldsymbol{x}\in \partial D
\end{equation}
and
\begin{equation}\label{double-layer potential}
	\mathcal{D}(\varphi)(\boldsymbol{x}):=\int_{\partial D}\varphi(\boldsymbol{y})\frac{\partial \Phi(\boldsymbol{x},\boldsymbol{y})}{\partial \nu(\boldsymbol{y})}ds(\boldsymbol{y}),\qquad \boldsymbol{x}\in \partial D.
\end{equation}
These are called, respectively, single-layer and double-layer potential with layer potential $\varphi$ defined on $\partial D$. We also define the operators $\mathcal{D^\prime}$ and $\mathcal{\widetilde{D}}:L^2(\partial D)\to L^2(\partial D)$ as follows
\begin{equation}\label{Dtild}
	\mathcal{D}^\prime(\varphi)(\boldsymbol{x}):=\int_{\partial D}\varphi(\boldsymbol{y})\frac{\partial \Phi(\boldsymbol{x},\boldsymbol{y})}{\partial \nu(\boldsymbol{x})}ds(\boldsymbol{y}),\qquad \boldsymbol{x}\in \partial D
\end{equation}
and
\begin{equation}\label{modified double-layer potential}
	\widetilde{\mathcal{D}}(\varphi)(\boldsymbol{x}):=\int_{\partial D}\varphi(\boldsymbol{y})\left(\frac{\partial \Phi(\boldsymbol{x},\boldsymbol{y})}{\partial \nu(\boldsymbol{y})}+1\right)ds(\boldsymbol{y}),\qquad \boldsymbol{x}\in \partial D.
\end{equation}
The relationships between the layer potentials and the solutions of the LBVPs are provided in the first column of Table \ref{table_layer list}. The second column of this table lists the BIEs that the potential density function layer potential $\varphi$ satisfies. More details can be found in \cite{Kress2014}.
\begin{table}[htb]
	\centering
	\caption{Layer potentials estimated for the LBVPs.}
	\label{table_layer list}
	
	\begin{tabular}{@{}l|l@{}}
		\midrule
		LBVP & BIE \\ \midrule
		
		interior problem with NBC, $u=\mathcal{S}\varphi $ &$\left(\mathcal{I} + 2\mathcal{D}^\prime\right)\varphi = 2f$
		\\
		
		interior problem with DBC, $u=\mathcal{D}\varphi$  & $\left(\mathcal{I} -2 \mathcal{D}\right)\varphi = -2f$ \\
		
		exterior problem with NBC, $u=\mathcal{S}\varphi$ &
		$\left(\mathcal{I}-2\mathcal{D}^\prime\right)\varphi = -2f$ \\
		
		exterior problem with DBC,  $u=\widetilde{\mathcal{D}}\varphi$ &$\left(\mathcal{I} + 2\widetilde{\mathcal{D}}\right)\varphi = 2f$\\
		
		\midrule
	\end{tabular}	
\end{table}

In this paper, we are not solely concerned with the boundary value problems of the Laplace equation and their corresponding BIEs discussed above. In fact, many other boundary value problems also have solutions that can be expressed as potential integrals. Having established the BIEs for solving boundary value problems, which utilize layer potentials and integral operators defined on a given boundary, we now transition to a more nuanced exploration.
From the definitions of operators $\mathcal{S}$, $\mathcal{D}$, $\mathcal{D}'$ and $\widetilde{\mathcal{D}}$, we recognize that the integral operators themselves are intimately linked to the nature of the boundary. We move towards a conceptual framework where the boundary is not just a passive parameter, but an active component of the integral operators. This shift in perspective leads to the formulation of a more dynamic integral equation, which incorporates the boundary as a variable, paving the way for more flexible and comprehensive solutions to complex problems.

For the sake of subsequent discussion, we define boundary integral operator $\mathcal{K}:L^2(\partial D)\to L^2(\partial D)$ by
\begin{equation}\label{BIO}
	(\mathcal{K}(\varphi))(\boldsymbol{x})=\int_{\partial D}\varphi(\boldsymbol{y})K(\boldsymbol{x},\boldsymbol{y})ds(\boldsymbol{y}),
\end{equation}
where $K$ is a kernel function defined on $\partial D \otimes \partial D$, and $\otimes$ denotes the Cartesian product. Using this representation, all second kind BIEs can be rewritten as the equation of the operator
\begin{equation}\label{BIO equation}
	(\mathcal{I-K})\varphi=\widetilde{f},
\end{equation}
where $\mathcal{I}$ denote the identity operator, $\widetilde{f}$ denote the right-hand side of the BIEs, and $\varphi$ is unknown. For a fixed boundary, there are many numerical methods \cite{atkinson_1997} for solving operator equation $\eqref{BIO equation}$. In many application scenarios, it is necessary to solve the same boundary value problem for elliptic PDEs defined on different domains. In this case, these boundary value problems with different boundaries correspond to different integral operators $\mathcal{K}$, which increases computational time during numerical discretization. From the form of \eqref{BIO}, it can also be seen that the defined integral operator is related not only to the potential density function $\varphi$ but also to the boundary $\partial D$. This motivates us to consider the form of boundary integral operators (BIO) introduced below,  that treats the boundary $\partial D$ as an independent input.

Let $J:=\{\boldsymbol{\gamma}\vert \boldsymbol{\gamma}\:\mathrm{is\:simple\:closed\:curve},\boldsymbol{\gamma}\in C^2\}$. For convenience, we use the symbol $\boldsymbol{\gamma}$ to represent the parametric equation of the curve $\boldsymbol{\gamma}$ defined on the interval $I:=[0, 2\pi]$, and use $\varphi$ and $\widetilde{f}$ to represent $\varphi\circ \boldsymbol{\gamma}$ and $\widetilde{f}\circ \boldsymbol{\gamma}$, respectively. We define integral operator $\mathcal{H}: J\times L^2(I)\to L^2(I)$  as
\begin{equation}\label{BIO_2}(\mathcal{H}(\boldsymbol{\gamma},\varphi))(t)=\int_{I}\varphi(s)K(\boldsymbol{\gamma}(t),\boldsymbol{\gamma}(s))\sqrt{(\gamma'_1)^2(s)+(\gamma'_2)^2(s)}ds,
\end{equation}
where $\boldsymbol{\gamma}:=(\gamma_1,\gamma_2)$.
It is evident that $\mathcal{H}$ is a nonlinear operator. For each fixed boundary $\boldsymbol{\gamma}$, it becomes the linear operator $\mathcal{K}$ defined by \eqref{BIO}. Then the equation $\eqref{BIO equation}$ can be rewritten as
\begin{equation}\label{operator equation}
(\mathcal{I}-\mathcal{H}(\boldsymbol{\gamma},\cdot))\varphi=\widetilde{f}.
\end{equation}
It is well known \cite{Kress2014} that for every fixed boundary $\boldsymbol{\gamma}$, $\mathcal{H}(\boldsymbol{\gamma},\cdot)$ is a compact operator, and $\varphi$ can be obtained from the following equation
\begin{equation}\label{BIO equation_2}
\varphi=(\mathcal{I}-\mathcal{H}(\boldsymbol{\gamma},\cdot))^{-1}\widetilde{f}.
\end{equation}

When the boundary is given, we can use classical numerical methods to represent equation \eqref{operator equation} as a discrete linear system, such as the boundary element method, Nystr\"{o}m method, collocation method, and spectral method \cite{atkinson_1997}. These systems can then be solved using direct or iterative methods. However, in many scenarios, the boundary is not fixed, such as in fluid-structure interaction problems. In such cases, if we still use classical numerical methods, we often need to frequently update the coefficients of the discretized linear system. This increases the computational complexity of the algorithm, necessitating more flexible approaches. Fortunately, operator learning methods, which are well-developed \cite{luludeeponet,jin2022,li2021fno}, excel in such environments due to their computational efficiency and the ability for dealing with nonlinear operators like $(\mathcal{I}-\mathcal{H}(\boldsymbol{\gamma},\cdot))^{-1}$
in equation \eqref{BIO equation_2}. These methods are skilled at learning from data and can generalize to address a wide spectrum of problems with variable boundaries without the need for retraining. This capability makes them particularly suitable for solving BIEs with changing boundaries as shown in \eqref{BIO equation_2}.

Although our discussions in this section and subsequent ones primarily focus on the second kind BIE $\eqref{BIO equation_2}$, it is recognized that solutions to elliptic PDEs, such as LBVPs, can be expressed through layer potential integrals of layer potential $\varphi$ that satisfy first kind BIEs \cite{atkinson_1997}. The experiments detailed in Section 4 will showcase that our proposed operator learning models exhibit remarkable capabilities in learning the  BIOs for the first kind BIEs.

Additionally, when solving PDEs using the boundary integral equation method (BIEM), such as LBVPs introduced at the beginning of this section, computing potential integrals is also necessary. Currently, numerous fast algorithms are available for these potential integral calculations, marking a significant advancement over traditional techniques. Therefore, the primary goal of this article is to develop an operator learning model for solving BIEs and facilitate efficient solutions to boundary value problems related to elliptic PDEs.

\section{Optimizing Operator Learning Frameworks Through the Integration of Boundary Integral Equations}

To clarify the expression, we denote$$\overline{\mathcal{H}}(\boldsymbol{\gamma},\widetilde{f}):=(\mathcal{I-H(\boldsymbol{\gamma},\cdot)})^{-1}\widetilde{f},$$ and subsequently rewrite equation \eqref{BIO equation_2} as follows
\begin{equation}
	\begin{aligned}\label{compose BIO equation_2}
			\varphi=\overline{\mathcal{H}}(\boldsymbol{\gamma},\widetilde{f}).
	\end{aligned}
\end{equation}	
In this section, we initially demonstrate the application of the DeepONet framework to learn the nonlinear operator $\overline{\mathcal{H}}$. Then, we propose a novel framework for operator learning, aimed at modeling $\overline{\mathcal{H}}$, which is detailed in the second subsection. Lastly, we outline the methodology employed for dataset generation used in this article.

\subsection{Boundary integral type DeepOnet (BI-DeepONet)}

DeepONet is a neural network architecture designed for learning operators in the context of solving PDEs or other equations. It consists of two primary components: the trunk net and the branch net. The trunk net is responsible for encoding the points in the domain at which the outputs of the operators are evaluated. It processes these domain points to extract relevant features, which are necessary for the network to effectively learn the mapping defined by the operators. The branch net is specifically designed to process the input functions or parameters associated with the problem. For example, in the context of solving PDEs, these inputs might include initial conditions, boundary conditions, or any other parameterized function that defines the problem. Just like the trunk net, the branch net processes its inputs to extract relevant features. However, these features are related to the functional aspects of the problem rather than the spatial or domain aspects. It learns to understand how these input functions influence the output of the operator.

\begin{figure}[htbp]
		\centering
		\includegraphics[width=0.7\linewidth]{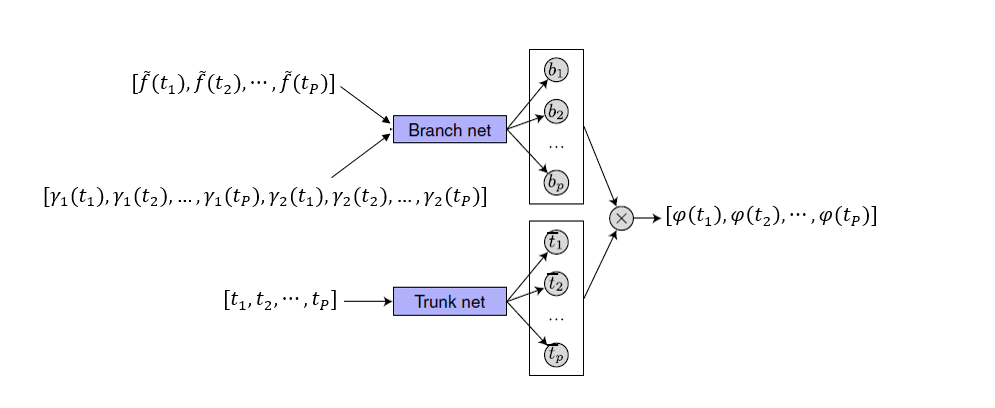}
		\caption{The network structure of BI-DeepONet.}
		\label{BI-DeepONet}
\end{figure}

Therefore, when using DeepONet to learn the operator $\overline{\mathcal{H}}$ in equation \eqref{compose BIO equation_2}, we design the input of the trunk net as uniformly sampled points $[t_1, t_2, \ldots, t_p]$ on the interval $I$, where $p\in\mathbb{N}$, and the input of the branch net as the sample of the boundary parameter function $\boldsymbol{\gamma}$ and the right-hand side function $\widetilde{f}$ at the sampled points $[t_1, t_2, \ldots, t_p]$. We call the DeepONet for learning $\overline{\mathcal{H}}$ as boundary integral type DeepOnet (BI-DeepONet). The structure of BI-DeepONet show as in Figure $\ref{BI-DeepONet}$.

BI-DeepONet utilizes the DeepONet framework to model the nonlinear operator $\overline{\mathcal{H}}$. This approach, while innovative, faces two significant challenges. The primary concern is that employing sampling points on the interval $[0,2\pi]$ may fail to capture the high-frequency components of the input function adequately. A possible remedy is to integrate Fourier feature networks (see \cite{zhu2023}) into the inputs of both the branch net and the trunk net, which is aimed at bolstering the model's capability to extract and utilize high-frequency information. However, even with this enhancement, there remains a necessity to increase the number of sampling points to improve accuracy. Moreover, attaching Fourier features, while beneficial for handling high frequencies, does not inherently ensure that the model's output will maintain the essential periodicity of the operator. This aspect needs careful consideration in the model's design and training strategy to ensure that the periodic nature of the problem is accurately reflected in the outputs.

\subsection{Boundary Integral Trigonometric Deep Operator Network (BI-TDONet)}
In this article, we tackle two significant challenges encountered in BI-DeepONet by adopting trigonometric series representations for $\boldsymbol{\gamma}$ and $\widetilde{f}$, and utilizing their coefficients as inputs for the branch net. This strategy not only facilitates the capture of high-frequency information more effectively but also ensures that the outputs, expressed as coefficients of trigonometric series, maintain the periodicity required for accurately approximating the left-hand side term $\varphi$ in \eqref{compose BIO equation_2}. We avoid using Fourier coefficients despite their relevance because they involve complex numbers, which would complicate the training of neural networks due to the additional computational overhead.

Furthermore, these modifications in the model's input-output structure have driven us to consider a new architectural direction for DeepONet. Originally designed based on the universal approximation theorem, which presumes function continuity, DeepONet struggles with the ``discontinuity" challenges presented by the frequency domain representations of $\boldsymbol{\gamma}$ and $\widetilde{f}$. To address these issues more effectively, we are exploring the development of a new network architecture optimized for using the coefficients of trigonometric series as inputs, aiming to enhance precision and adaptability in modeling nonlinear operators.

Next, after reviewing the classic Fourier-Galerkin method, we will introduce the operator learning framework discussed in this subsection. Let $L^2(I)$ be the space of square integrable functions on $I$ equipped with the inner product $\langle\cdot,\cdot\rangle$, defined by
\begin{equation*}
	\langle\phi,\psi\rangle:=\int_{I} \phi(t) \overline{\psi}(t)dt, \quad \phi, \psi\in L^2(I).
\end{equation*}
For each $n\in \mathbb{N}$, let $\mathbb{Z}_n:=\{1,2,\cdots,n\}$, and $X_n$ be the finite dimensional subspace generated by the basis $\{\sin(k\cdot),\cos(k\cdot): k\in\mathbb{Z}_n\}\cup\{1\}$. Also let $\mathcal{P}_n$ be the orthogonal projection operator of $L^2(I)$ onto $X_n$, which is expressed as
\begin{equation}\label{trigonometic}
(\mathcal{P}_n \phi)(t) = \phi_0+ \sum_{k \in \mathbb{Z}_n} (\phi^c_k \cos(kt)+\phi^s_k \sin(kt)),\quad t\in I,
\end{equation}
for all $\phi \in L^2(I)$. Here, $\phi_0, \phi^c_k$ and $\phi^s_k$ are the coefficients of the trigonometric series of function $\phi$, defined by
\begin{equation}\label{fourier transform}
	\phi_0:=\frac{1}{2\pi}\int_{I} \phi(s)ds,~~ \phi^c_k:=\frac{1}{\pi}\int_{I} \phi(s) \cos(ks)ds,\quad{~\rm and~}\quad \phi^s_k:=\frac{1}{\pi}\int_{I} \phi(s) \sin(ks)ds.
\end{equation}
Therefore, when applying the Fourier-Galerkin method to solve \eqref{operator equation}, we seek $\varphi_n\in X_n$ such that
\begin{equation}\label{finite BIO0}
	\mathcal{P}_n(\mathcal{I}-\mathcal{H}(\boldsymbol{\gamma},\cdot))\varphi_n=\mathcal{P}_n \widetilde{f}.
\end{equation}

To facilitate the introduction of the operator learning framework BI-TDONet, we denote $\boldsymbol{\varphi}_n$ and ${\mathbf{f}}_n$ as the vectors composed of coefficients of the trigonometric series of  $\varphi_n$ and $\mathcal{P}_n\widetilde{f}$, respectively, i.e.,
$\boldsymbol{\varphi}_n := [(\varphi_n)_0, (\varphi_n)_1^c, (\varphi_n)_2^c,\ldots,
(\varphi_n)_n^c,
(\varphi_n)_1^s, (\varphi_n)_2^s,\ldots,   (\varphi_n)_n^s]$ and
${\mathbf{f}}_n := [\widetilde{f}_0, \widetilde{f}_1^c, \widetilde{f}_2^c,\ldots,
\widetilde{f}_n^c,
\widetilde{f}_1^s,
\widetilde{f}_2^s, \ldots,  \widetilde{f}_n^s]$.
We also define $$
\boldsymbol{\gamma}_n := [(\gamma_1)_0,  (\gamma_1)_1^c, (\gamma_1)_2^c,\ldots,
(\gamma_1)_n^c,
(\gamma_1)_1^s, (\gamma_1)_2^s,\ldots,
(\gamma_1)_n^s,
(\gamma_2)_0,
(\gamma_2)_1^c,
(\gamma_2)_2^c,\ldots,
(\gamma_2)_n^c,
(\gamma_2)_1^s, (\gamma_2)_2^s,\ldots, (\gamma_2)_n^s],
$$
which is consisted by the coefficients of the trigonometric series of  $\boldsymbol{\gamma}$.
For ease of implementation, we truncate $\boldsymbol{\gamma}$, using only the trigonometric series coefficients in $\boldsymbol{\gamma}_n$ to approximate $\boldsymbol{\gamma}$. This means that $\boldsymbol{\varphi}_n$, $\mathbf{f}_n$, and $\boldsymbol{\gamma}_n$ satisfy the following equation
\begin{equation}\label{finite BIO01}	\left(\mathbf{I}_n-\frac{1}{\pi}\mathbf{H}_n({\boldsymbol{\gamma}}_n)\right)\boldsymbol{\varphi}_n=\mathbf{f}_n,
\end{equation}
where $\mathbf{I}_n$ is the identity matrix of size $n$, $\mathbf{H}_n({\boldsymbol{\gamma}}_n)$ is a Fourier-Galerkin matrix, the details of which will be described in Subsection 3.3.

As mentioned earlier, when $\boldsymbol{\gamma}\in J$ is given, the operator $\overline{\mathcal{H}}(\boldsymbol{\gamma},\cdot):L^2(I)\to L^2(I)$ is a bounded linear operator. Thus, it can be expanded into the form of the singular value expansion (SVE) by following Lemma in \cite{crane2020}, which is reviewed as follows.
\begin{Lemma}
	Let $X$ and $Y$ be real separable Hilbert spaces and let $T : X \to Y$ be a bounded linear operator. Then there
	exist a Borel space $(M,\mathcal{A}, \mu)$, isometries $V:L^2(M,\mathcal{A}, \mu)\to X$, $U:L^2(M,\mathcal{A}, \mu)\to Y$,
	and an essentially bounded measurable function $\sigma:M \to \mathbb{R}$ such that
	$$T = Um_{\sigma}V^{-1},$$
	where $V^{-1}$ is the generalized inverse of $V$ and $m_{\sigma}$ is the multiplication operator defined by
	\begin{equation}\label{eq_SVE}
		\begin{aligned}
			m_{\sigma}&:L^2(M,\mathcal{A}, \mu)\to L^2(M,\mathcal{A}, \mu),\\
			m_{\sigma}(g)&=\sigma g, \quad {\rm for~all}~ g\in L^2(M,\mathcal{A}, \mu).
		\end{aligned}
	\end{equation}
	Moreover, $\sigma>0$ a.e.
\end{Lemma}

For given $\boldsymbol{\gamma}\in J$, operator $\overline{\mathcal{H}}(\boldsymbol{\gamma},\cdot)$ is a bounded linear operator by the Fredholm theorem (see \cite{Kress2014}). Hence, based on the above lemma, $\overline{\mathcal{H}}(\boldsymbol{\gamma},\cdot)$ can be represented in the form of the SVE \eqref{eq_SVE} where the isometries $V$ and $U$, and the essentially bounded measurable function $\sigma$ do all dependent on $\boldsymbol{\gamma}$. These are denoted as $V(\boldsymbol{\gamma})$, $ U(\boldsymbol{\gamma})$ and $\sigma(\boldsymbol{\gamma})$, respectively. The operator $\overline{\mathcal{H}}(\boldsymbol{\gamma},\widetilde{f})$ can thus be expressed as
\begin{equation}\label{SVE of compact}	
\overline{\mathcal{H}}(\boldsymbol{\gamma},\widetilde{f})=U(\boldsymbol{\gamma})m_{\sigma(\boldsymbol{\gamma})}V(\boldsymbol{\gamma})^{-1}\widetilde{f}.
\end{equation}
Inspired by the structure of \eqref{SVE of compact}, we propose approximate operator $\overline{\mathcal{H}}$ using  the following neural network architecture,
\begin{equation}\label{approx KBK1}
\overline{\mathcal{H}}(\boldsymbol{\gamma},\widetilde{f})\approx\mathcal{N}\left(\boldsymbol{\theta}_U,\left[ \boldsymbol{\gamma}_n,\left(\mathcal{N}\left( \boldsymbol{\theta}_V,[\boldsymbol{\gamma}_n,{\mathbf{f}}_n]\right)\odot\mathcal{N}\left( \boldsymbol{\theta}_\sigma,\boldsymbol{\gamma}_n\right)   \right)\right]  \right) ,
\end{equation}
where $\mathcal{N}(\boldsymbol{\theta},\cdot)$ represents a fully connected neural network (FNN) with parameters $\boldsymbol{\theta}$, and ``$\odot$" indicates the Hadamard product.

\begin{figure}[H]
	\centering
	\includegraphics[width=0.6\linewidth]{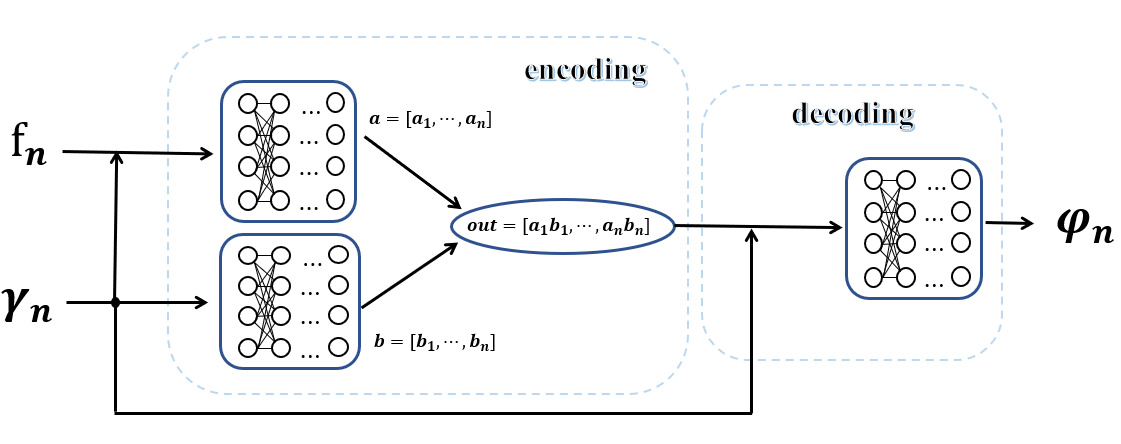}
	\caption{The network structure of BI-TDONet.}
	\label{BITDONet structure}
\end{figure}

In equation \eqref{approx KBK1}, $\mathcal{N}\left(\boldsymbol{\theta}_\sigma,\cdot\right)$ serves as an approximation for $m_{\sigma(\cdot)}$ from equation \eqref{SVE of compact}, dependent solely on $\boldsymbol{\gamma}_n$. Additionally, $\mathcal{N}\left( \boldsymbol{\theta}_V, [\cdot, \diamond]\right)$ approximates the mapping $V(\cdot)^{-1}\diamond$ from equation \eqref{SVE of compact}, influenced by both ${\boldsymbol{\gamma}}_n$ and ${\mathbf{f}}_n$. Both networks, $\mathcal{N}\left( \boldsymbol{\theta}_\sigma,\cdot\right)$ and $\mathcal{N}\left( \boldsymbol{\theta}_V,\cdot\right)$, can be considered as encoding processes. The FNN $\mathcal{N}(\boldsymbol{\theta}_U, [\cdot,\diamond])$, which learns the mapping $U(\cdot)\diamond$ from \eqref{SVE of compact}, is treated as the decoding process. We refer to this architecture as BI-TDONet. The schematic diagram of the network architecture is illustrated in Figure $\ref{BITDONet structure}$.

\subsection{Generate datasets}

This subsection provides a detailed explanation of the method of generating the dataset used to train the neural network discussed in this article.
Specifically, we will show how to generate the dataset ${(\boldsymbol{\varphi}_n,\mathbf{f}_n, \boldsymbol{\gamma}_n)}$. Here, $\boldsymbol{\varphi}_n$, $\mathbf{f}_n$ and $\boldsymbol{\gamma}_n$ satisfy \eqref{finite BIO01}, and the matrix $\mathbf{H}_n(\boldsymbol{\gamma}_n)$ in \eqref{finite BIO01} has the following form
\begin{align*}
	\footnotesize
	\mathbf{H}_n({\boldsymbol{\gamma}}_n):=
	\begin{bmatrix}
		\mathbf{H}_n^{1,1}({\boldsymbol{\gamma}}_n)&\mathbf{H}_n^{c,1}({\boldsymbol{\gamma}}_n)&\mathbf{H}_n^{s,1}({\boldsymbol{\gamma}}_n)\\
		\mathbf{H}_n^{1,c}({\boldsymbol{\gamma}}_n)&\mathbf{H}_n^{c,c}({\boldsymbol{\gamma}}_n)&\mathbf{H}_n^{s,c}({\boldsymbol{\gamma}}_n)\\
\mathbf{H}_n^{1,s}({\boldsymbol{\gamma}}_n)&\mathbf{H}_n^{c,s}({\boldsymbol{\gamma}}_n)&\mathbf{H}_n^{s,s}({\boldsymbol{\gamma}}_n)
	\end{bmatrix},
\end{align*}
where
\begin{align*}
	\mathbf{H}_n^{1,1}({\boldsymbol{\gamma}}_n):&=\frac{1}{2}\left\langle  {\mathcal{H}}({\boldsymbol{\gamma}}_n,1),1\right\rangle ,\\
	 \mathbf{H}_n^{c,1}({\boldsymbol{\gamma}}_n):&=\frac{1}{2}\left[ \left\langle  {\mathcal{H}}({\boldsymbol{\gamma}}_n,\cos(k\cdot)),1\right\rangle :k\in\mathbb{Z}_n\right] ,\\ \mathbf{H}_n^{s,1}({\boldsymbol{\gamma}}_n):&=\frac{1}{2}\left[ \left\langle  {\mathcal{H}}({\boldsymbol{\gamma}}_n,\sin(k\cdot)),1\right\rangle :k\in\mathbb{Z}_n\right] ,\\
	\mathbf{H}_n^{1,c}({\boldsymbol{\gamma}}_n):&=\left[ \left\langle  {\mathcal{H}}({\boldsymbol{\gamma}}_n,1),\cos(k\cdot)\right\rangle :k\in\mathbb{Z}_n\right]^\top ,\\
	\mathbf{H}_n^{1,s}({\boldsymbol{\gamma}}_n):&=\left[ \left\langle  {\mathcal{H}}({\boldsymbol{\gamma}}_n,1),\sin(k\cdot)\right\rangle :k\in\mathbb{Z}_n\right]^\top ,\\
	\mathbf{H}_n^{c,c}({\boldsymbol{\gamma}}_n):&=\left[ \left\langle  {\mathcal{H}}({\boldsymbol{\gamma}}_n,\cos(k_1\cdot)),\cos(k_2\cdot)\right\rangle :k_1,k_2\in\mathbb{Z}_n\right],\\	
\mathbf{H}_n^{c,s}({\boldsymbol{\gamma}}_n):&=\left[ \left\langle  {\mathcal{H}}({\boldsymbol{\gamma}}_n,\cos(k_1\cdot)),\sin(k_2\cdot)\right\rangle :k_1,k_2\in\mathbb{Z}_n\right],\\
	\mathbf{H}_n^{s,c}({\boldsymbol{\gamma}}_n):&=\left[ \left\langle  {\mathcal{H}}({\boldsymbol{\gamma}}_n,\sin(k_1\cdot)),\cos(k_2\cdot)\right\rangle :k_1,k_2\in\mathbb{Z}_n\right],\\	
\mathbf{H}_n^{s,s}({\boldsymbol{\gamma}}_n):&=\left[ \left\langle  {\mathcal{H}}({\boldsymbol{\gamma}}_n,\sin(k_1\cdot)),\sin(k_2\cdot)\right\rangle :k_1,k_2\in\mathbb{Z}_n\right].
\end{align*}

To generate the dataset used in this article, we generate $\boldsymbol{\gamma}_n$ and $\boldsymbol{\varphi}_n$. Then, we compute $\mathbf{f}_n$ using equation \eqref{finite BIO01}. Here, the values of $(\gamma_1)_0, (\gamma_2)_0$, and $\varphi_0$ are obtained by sampling the standard normal distribution. The values of $(\gamma_1)_k^c$, $(\gamma_2)_k^c$, $(\gamma_1)_k^s$, and $(\gamma_2)_k^s$, $1\leq k\leq n$, are obtained by sampling the Gaussian distribution with a mean of $0$ and a variance of $(\rho^{k-1})^2$, where $0<\rho<1$ denotes the rate of decay of the coefficients. Additionally, $\varphi_k^c$ and $\varphi_k^s$ for $1 \leq k \leq n$ are derived by sampling from a Gaussian distribution with a mean of $0$ and a variance of $k^{-m}$.

\begin{figure}[h]
	\centering
	\begin{subfigure}{0.4\textwidth}
		\centering
		\includegraphics[width=\textwidth]{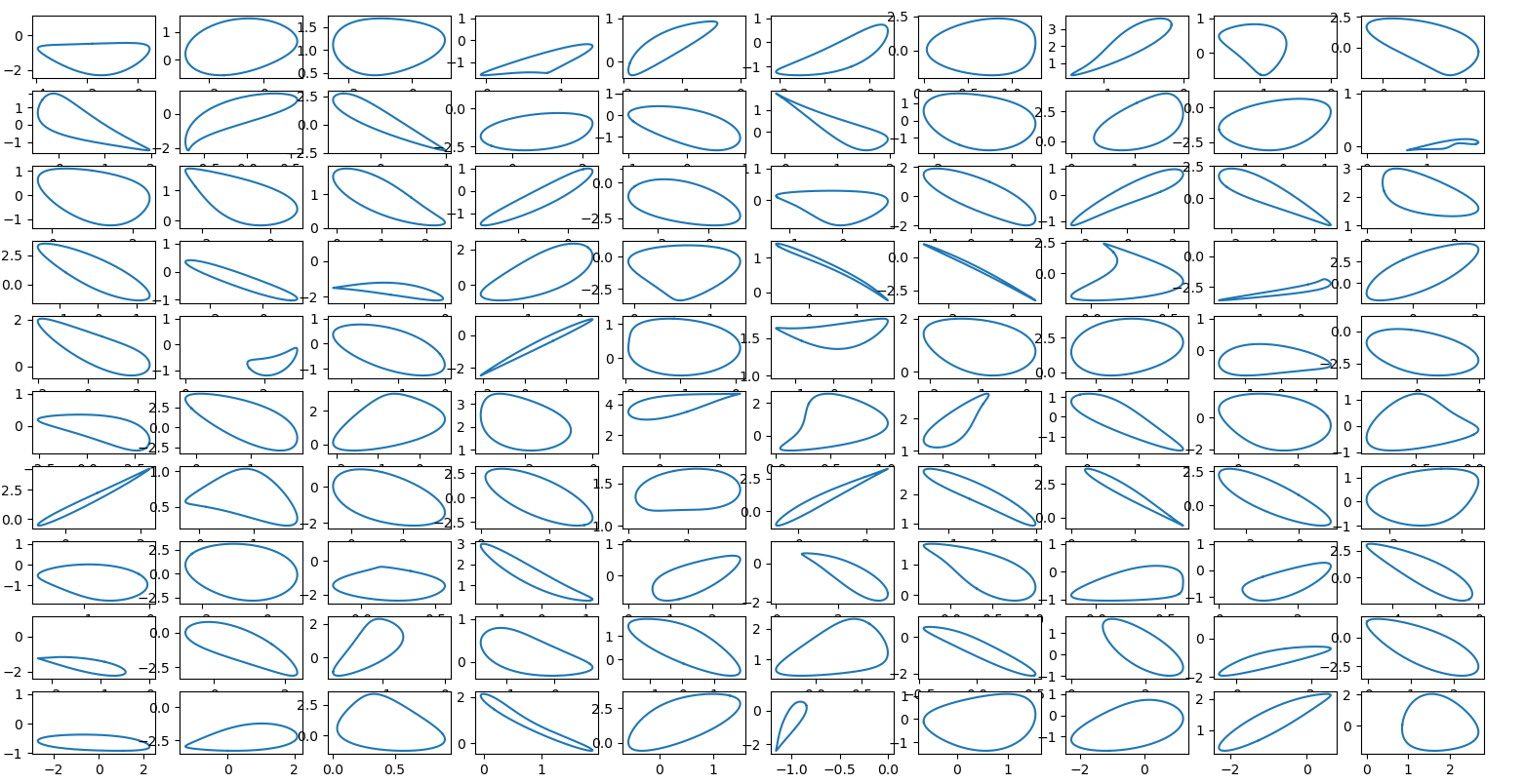}
		\subcaption{Decreasing weight $\rho$=0.1\\}
	\end{subfigure}
	\begin{subfigure}{0.4\textwidth}
		\centering
		\includegraphics[width=\textwidth]{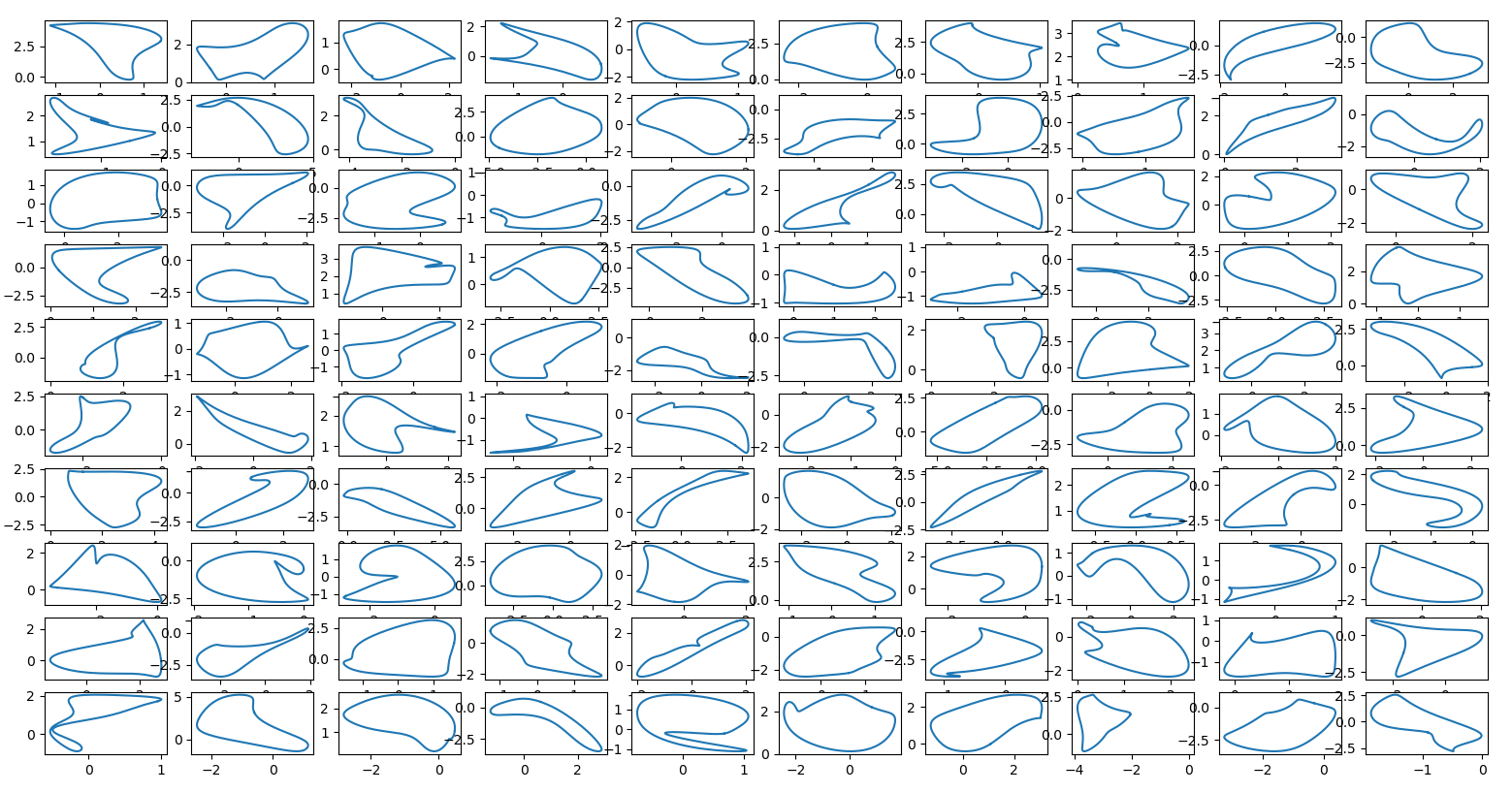}
		\subcaption{Decreasing weight $\rho$=0.4\\ }
	\end{subfigure}\\
	\begin{subfigure}{0.4\textwidth}
		\centering
		\includegraphics[width=\textwidth]{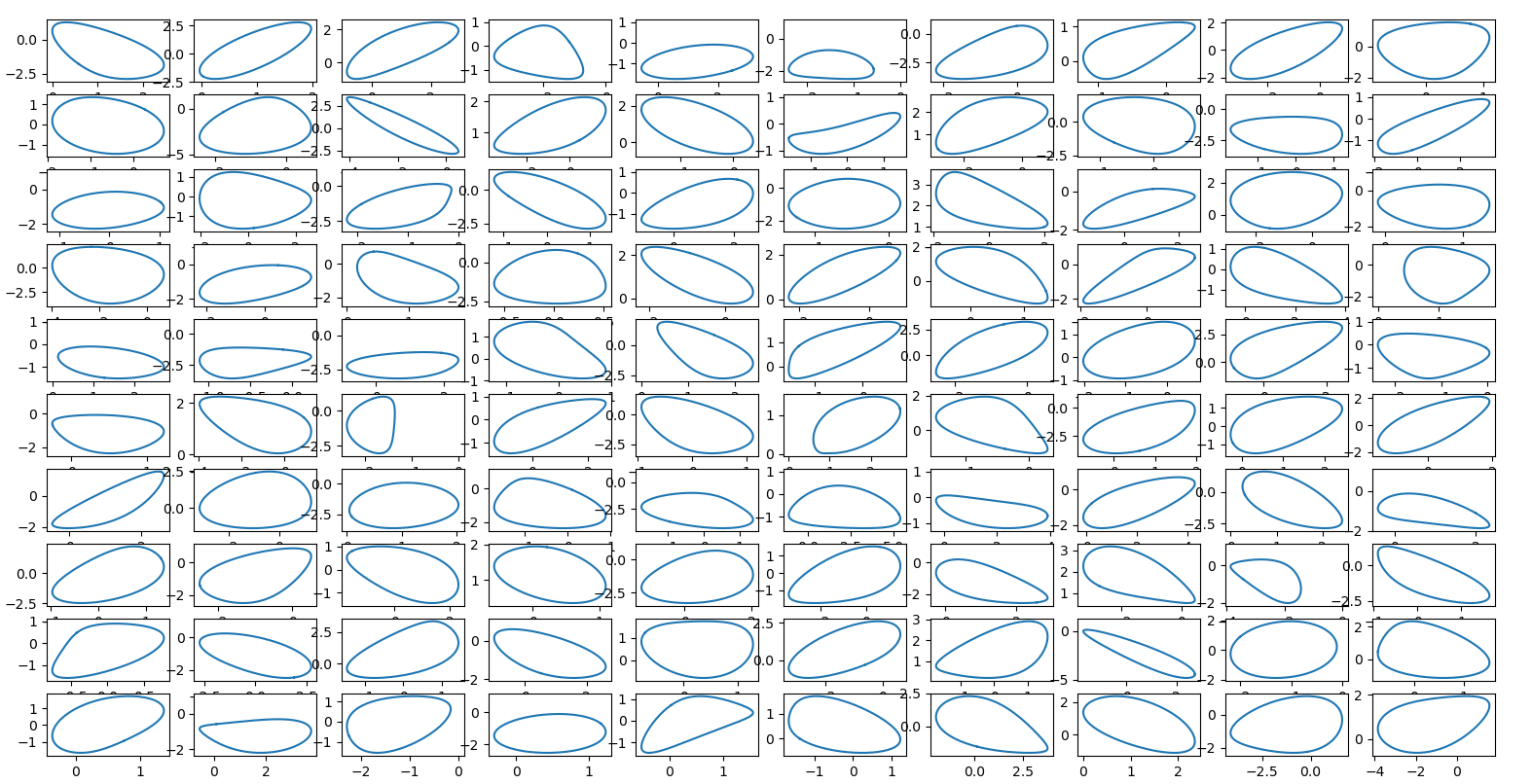}
		\subcaption{Decreasing weight $\rho$=0.1 and limiting maximum absolute curvature to less than 10}
	\end{subfigure}
	\caption{Partial generated boundary data. (a) and (b) are generated by with $\rho=0.1$ and $\rho=0.4$, respectively. (c) generated by with $\rho=0.1$ and limiting maximum curvature to less than 10. }
	\label{boundary}
\end{figure}

It is noteworthy that the generated boundary $\tilde{\boldsymbol{\gamma}}_n$ described by $\boldsymbol\gamma_n$ may not a simple closed curves, where
$$
\tilde{\boldsymbol{\gamma}}_n:=\left((\gamma_1)_0+ \sum_{k \in \mathbb{Z}_n} ((\gamma_1)^c_k \cos(kt)+(\gamma_1)^s_k \sin(kt)), (\gamma_2)_0+ \sum_{k \in \mathbb{Z}_n} ((\gamma_2)^c_k \cos(kt)+(\gamma_2)^s_k \sin(kt))\right).
$$
Therefore, we need to judge wether each generated data vector $\boldsymbol{\gamma}_n$ describes a simple closed curve\cite{franco2012}.

Additionally, we filtered the generated boundaries based on their curvature. As depicted in images (a) and (b) of Figure \ref{boundary}, some boundaries enclose narrowly confined regions. Generating training data and conducting potential integral calculations for such areas demand substantial computational resources and specially designed algorithms to ensure data quality. Given that the primary objective of this article is to validate the effectiveness of operator learning methods for solving PDEs using BIEs, we selectively included boundaries in our dataset based on their curvature to avoid overly narrow regions. Specifically, we calculate the curvature of the vector-valued function $\boldsymbol{\tilde{\gamma}}_n$ using the following formula:
\begin{equation}\label{curvature}
\text{cur}(t) := \frac{\gamma_1^\prime(t) \gamma_2^{\prime\prime}(t) - \gamma_1^{\prime\prime}(t) \gamma_2^\prime(t)}{\left(\gamma_1^\prime(t)^2 + \gamma_2^\prime(t)^2\right)^{3/2}}
\end{equation}
Using this formula, we computed the curvature of the boundaries and excluded from the dataset any boundaries where the absolute curvature exceeds $10$. This approach ensures that the boundaries included in the dataset do not enclose excessively narrow areas, thus facilitating more efficient experimental procedures.

\begin{figure}[htbp]
	\centering
	\begin{subfigure}{0.49\textwidth}
		\centering
		\includegraphics[width=\textwidth]{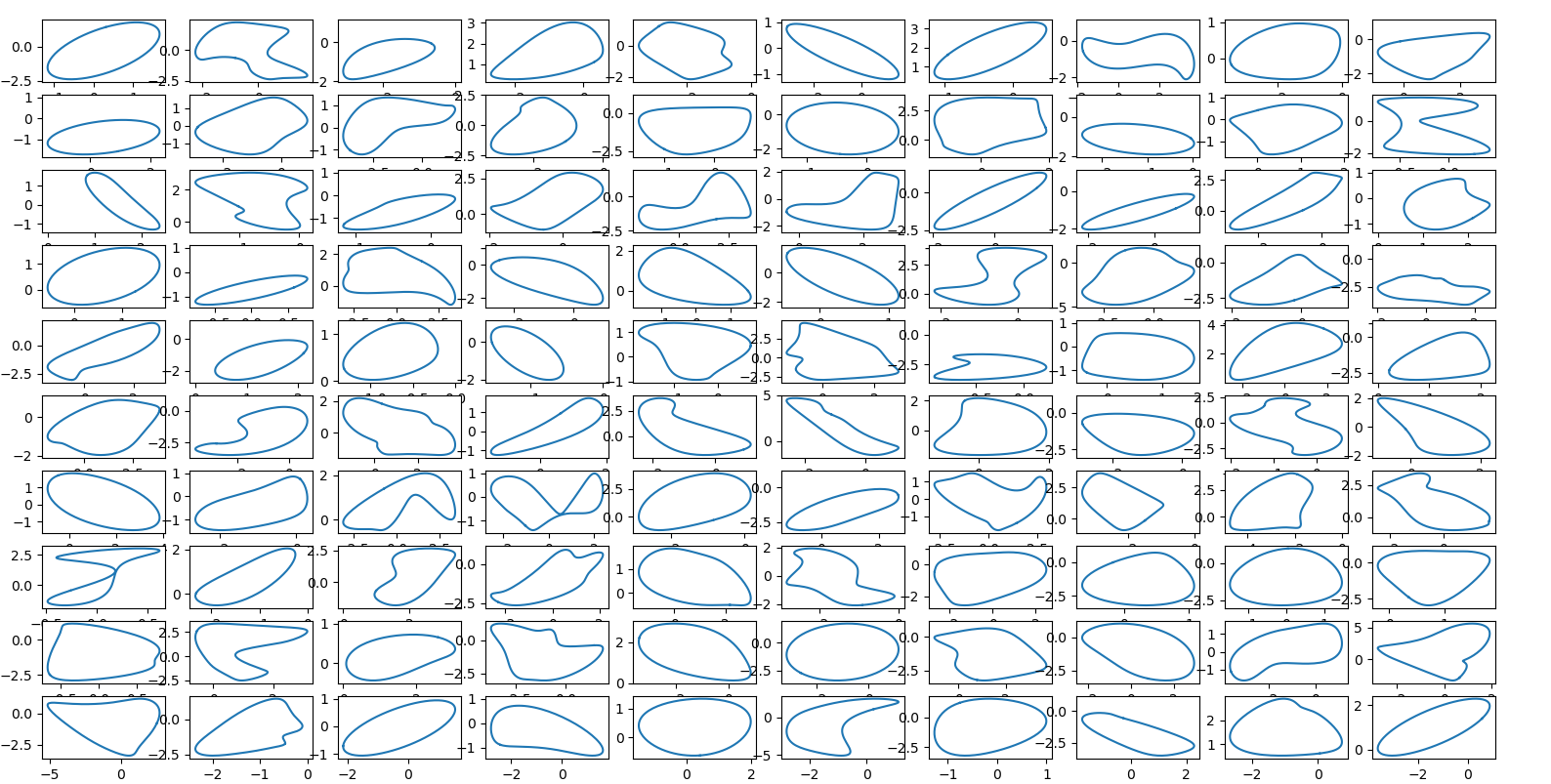}
		\subcaption{Partial boundary in the boundary datasets}
	\end{subfigure}
	\begin{subfigure}{0.49\textwidth}
		\centering
		\includegraphics[width=\textwidth]{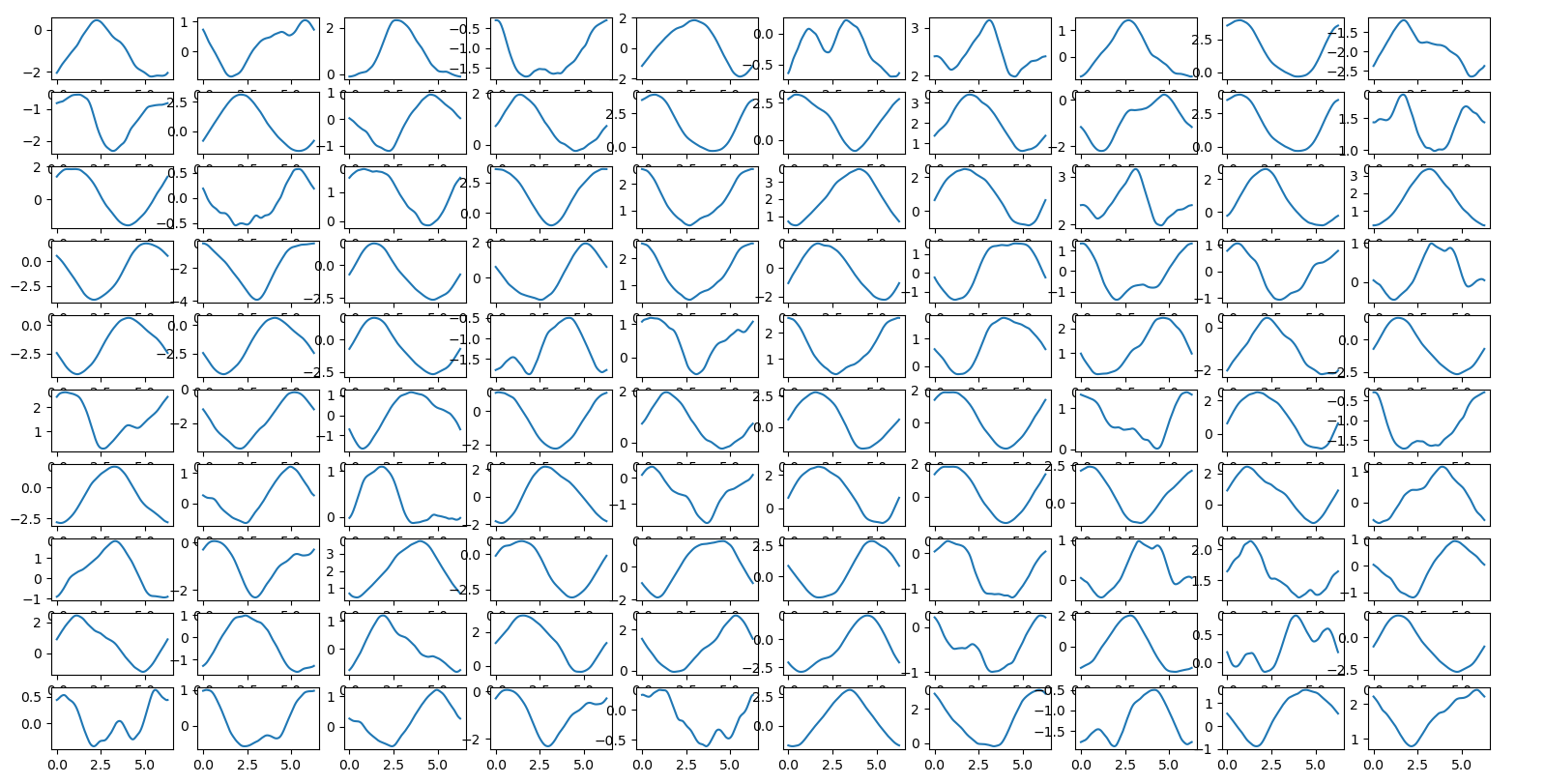}
		\subcaption{The visualization of  $\varphi_n$}
	\end{subfigure}
	\caption{(a) The diversity of boundaries in the boundary dataset. (b) The function $\varphi_n$ generated with $m=5$.}
	\label{data}
\end{figure}

We set $n=20$ and ensure that the maximum absolute curvature of the boundary remains below $10$ when generating boundaries. Additionally, we specify $\rho=0.1, 0.2, \ldots, 0.6$, and generate a total of $5998$ vectors $\boldsymbol{\gamma}_n$. These vectors describe $5998$ distinct boundaries.
The $100$ of $5998$ boundaries are shown in Figure \ref{data} (a).  When generating $\boldsymbol{\varphi}_n$, we set $m = 5$, and the resulting $\varphi_n$ is shown in Figure \ref{data} (b).

\section{Experiments}

In this section, we apply BI-TDONet and BI-DeepONet to solve LBVPs, potential flow issues, elastostatic problems, and the obstacle scattering problem for acoustic waves, in order to evaluate the performance of these two operator learning architectures. In all experiments, we use the mean $l_2$ norm error (MNE) and the mean $l_2$ norm relative error (MRE) of trigonometric coefficients as metrics to assess model performance. Additionally, the loss function used during the training process of all models is MRE. Given the distinct outputs of the BI-DeepONet and BI-TDONet networks—with BI-DeepONet outputting function values of the layer potentials on the trunk net, and BI-TDONet outputting the trigonometric coefficients of the layer potentials—we delineate the loss functions for each network individually. To achieve a more concise representation, we denote the BI-DeepONet by $\overline{\mathcal{N}}(\boldsymbol{\theta},\cdot)$. We also define vectors $\overline{\mathbf{\boldsymbol{\gamma}}}_p$, $\overline{\mathbf{f}}_p$, and $\overline{\boldsymbol{\varphi}}_p$ representing the vectors $[\gamma_1(t_1), \gamma_1(t_2), \ldots, \gamma_1(t_p), \gamma_2(t_1), \gamma_2(t_2), \ldots, \gamma_2(t_p)]$, $[\widetilde{f}(t_1), \widetilde{f}(t_2), \ldots, \widetilde{f}(t_p)]$, and $[\varphi(t_1), \varphi(t_2), \ldots, \varphi(t_p)]$, respectively. The loss function for BI-DeepONet is defined as
\begin{equation*}
	\begin{aligned}
		Loss_{D} :=\sum_{j=1}^{M} \frac{\left\| \overline{\mathcal{N}}(\boldsymbol{\theta}, [\overline{\mathbf{\boldsymbol{\gamma}}}^j_p, \overline{\mathbf{f}}^j_p, [t_1, \ldots, t_p]]) - \overline{\boldsymbol{\varphi}}^j_p \right\|_2}{\left\| \overline{\boldsymbol{\varphi}}^j_p \right\|_2}.
	\end{aligned}
\end{equation*}
where $M$ represents the number of training samples. The loss function for the BI-DeepONet is then formulated as
\begin{equation*}
	Loss_{T} :=\sum_{j=1}^{M} \frac{\left\| \mathcal{N}\left(\boldsymbol{\theta}_U,\left[ \boldsymbol{\gamma}_n^j,\left(\mathcal{N}\left( \boldsymbol{\theta}_V,[\boldsymbol{\gamma}_n^j,{\mathbf{f}}_n^j]\right)\odot\mathcal{N}\left( \boldsymbol{\theta}_\sigma,\boldsymbol{\gamma}_n^j\right)   \right)\right]  \right) - \boldsymbol{\varphi}_n^j \right\|_2}{\left\| \boldsymbol{\varphi}_n^j \right\|_2},
\end{equation*}
The construction and training of the networks were conducted on a Linux system equipped with an NVIDIA GeForce RTX 3090 graphics card, which has 24GB of video memory.

\subsection{LBVPs}

%
%
%
%
%
%

In this subsection, we present five illustrative cases: the interior and exterior Laplace Boundary Value Problems (LBVPs) with Dirichlet and Neumann boundary conditions, and an interior Dirichlet problem characterized by slowly decaying trigonometric coefficients $\mathbf{f}_n$ as frequency increases. For the initial four scenarios, we use the generated $5,998$ boundaries detailed in Subsection 3.3. Additionally, by setting $m = 5$, we generated a total of $200$ vectors $\boldsymbol{\varphi}_n$ by the method describe in Subsection 3.3. By tensor-producing the sets ${\boldsymbol{\gamma}_n}$ and ${\boldsymbol{\varphi}_n}$, and computing $\mathbf{f}_n$ by equation \eqref{finite BIO01}, we compiled a comprehensive dataset containing $1,199,600$ samples.

For the final scenario involving slowly decaying coefficients, we again utilized the $5,998$ boundaries. However, with $m=0.1$ for generating $\boldsymbol{\varphi}_n$, the dataset similarly comprised $1,199,600$ samples. In all experiments, $80\%$ of the data was used for training, while the remaining $20\%$ was reserved for testing.

Within the BI-DeepONet framework, both inputs and outputs are treated as function values. We set $p=128$
and defined the input for the trunk net as uniformly distributed points $[t_1, t_2, \ldots, t_p]$ over interval $I$, and the input for the branch net as samples of $\boldsymbol{\gamma}$ and $\widetilde{f}$ at these points. This setup results in variations in the dimensions of the input and output vectors between BI-TDONet and BI-DeepONet. To clarify the architecture of BI-DeepONet, we describe its structure as $[[256,300,300,300,300]$, $[128,300,300,300,300]$, $[1,300,300,300,300]]$, where the first two sequences of numbers represent the branch net's structure, and the last sequence details the trunk net's configuration. Notably, BI-DeepONet does not require specifying the output layer's size, as its output consists of the function values sampled at $[t_1, t_2, \ldots, t_p]$.
The BI-TDONet architecture is outlined as $[[123, 300, 300, 300,300, 41]$, $[82, 300, 300, 300,300, 41]$, $[123, 300, 300, 300, 300, 41]]$, corresponding to the network structures of $\mathcal{N}\left( \boldsymbol{\theta}_V,\cdot\right)$, $\mathcal{N}\left( \boldsymbol{\theta}_\sigma,\cdot\right)$, and $\mathcal{N}\left( \boldsymbol{\theta}_U,\cdot\right)$, respectively.

During the training of BI-DeepONet, we configured the batch size to $8,192$ and set the number of iterations at $3,000,000$, equivalent to roughly $25,000$ epochs. We initiated the training with a learning rate of $0.001$, employing the Adam optimization algorithm. The learning rate was scheduled to decay on an inverse-time basis, with a decay interval of $1/100$ of the total iterations and a decay factor of $0.5$.

For BI-TDONet, we also set the batch size to $8,192$ but limited the number of epochs to $5,000$. The initial learning rate was fixed at $0.001$, again using the Adam optimization algorithm. Learning rate decay was dynamically managed; specifically, if there was no decrease in the loss over a span equal to $1/100$ of the total epochs, the learning rate was halved. This approach ensured that adjustments in the learning rate were responsive to the training progress.

\begin{table}[!ht]
	\centering
	
		\small
		\begin{tabular}{|c|c|c|c|c|c|c|} \hline
			Problems &Model
			& MNE
			& MRE 
			&variance-MNE 
			&variance-MRE
			&Mean-Time /ms \\ \hline
			\multirow{2}{*}{\tabincell{c}{IDP}}
			&BI-DeepONet
			&$1.0284\times10^{-2}$
			&$6.9597\times10^{-3}$
			&$4.0761\times10^{-4}$
			&$2.2122\times10^{-4}$
			&$3.3181\times10^{-4}$\\
			\cline{2-7}
			&BI-TDONet
			&$1.0216\times10^{-3}$
			&$6.4433\times10^{-4}$
			&$1.2698\times10^{-4}$
			&$5.1572\times10^{-5}$
			&$5.6207\times10^{-3}$\\
			\hline
			\multirow{2}{*}{\tabincell{c}{EDP}}
			&BI-DeepONet
			&$1.8361\times10^{-2}$
			&$1.1384\times10^{-2}$
			&$4.8651\times10^{-4}$
			&$2.1037\times10^{-4}$
			&$4.4793\times10^{-3}$\\
			\cline{2-7}
			&BI-TDONet
			&$1.1721\times10^{-3}$
			&$7.3597\times10^{-4}$
			&$1.8453\times10^{-4}$
			&$7.8372\times10^{-5}$
			&$5.6645\times10^{-3}$\\
			\hline
			\multirow{2}{*}{\tabincell{c}{INP}}
			&BI-DeepONet
			&$1.9512\times10^{-2}$
			&$1.6894\times10^{-2}$ &$2.4327\times10^{-4}$
			&$2.8173\times10^{-4}$
			&$3.1916\times10^{-4}$\\
			\cline{2-7}
			&BI-TDONet
			&$5.3335\times10^{-4}$
			&$4.2781\times10^{-4}$
			&$5.1137\times10^{-5}$
			&$7.9158\times10^{-5}$
			&$5.3468\times10^{-3}$\\
			\hline
			\multirow{2}{*}{\tabincell{c}{ENP}}
			&BI-DeepONet
			&$1.7834\times10^{-2}$
			&$1.6536\times10^{-2}$
			&$2.0887\times10^{-3}$
			&$2.3144\times10^{-4}$
			&$2.6391\times10^{-4}$\\
			\cline{2-7}
			&BI-TDONet
			&$2.5623\times10^{-4}$
			&$4.5155\times10^{-4}$
			&$1.0551\times10^{-5}$
			&$4.1330\times10^{-5}$
			&$5.4866\times10^{-3}$\\
			\hline	
			\multirow{2}{*}{\tabincell{c}{\footnotesize HFIDP }}
			&BI-DeepONet
			&$2.2715\times10^{-2}$
			&$3.9779\times10^{-3}$
			&$3.7042\times10^{-4}$
			&$5.3401\times10^{-3}$
			&$1.4715\times10^{-3}$\\
			\cline{2-7}
			&BI-TDONet
			&$1.5330\times10^{-3}$
			&$2.6734\times10^{-4}$
			&$1.3080\times10^{-4}$
			&$3.5746\times10^{-6}$
			&$6.7715\times10^{-3}$\\
			\hline	
		\end{tabular}
\caption{Errors of BI-DeepONet and BI-TDONet for the LBVPs on the test set}
	\label{testing error}
\end{table}

Table \ref{testing error} presents the  MNE and MRE for different test cases using two models after adequate training. The first four rows of the first column correspond to the abbreviations for the interior Dirichlet problem (IDP), exterior Dirichlet problem (EDP), interior Neumann problem (INP), and exterior Neumann problem (ENP). The final row of this column refers to the interior Dirichlet problem where the trigonometric coefficients $\mathbf{f}_n$ decay slowly with increasing frequency, leading to richer high-frequency information, abbreviated as HFIDP. The last column of Table \ref{testing error} shows the inference time spent on the test sets of various cases by the two operator learning frameworks described in this article after training. Notably, even without specific optimizations for operator inference, the average time spent on operator inference for each case is less than $0.01$ millisecond, which is significantly faster than many established fast algorithms for BIEs that have been enhanced over decades \cite{jiang2010,jiang2018,cai2008}. Additionally, a series of refined and accurate fast algorithms have been developed for computing potential integrals, which are essential for solving PDEs using the BIEM \cite{Andreas2013,Ding2021,chen2023}.

The results displayed in Table \ref{testing error} show that BI-TDONet achieves an  MNE ranging from $5.3335\times10^{-4}$ to $1.5330\times10^{-3}$ and an  MRE of less than $10^{-3}$ across all experimental test sets for each sample. The variances of MNE and MRE in Table \ref{testing error} indicate minimal fluctuations in errors across samples within the test sets, clearly demonstrating BI-TDONet's remarkable effectiveness in handling various LBVPs. In contrast, BI-DeepONet records an  MNE of less than $3\times10^{-2}$ and an  MRE of less than $2\times10^{-2}$.

The superior accuracy of BI-TDONet compared to BI-DeepONet can be attributed to two primary factors. First, while BI-DeepONet's inputs and outputs consist of sampled function values, BI-TDONet utilizes coefficients of trigonometric series, ensuring the periodic continuity of inputs and outputs. This method more effectively captures characteristics of functions with oscillatory features or pulse signals than multipoint sampling, resulting in lower errors for BI-TDONet after sufficient training. Second, the architecture of BI-TDONet is meticulously designed according to the Singular Value Expansion (SVE) of the inverse of second-kind integral operators, ensuring a perfect alignment between the model components and the inputs, particularly in establishing relationships between boundaries and various aspects of the SVE.

Although BI-DeepONet can also be seen as an extension of the Singular Value Decomposition (SVD) approach to nonlinear operator learning problems \cite{VENTURI2023}, as a general framework for nonlinear operator learning, it does not integrate the specific characteristics of BIOs. The nonlinearity of BIOs is intrinsically dependent on the properties of the boundaries, a feature that BI-DeepONet does not address.

Overall, both BI-DeepONet and BI-TDONet exhibit excellent performance in solving LBVPs across various domains.

\begin{figure}[htbp]
	\centering
	\begin{subfigure}{0.195\textwidth}
		\centering
		\includegraphics[width=\textwidth]{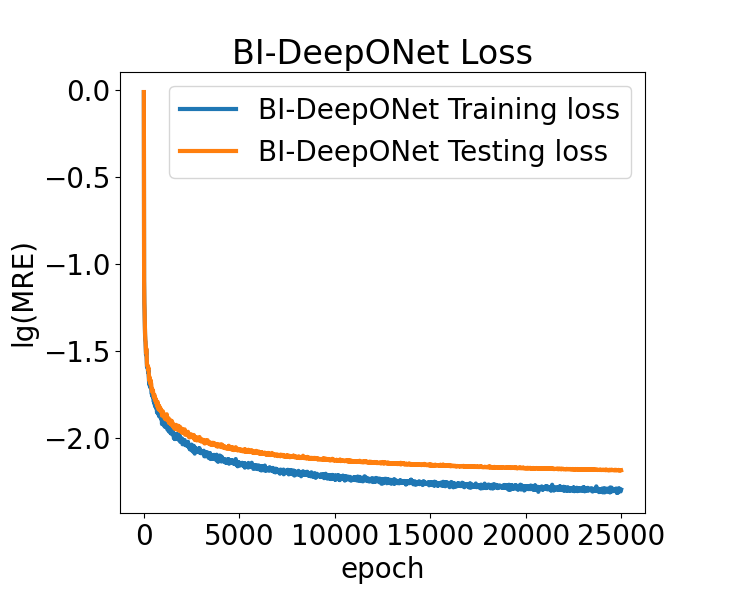}
		\subcaption{IDP}
	\end{subfigure}
	\begin{subfigure}{0.195\textwidth}
		\centering
		\includegraphics[width=\textwidth]{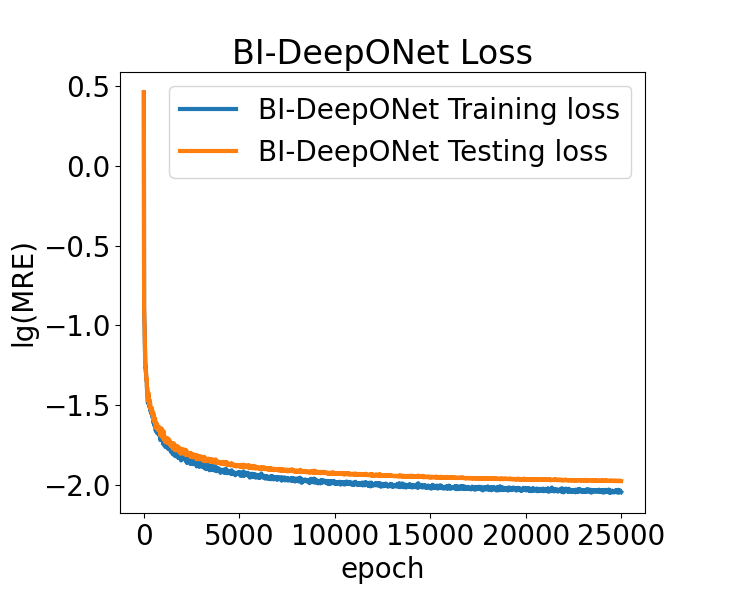}
		\subcaption{EDP}
	\end{subfigure}
	\begin{subfigure}{0.195\textwidth}
		\centering
		\includegraphics[width=\textwidth]{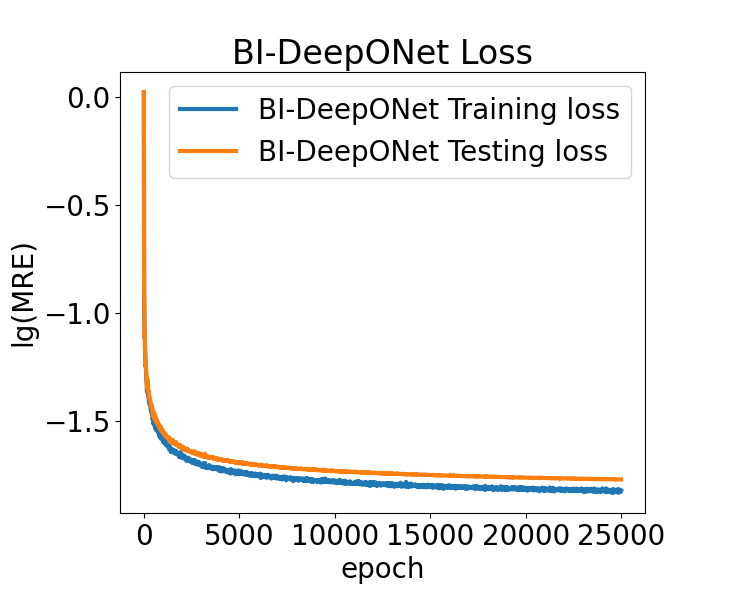}
		\subcaption{INP}
	\end{subfigure}
	\begin{subfigure}{0.195\textwidth}
		\centering
		\includegraphics[width=\textwidth]{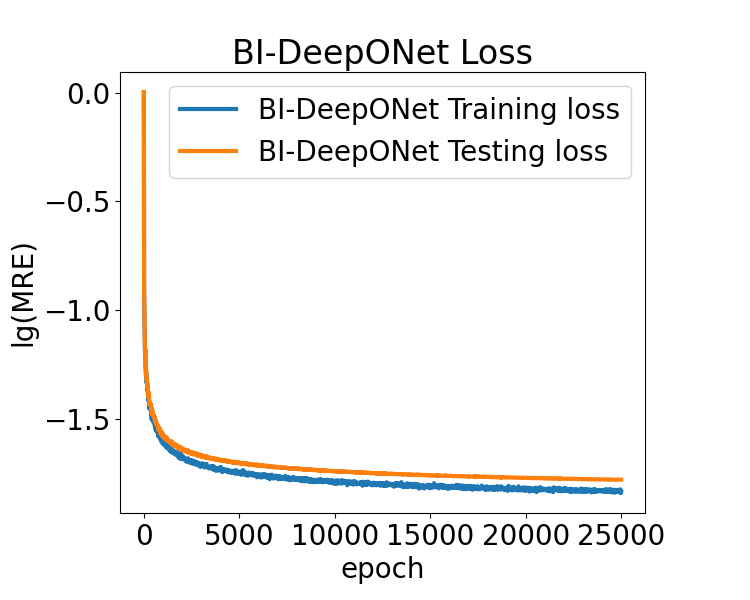}
		\subcaption{ENP}
	\end{subfigure}
	\begin{subfigure}{0.195\textwidth}
		\centering
		\includegraphics[width=\textwidth]{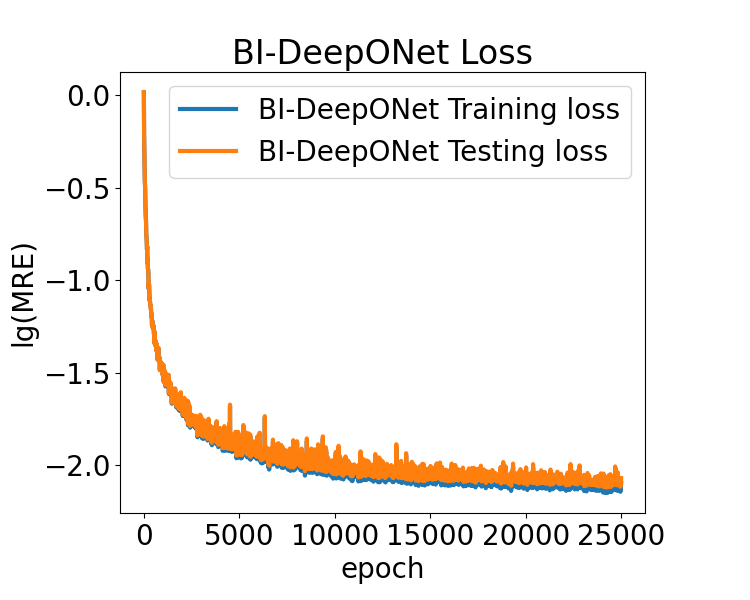}
		\subcaption{HEIDP}
	\end{subfigure}
	
	\begin{subfigure}{0.195\textwidth}
		\centering
		\includegraphics[width=\textwidth]{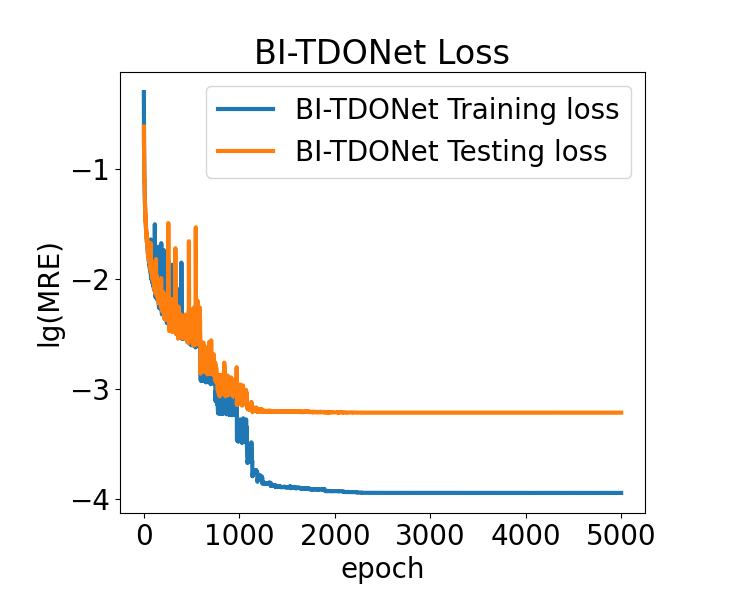}
		\subcaption{IDP}
	\end{subfigure}
	\begin{subfigure}{0.195\textwidth}
		\centering
		\includegraphics[width=\textwidth]{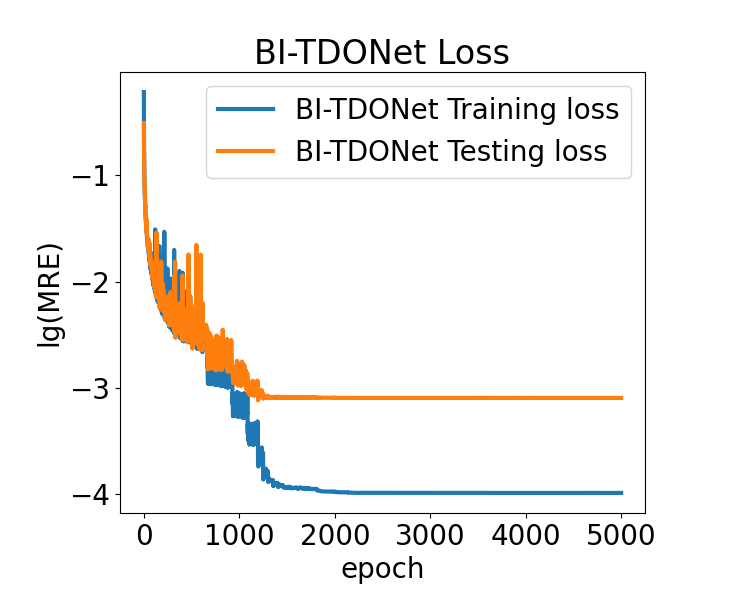}
		\subcaption{EDP}
	\end{subfigure}
	\begin{subfigure}{0.195\textwidth}
		\centering
		\includegraphics[width=\textwidth]{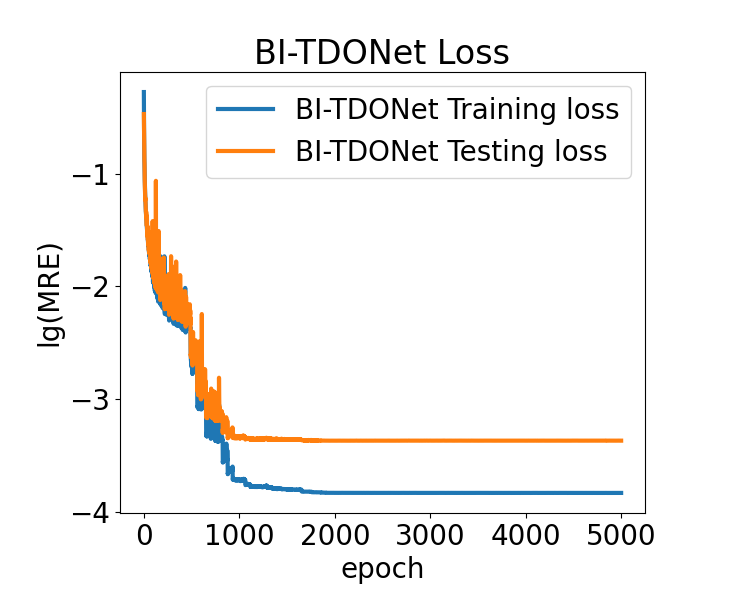}
		\subcaption{INP}
	\end{subfigure}
	\begin{subfigure}{0.195\textwidth}
		\centering
		\includegraphics[width=\textwidth]{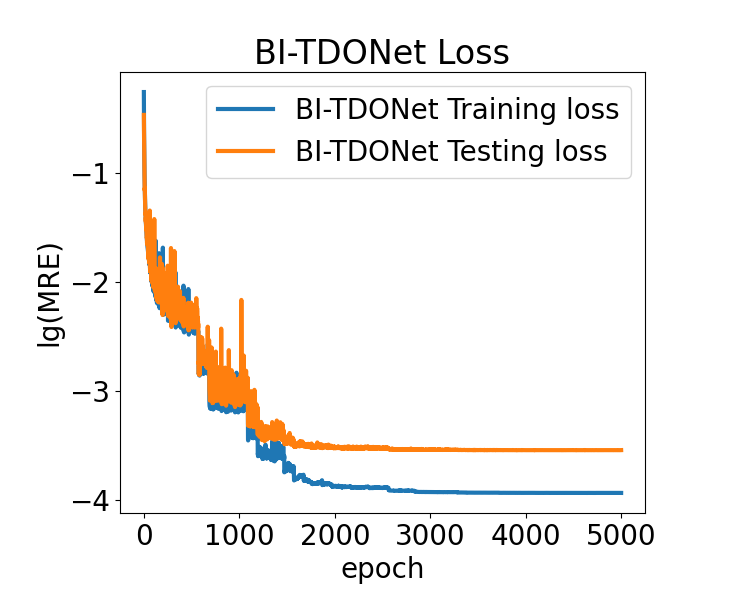}
		\subcaption{ENP}
	\end{subfigure}
	\begin{subfigure}{0.195\textwidth}
		\centering
		\includegraphics[width=\textwidth]{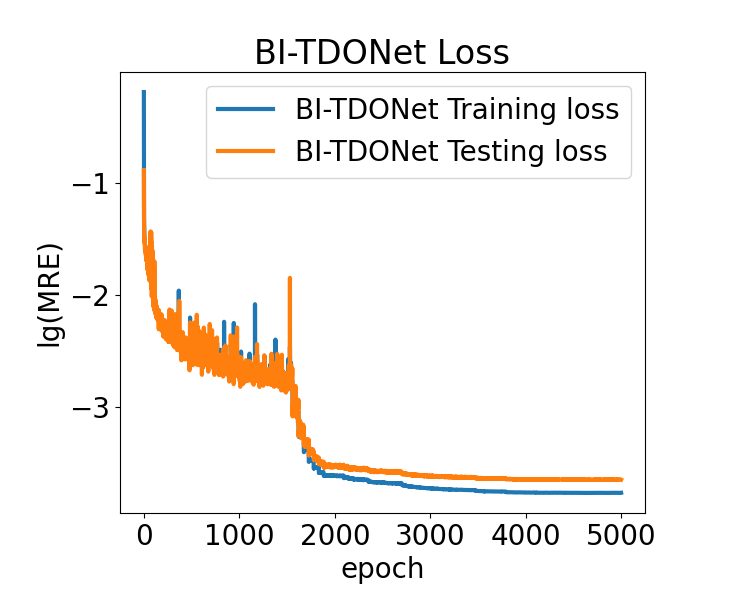}
		\subcaption{HFIDP}
	\end{subfigure}
	\caption{The logarithmic loss trajectories for BI-DeepONet and BI-TDONet.}
	\label{lbvp_loss}
\end{figure}

Figure \ref{lbvp_loss} demonstrates the logarithmic loss trajectories of the models throughout their training and testing cycles. It can be observed that BI-DeepONet’s loss tends to plateau when the epoch count reaches $15,000$, while BI-TDONet’s loss flattens at a significantly earlier stage, before reaching $2,000$ epochs. Additionally, both the training and testing losses for BI-TDONet are consistently lower than those observed for BI-DeepONet.

%

To further assess the performance of BI-DeepONet and BI-TDONet, we randomly selected two samples from the test set of the LBVPs and presented the predicted solutions derived from the outputs of both models. We use the true solution of the BIE (the trigonometric series corresponding to coefficients $\boldsymbol{\varphi_n}$ in the dataset) and the potential integrations from Table \ref{table_layer list} to calculate the solution of the LBVPs for points within the boundary-defined domain, treating this as the true solution.

The second column of Tables \ref{IDP result}-\ref{osc result} visualizes the boundary $\boldsymbol{\gamma}$ and the function $\widetilde{f}$. The third column displays the visual comparisons of the outcomes from both BI-TDONet and BI-DeepONet against the true BIE solution. The fourth column presents the true solutions of the LBVPs for these two randomly selected samples from the test set. The fifth column shows the predicted solutions of LBVPs derived from the outputs of both models through potential integration, as detailed in the second row of Table \ref{table_layer list}. In the sixth column, we plot the absolute differences between the true and predicted solutions.

The visual results in Tables \ref{IDP result}-\ref{osc result} indicate that BI-TDONet handles such input functions, while the performance of BI-DeepONet is notably inferior. It is important to highlight that the locations of maximum errors in the model outputs correspond to the locations of maximum errors in the computed solutions of the LBVPs, emphasizing the importance of accurately predicting the density functions.

\begin{table}[ht]
	\vspace{-0.3cm}
	\centering
	\resizebox{\textwidth}{!}{
		\begin{tabular}{|c|c|c|c|c|c|} \hline
			Model
			&Model Input
			&Model Output
			&True solution
			&Predict solution
			&Error \\
			\hline
			\raisebox{-6\height}{\tiny BI-DeepOnet}
			&\raisebox{-.9\height}{\includegraphics[width=3cm]{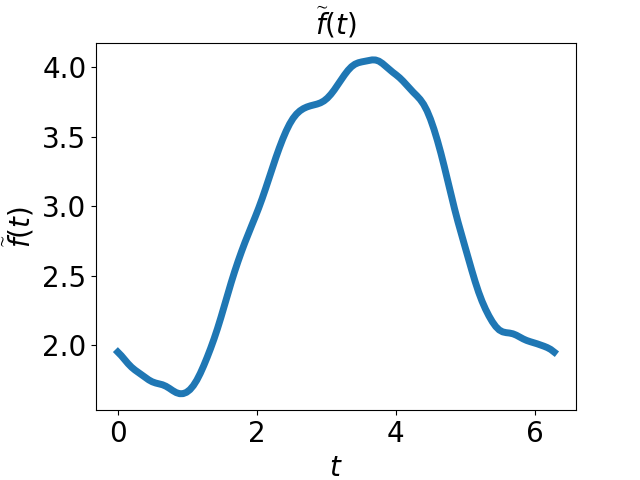}}
			&\raisebox{-.9\height}{\includegraphics[width=3cm]{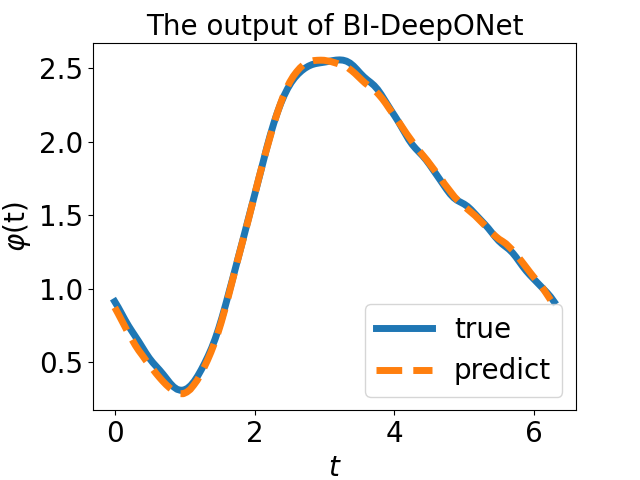}}
			&\multirow{2}*{\tabincell{c}{\raisebox{-1.15\height}{\includegraphics[width=4cm]{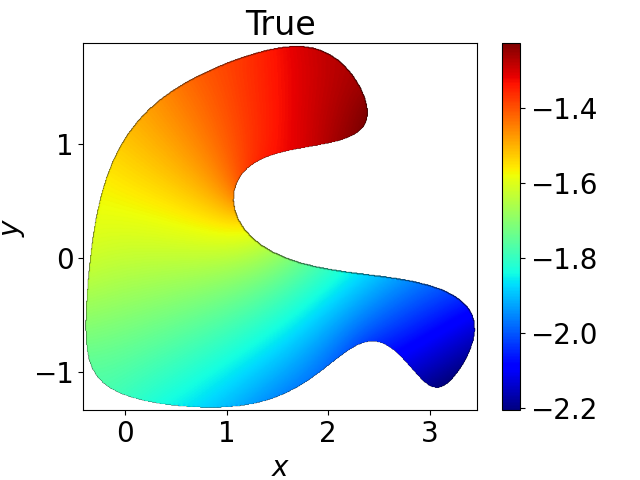}}
			}}
			&\raisebox{-.9\height}{\includegraphics[width=3cm]{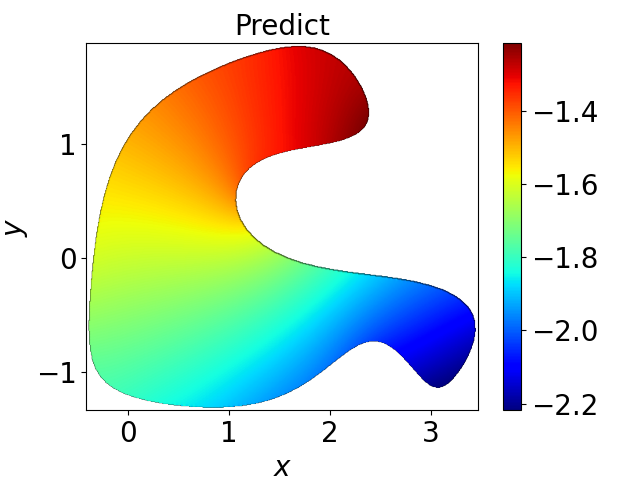}}
			&\raisebox{-.9\height}{\includegraphics[width=3cm]{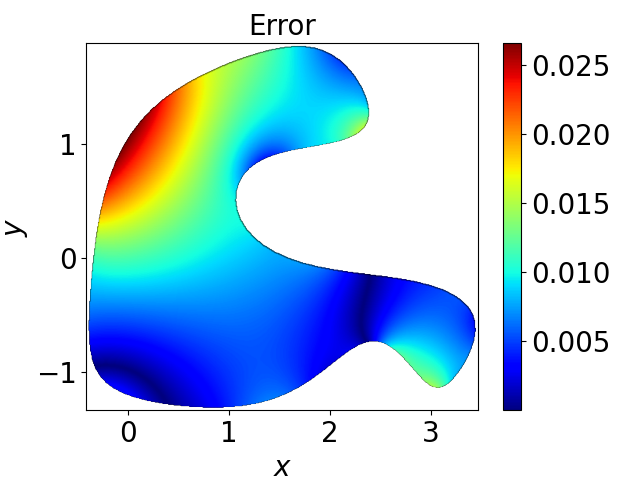}}\\
			\cline{1-1}\cline{3-3}\cline{5-6}
			\raisebox{-6\height}{\tiny{BI-TDONet }}
			&\raisebox{-.9\height}{\includegraphics[width=3cm]{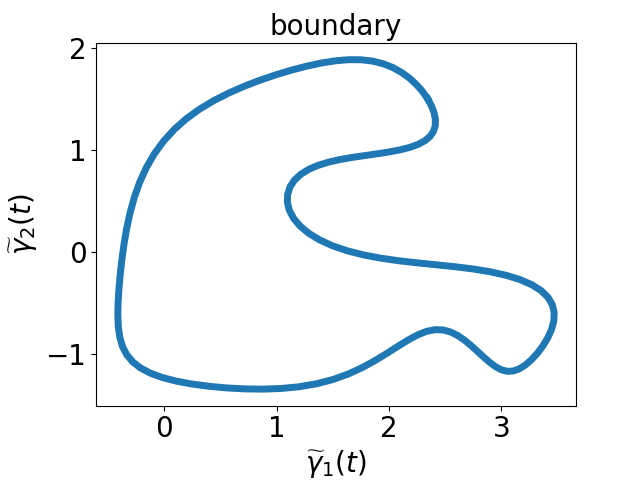}}
			&\raisebox{-.9\height}{\includegraphics[width=3cm]{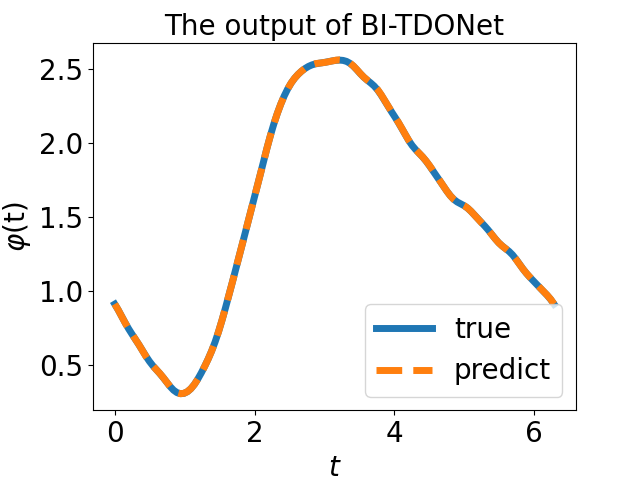}}
			&
			&\raisebox{-.9\height}{\includegraphics[width=3cm]{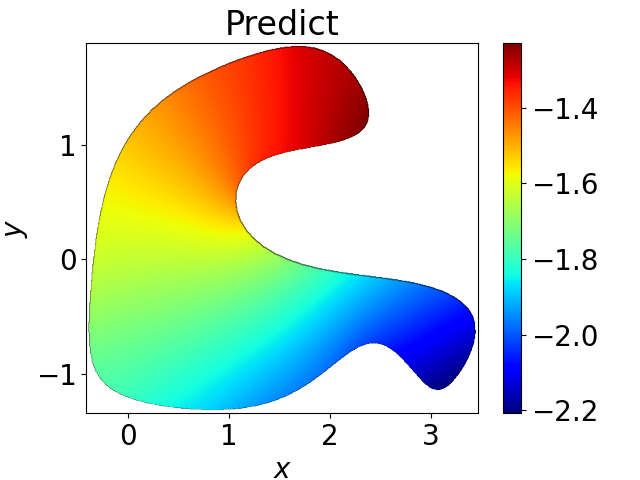}}
			&\raisebox{-.9\height}{\includegraphics[width=3cm]{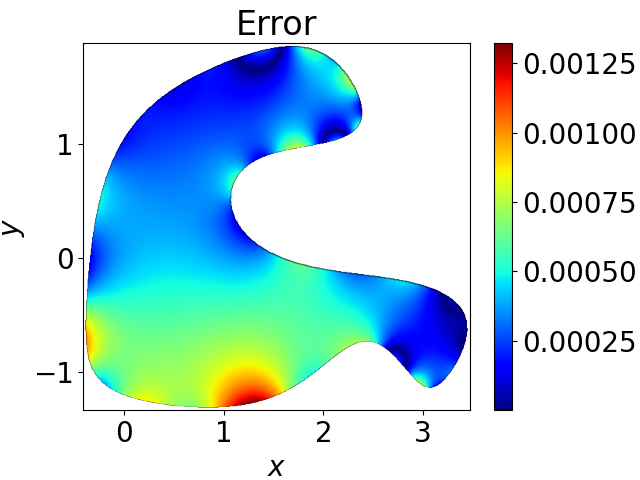}}\\
			\hline
			\raisebox{-6\height}{\tiny BI-DeepONet}
			&\raisebox{-.9\height}{\includegraphics[width=3cm]{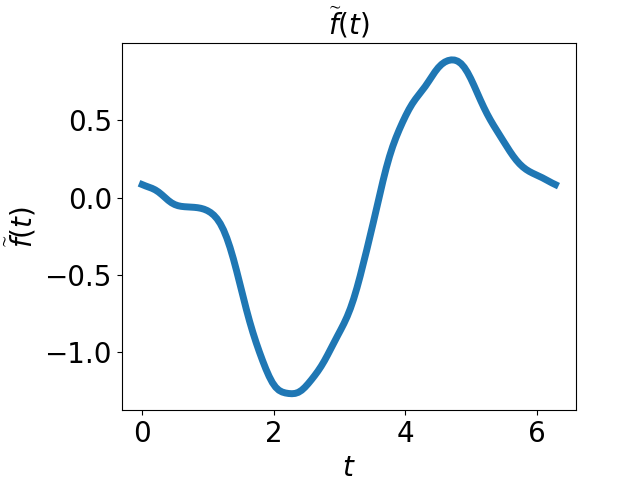}} &\raisebox{-.9\height}{\includegraphics[width=3cm]{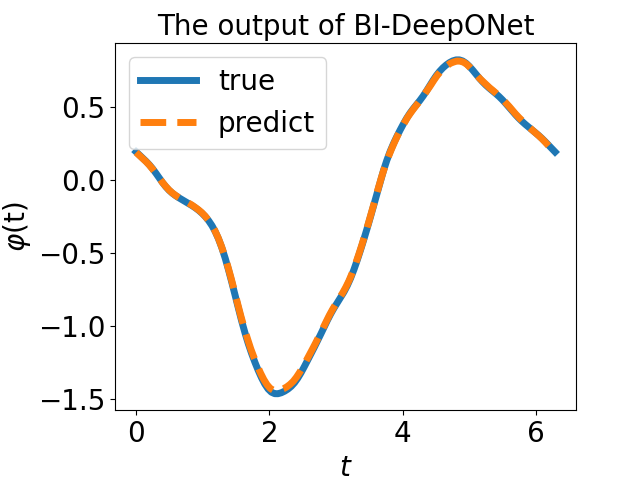}}
			&\multirow{2}*{\tabincell{c}{\raisebox{-1.2\height}{\includegraphics[width=4cm]{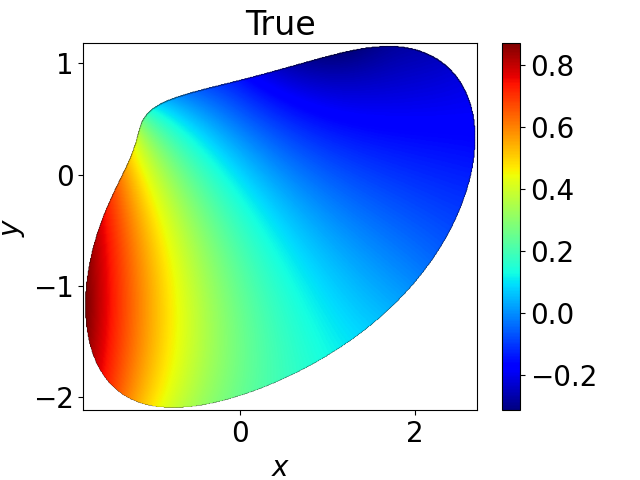}}}} &\raisebox{-.9\height}{\includegraphics[width=3cm]{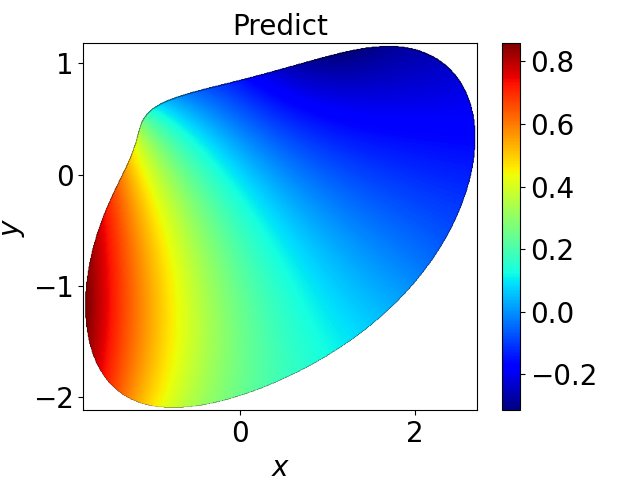}}
			&\raisebox{-.9\height}{\includegraphics[width=3cm]{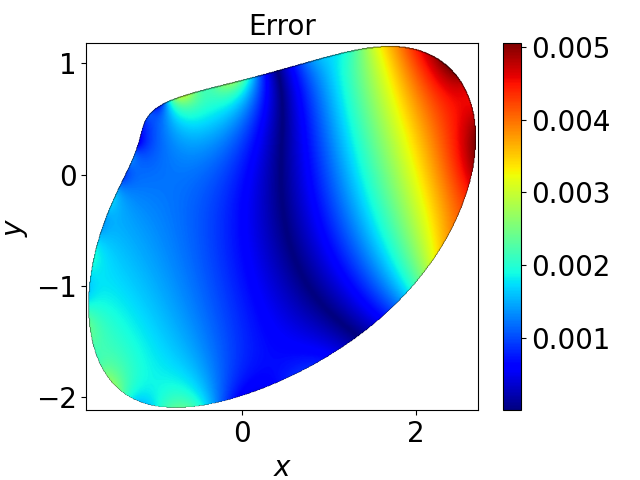}}\\
			\cline{1-1}\cline{3-3}\cline{5-6}
			\raisebox{-6\height}{\tiny BI-TDONet}
			&\raisebox{-.9\height}{\includegraphics[width=3cm]{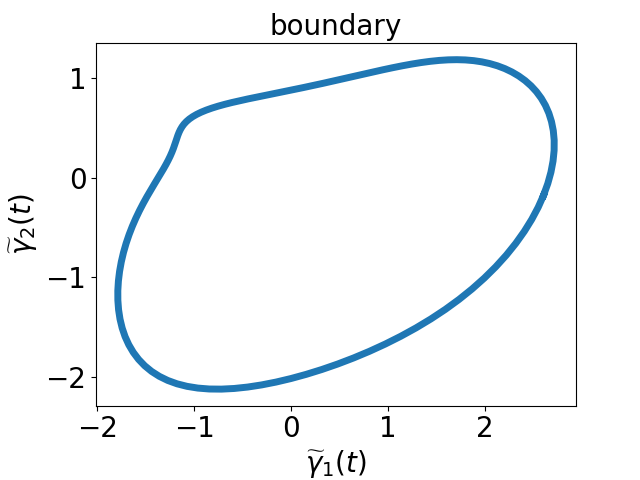}}
			&\raisebox{-.9\height}{\includegraphics[width=3cm]{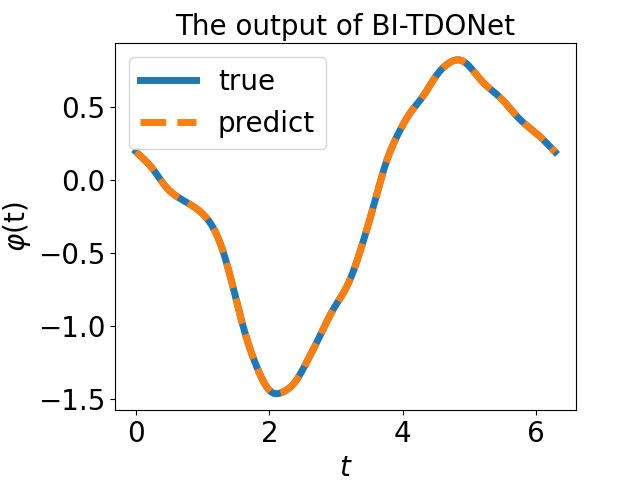}}
			&
			&\raisebox{-.9\height}{\includegraphics[width=3cm]{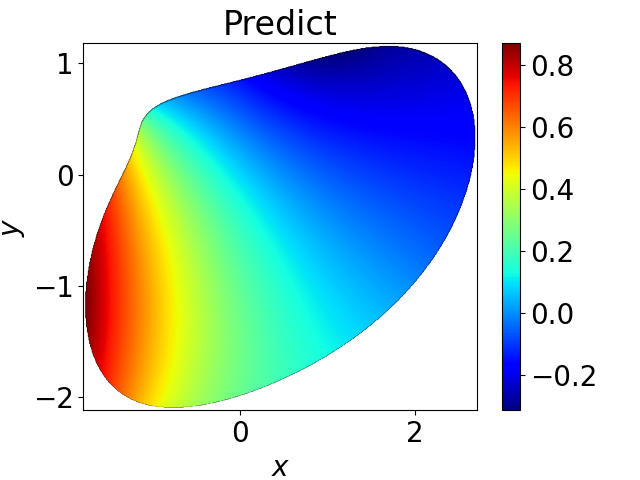}}
			&\raisebox{-.9\height}{\includegraphics[width=3cm]{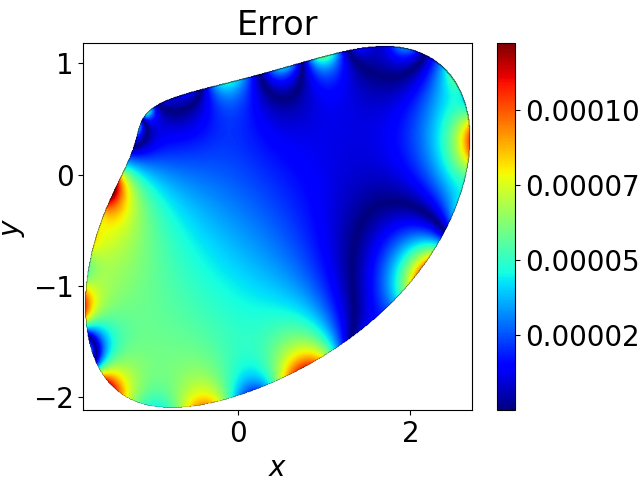}}\\
			\hline
	\end{tabular}}
    \caption{The solutions for two randomly selected samples of IDPs as processed by BI-DeepONet and BI-TDONet}
	\label{IDP result}
\end{table}

\begin{table}[ht]
	\vspace{-0.5cm}
	\centering
	\resizebox{\textwidth}{!}{
		\begin{tabular}{|c|c|c|c|c|c|} \hline
			Model
			&Model Input
			&Model Output
			&True solution
			&Predict solution
			& Error \\
			\hline
			\raisebox{-6\height}{\tiny BI-DeepONet}
			&\raisebox{-.9\height}{\includegraphics[width=3cm]{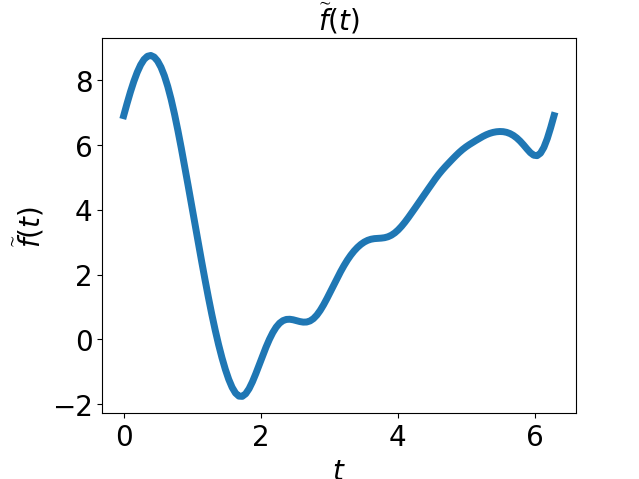}}
			&\raisebox{-.9\height}{\includegraphics[width=3cm]{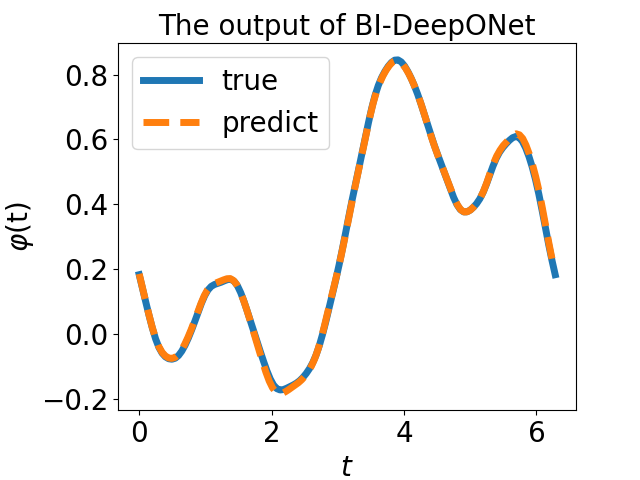}}
			&\multirow{2}*{\tabincell{c}{\raisebox{-1.15\height}{\includegraphics[width=4cm]{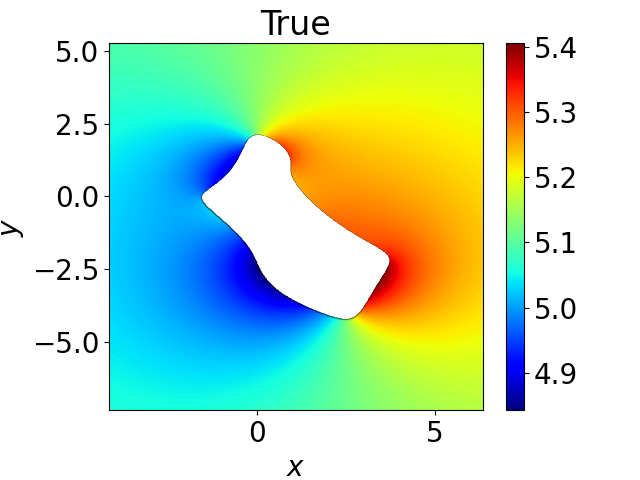}}}}
			&\raisebox{-.9\height}{\includegraphics[width=3cm]{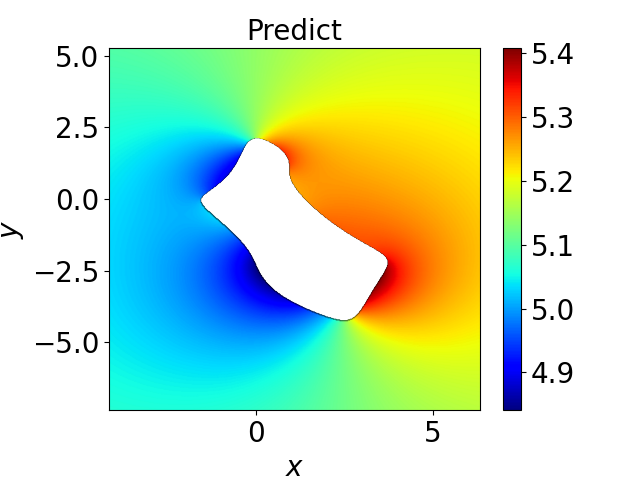}}
			&\raisebox{-.9\height}{\includegraphics[width=3cm]{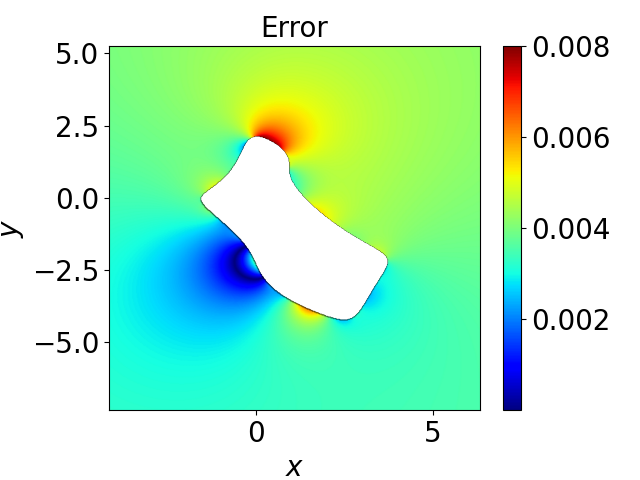}}\\
			\cline{1-1}\cline{3-3}\cline{5-6}
			\raisebox{-6\height}{\tiny BI-TDONet}
			&\raisebox{-.9\height}{\includegraphics[width=3cm]{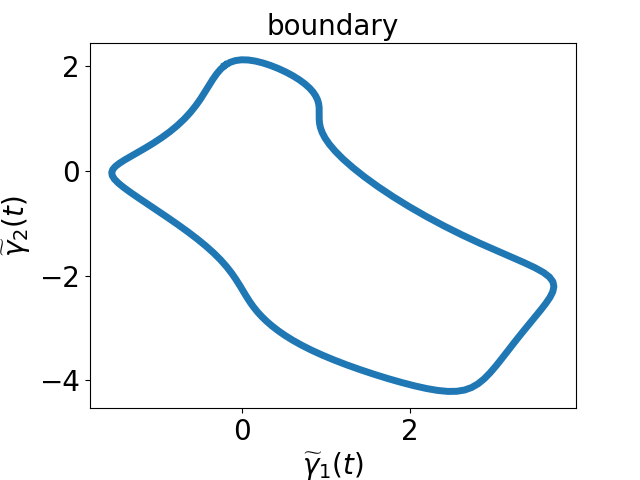}}
			&\raisebox{-.9\height}{\includegraphics[width=3cm]{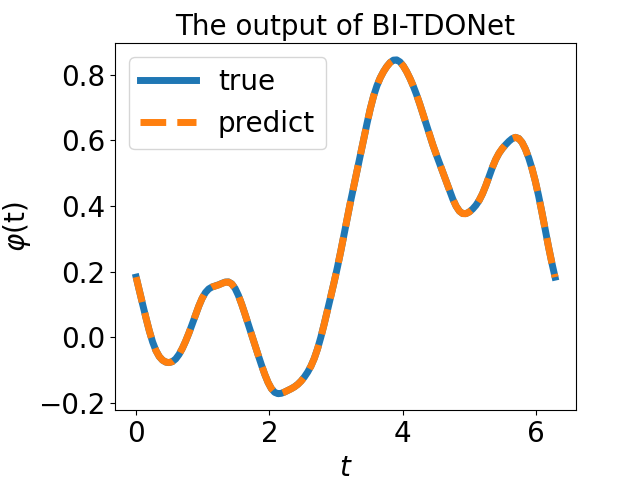}}
			&
			&\raisebox{-.9\height}{\includegraphics[width=3cm]{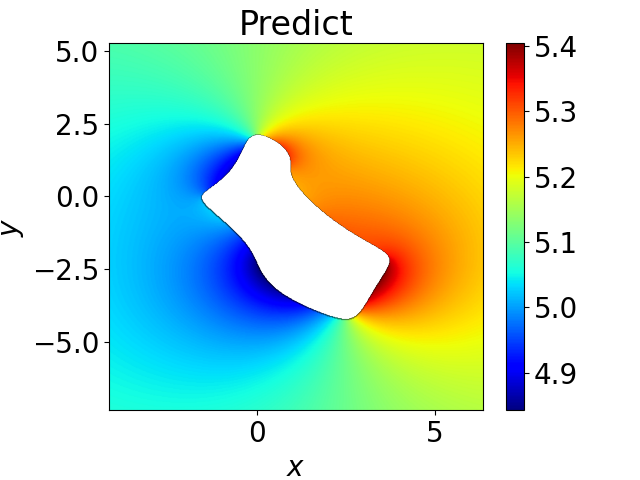}}
			&\raisebox{-.9\height}{\includegraphics[width=3cm]{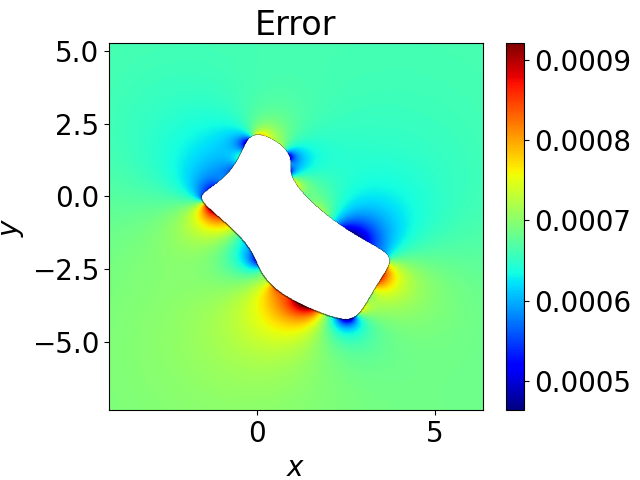}}\\
			\hline
			\raisebox{-6\height}{\tiny BI-DeepOnet}
			&\raisebox{-.9\height}{\includegraphics[width=3cm]{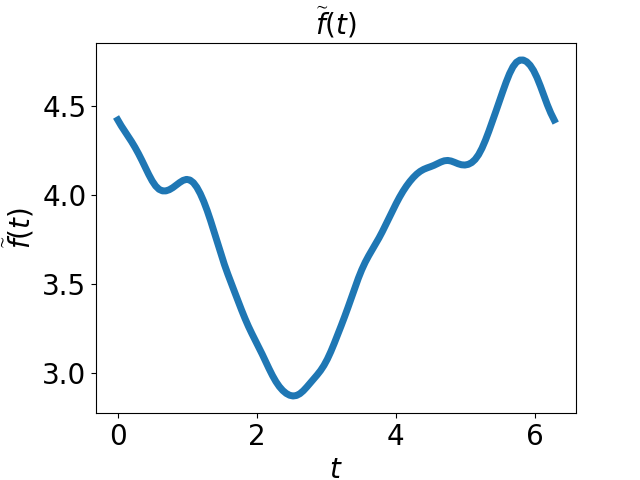}}
			&\raisebox{-.9\height}{\includegraphics[width=3cm]{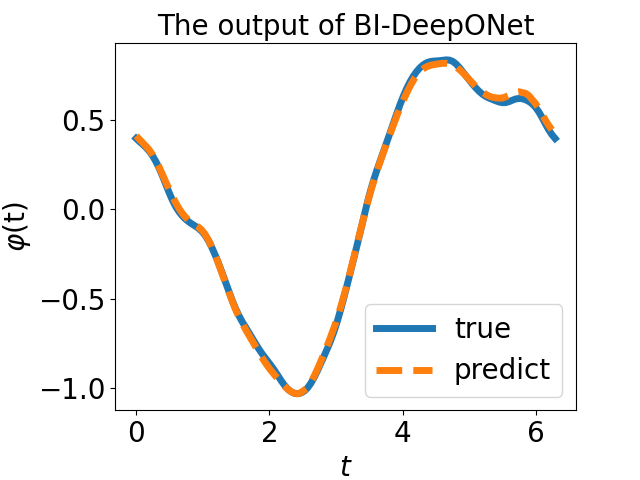}}
			&\multirow{2}*{\tabincell{c}{\raisebox{-1.15\height}{\includegraphics[width=4cm]{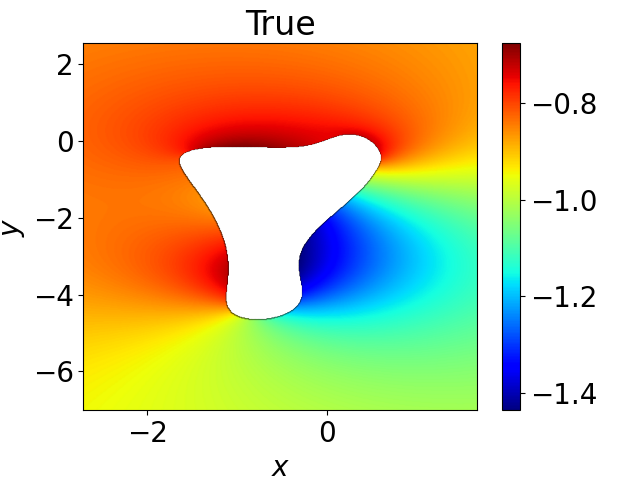}}}} &\raisebox{-.9\height}{\includegraphics[width=3cm]{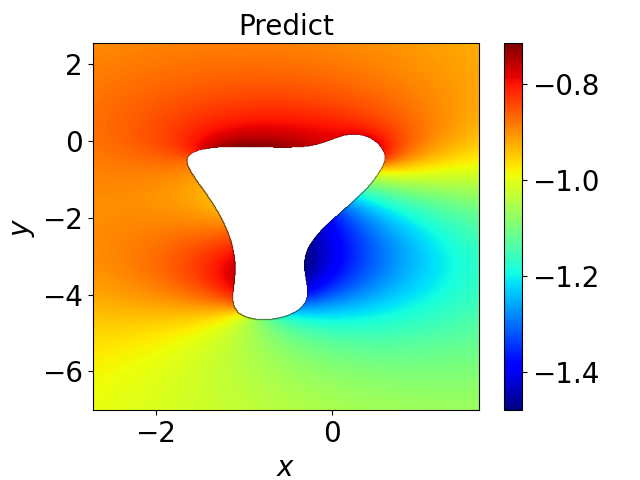}} &\raisebox{-.9\height}{\includegraphics[width=3cm]{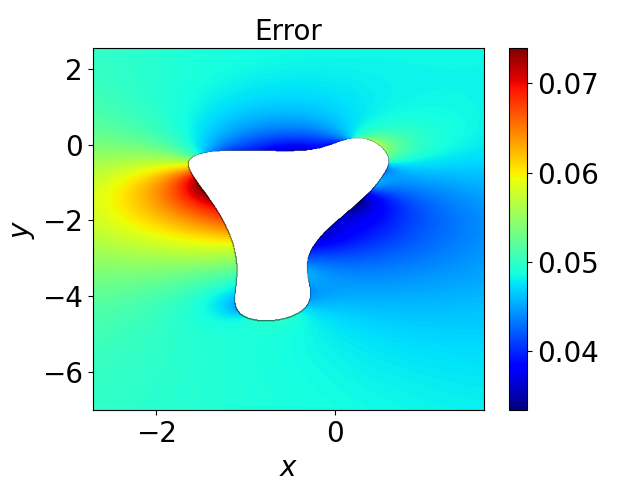}}\\
			\cline{1-1}\cline{3-3}\cline{5-6}
			\raisebox{-6\height}{\tiny BI-TDONet}
			&\raisebox{-.9\height}{\includegraphics[width=3cm]{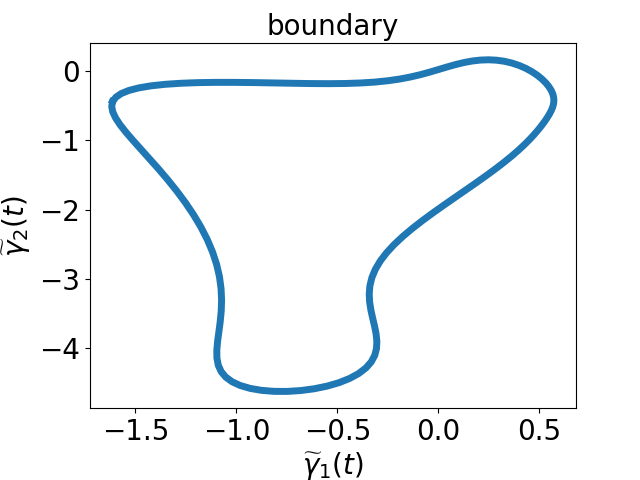}}
			&\raisebox{-.9\height}{\includegraphics[width=3cm]{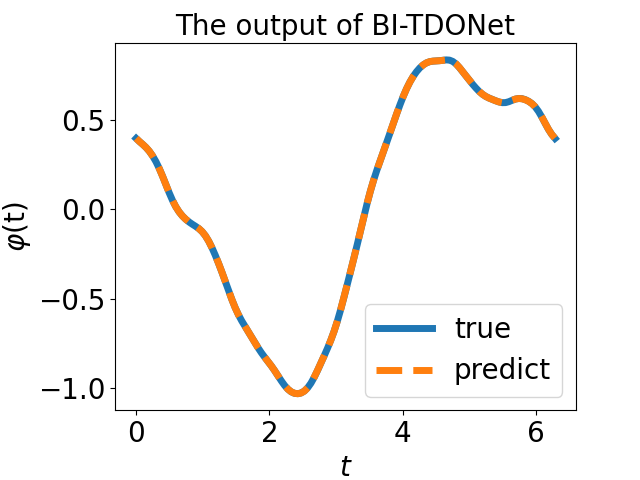}}
			&
			&\raisebox{-.9\height}{\includegraphics[width=3cm]{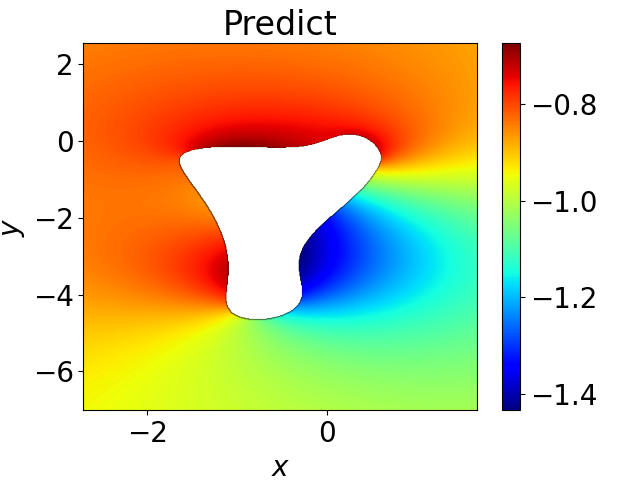}}
			&\raisebox{-.9\height}{\includegraphics[width=3cm]{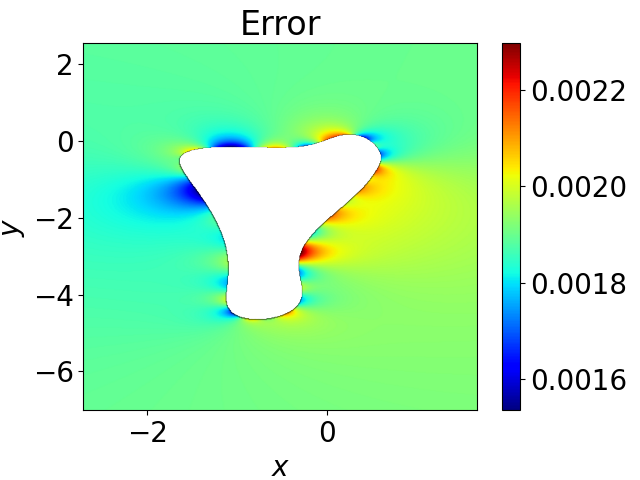}}\\ \hline
		\end{tabular}}
        \caption{The solutions for two randomly selected samples of EDPs as processed by BI-DeepONet and BI-TDONet}
		\label{EDP result}
\end{table}

\begin{table}[ht]
	\vspace{-0.5cm}
	\centering
	\resizebox{\textwidth}{!}{
		\begin{tabular}{|c|c|c|c|c|c|} \hline
			Model
			&Model Input
			&Model Output
			&True solution
			&Predict solution
			&Error\\
			\hline
			\raisebox{-6\height}{\tiny BI-DeepOnet}
			&\raisebox{-.9\height}{\includegraphics[width=3cm]{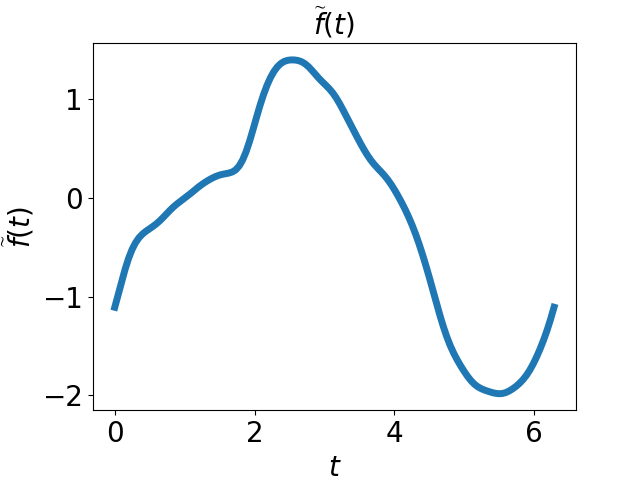}}			
			&\raisebox{-.9\height}{\includegraphics[width=3cm]{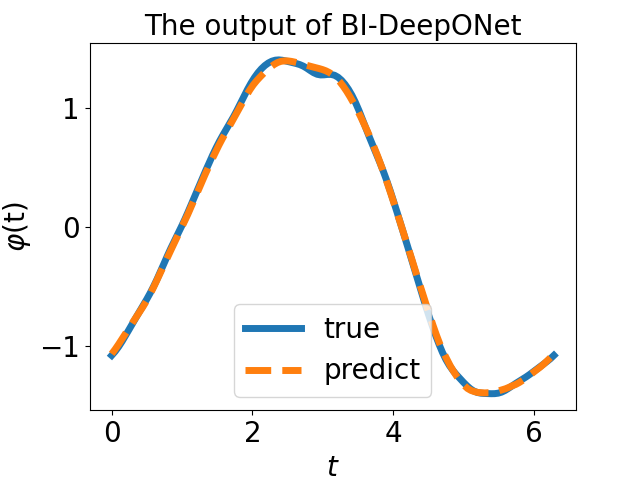}}
			&\multirow{2}*{\tabincell{c}{\raisebox{-1.25\height}{\includegraphics[width=4cm]{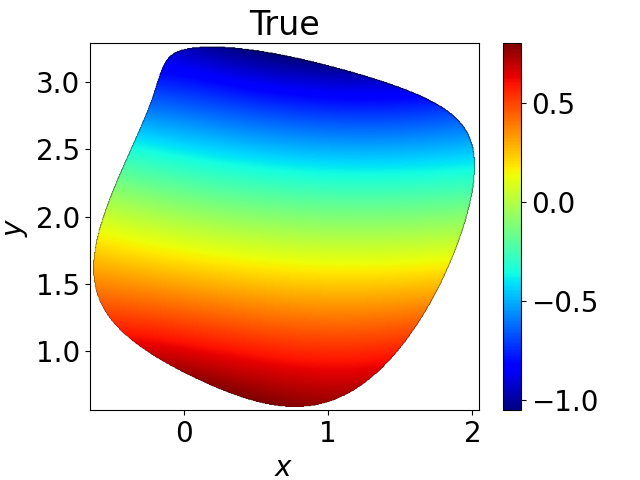}}}}
			&\raisebox{-.9\height}{\includegraphics[width=3cm]{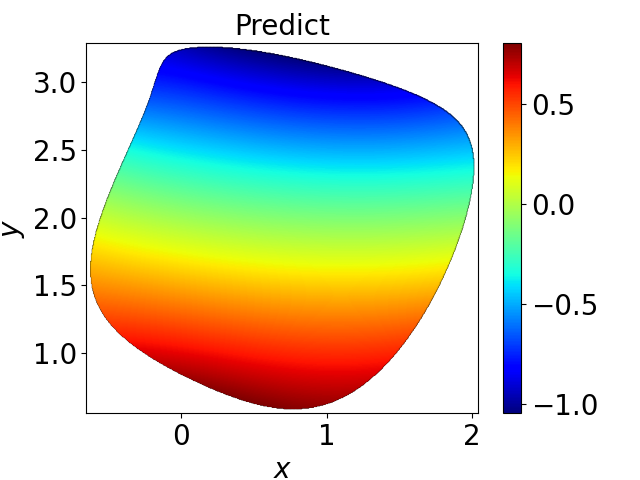}}
			&\raisebox{-.9\height}{\includegraphics[width=3cm]{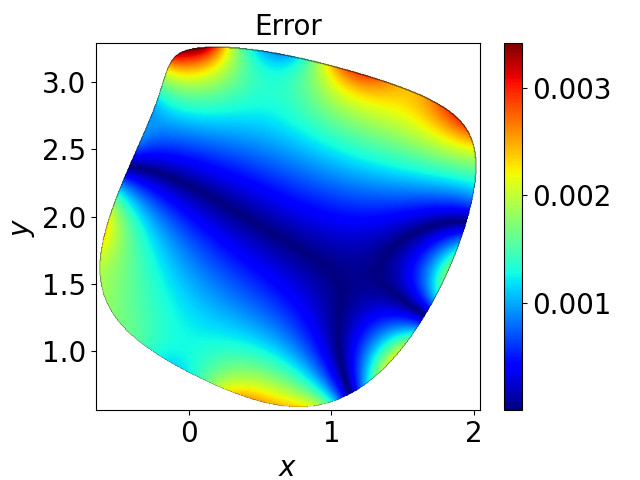}}\\
			\cline{1-1}\cline{3-3}\cline{5-6}
			\raisebox{-6\height}{\tiny{BI-TDONet }}
			&\raisebox{-.9\height}{\includegraphics[width=3cm]{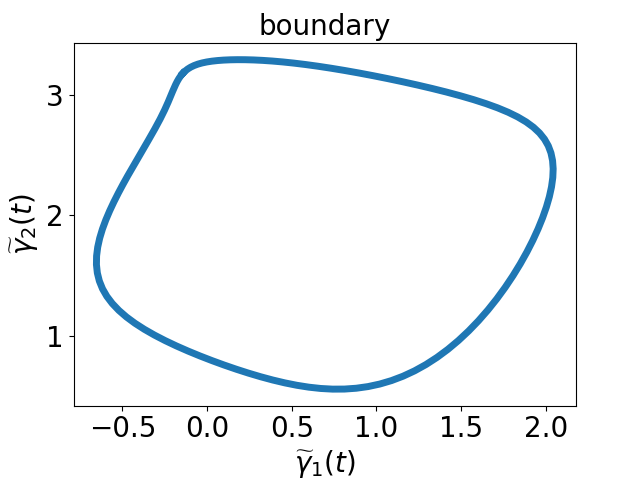}}
			&\raisebox{-.9\height}{\includegraphics[width=3cm]{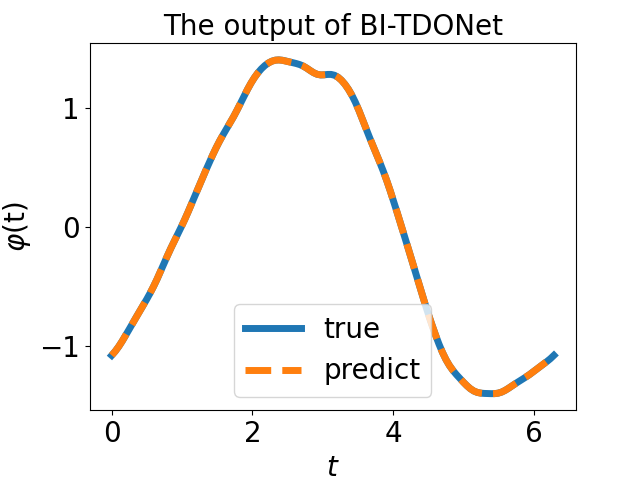}}
			&
			&\raisebox{-.9\height}{\includegraphics[width=3cm]{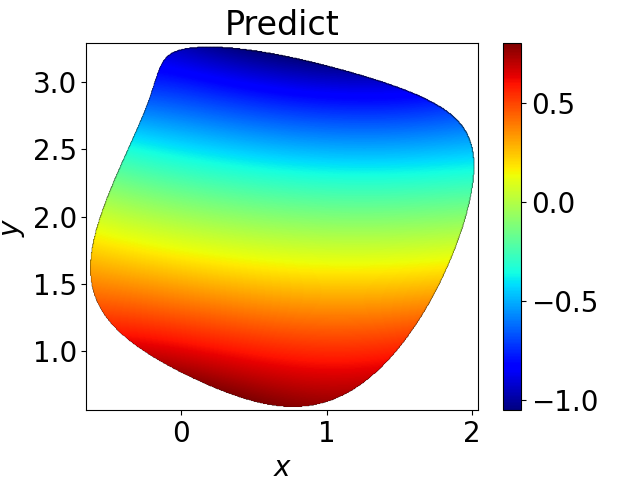}}
			&\raisebox{-.9\height}{\includegraphics[width=3cm]{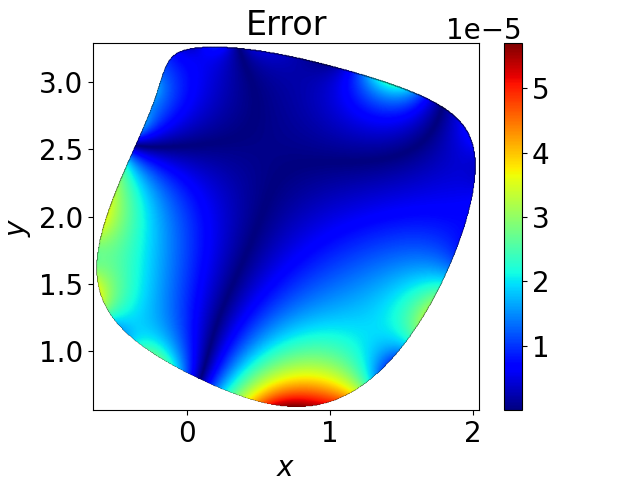}}\\
			\hline
			\raisebox{-6\height}{\tiny BI-DeepONet}
			&\raisebox{-.9\height}{\includegraphics[width=3cm]{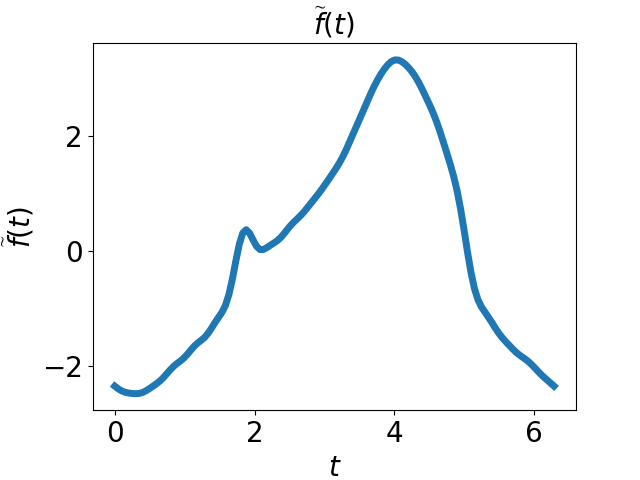}}
			&\raisebox{-.9\height}{\includegraphics[width=3cm]{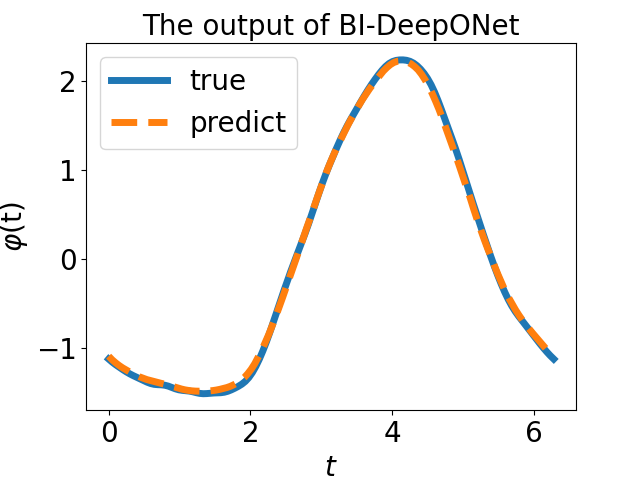}}
			&\multirow{2}*{\tabincell{c}{\raisebox{-1.2\height}{\includegraphics[width=4cm]{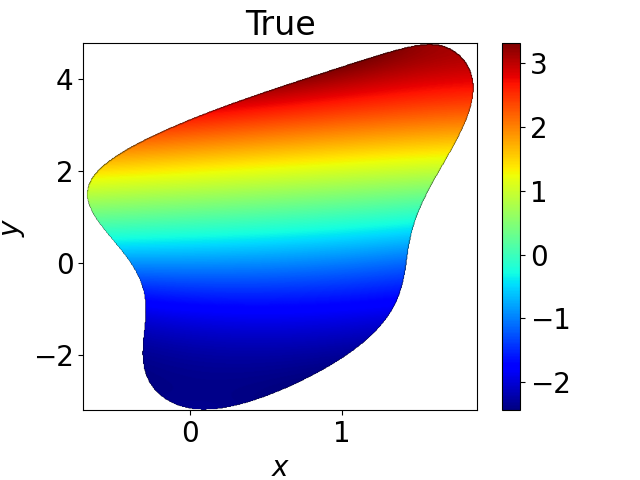}}}}  &\raisebox{-.9\height}{\includegraphics[width=3cm]{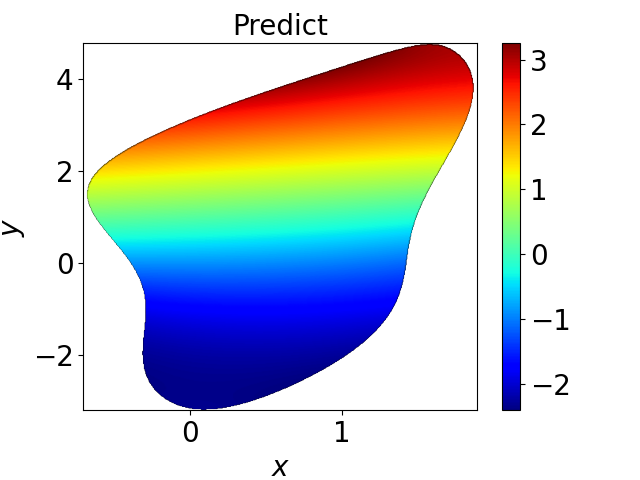}} &\raisebox{-.9\height}{\includegraphics[width=3cm]{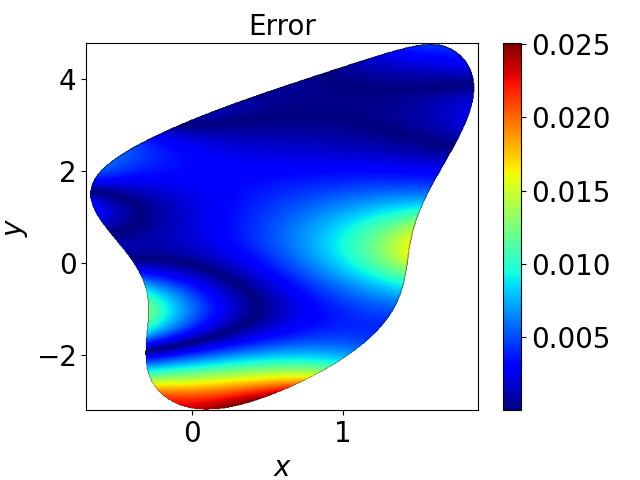}}\\
			\cline{1-1}\cline{3-3}\cline{5-6}
			\raisebox{-6\height}{\tiny BI-TDONet}
			&\raisebox{-.9\height}{\includegraphics[width=3cm]{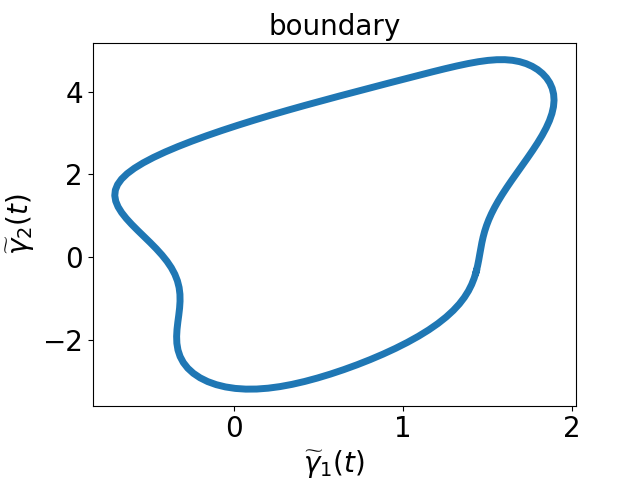}}
			&\raisebox{-.9\height}{\includegraphics[width=3cm]{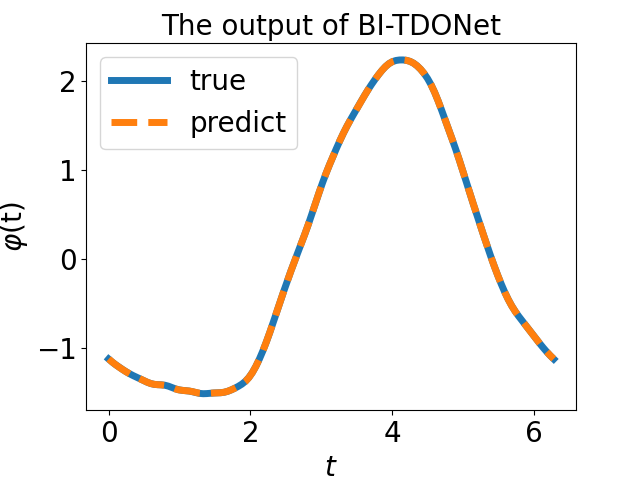}}
			&
			&\raisebox{-.9\height}{\includegraphics[width=3cm]{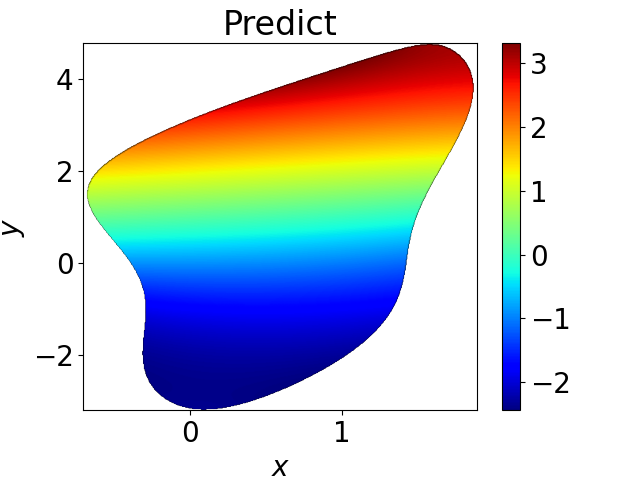}}
			&\raisebox{-.9\height}{\includegraphics[width=3cm]{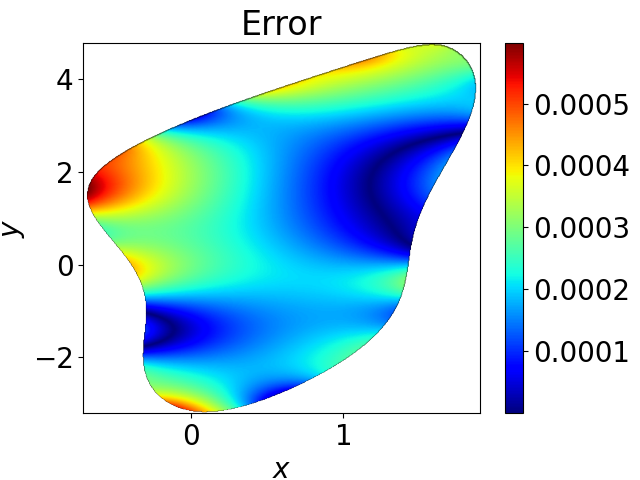}}\\ \hline
		\end{tabular}}
\caption{The solutions for two randomly selected samples of INPs as processed by BI-DeepONet and BI-TDONet}
		\label{INP result}
\end{table}

\begin{table}[ht]
	\vspace{-0.5cm}
	\centering
	\resizebox{\textwidth}{!}{
		\begin{tabular}{|c|c|c|c|c|c|} \hline
			Model
			&Model Input
			&Model Output
			&True solution
			&Predict solution
			&Error \\
			\hline
			\raisebox{-6\height}{\tiny BI-DeepOnet}
			&\raisebox{-.9\height}{\includegraphics[width=3cm]{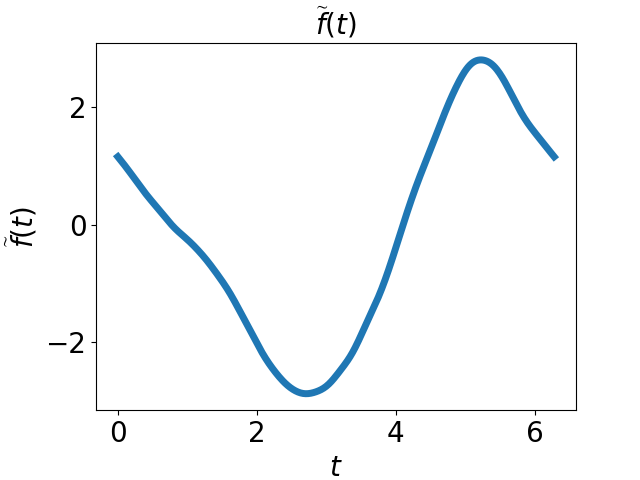}}
			&\raisebox{-.9\height}{\includegraphics[width=3cm]{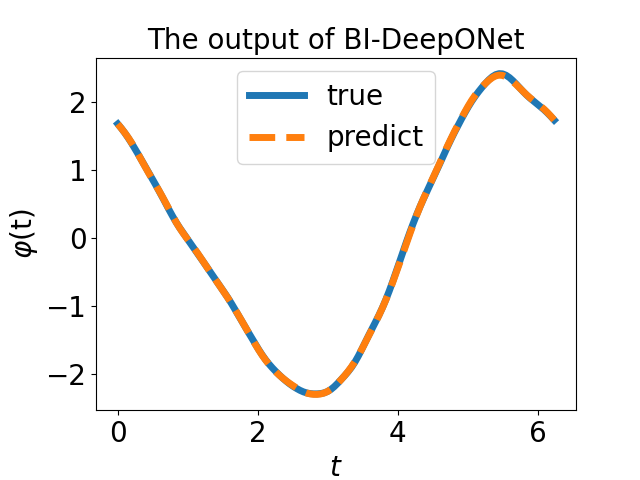}}
			&\multirow{2}*{\tabincell{c}{\raisebox{-1.15\height}{\includegraphics[width=4cm]{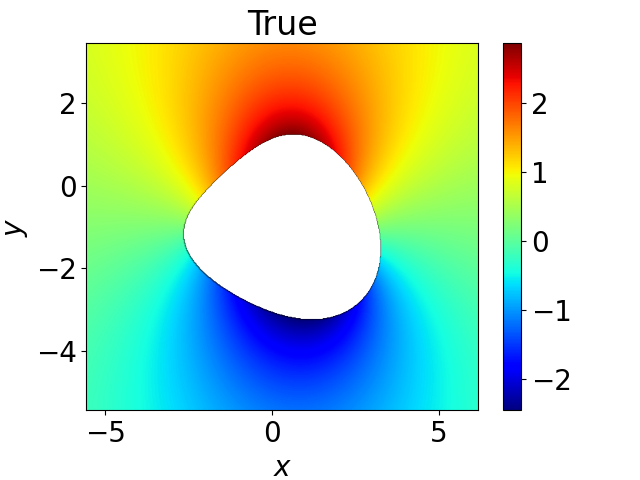}}}}
			&\raisebox{-.9\height}{\includegraphics[width=3cm]{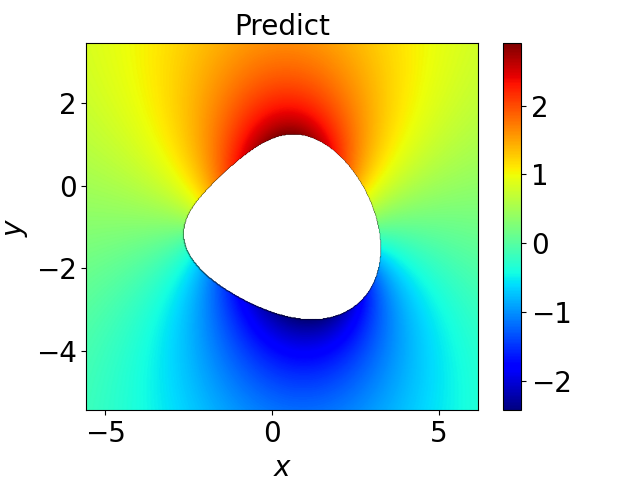}}
			&\raisebox{-.9\height}{\includegraphics[width=3cm]{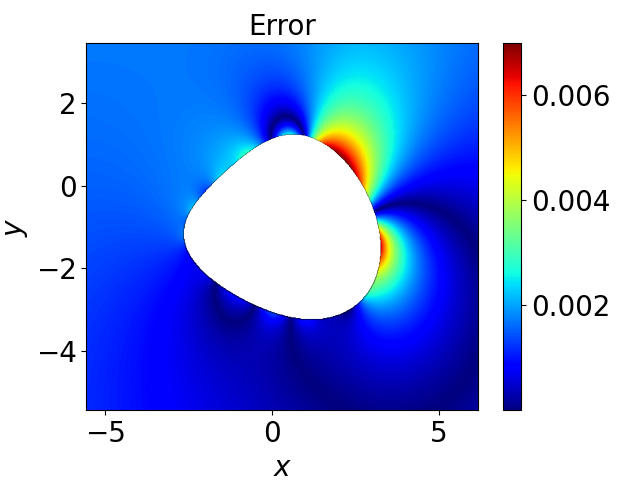}}\\
			\cline{1-1}\cline{3-3}\cline{5-6}
			\raisebox{-6\height}{\tiny{BI-TDONet }}
			&\raisebox{-.9\height}{\includegraphics[width=3cm]{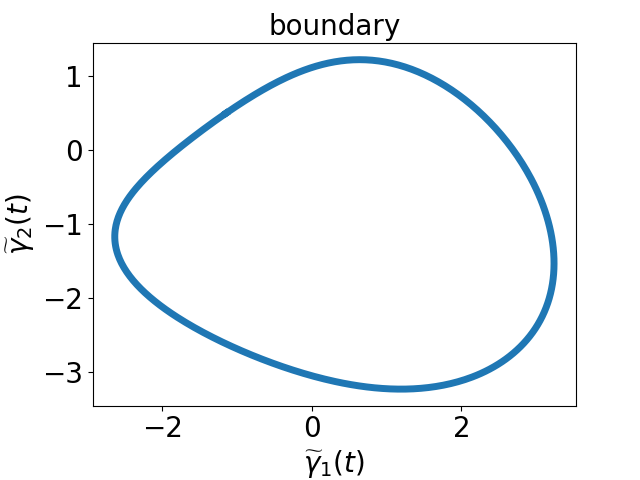}}
			&\raisebox{-.9\height}{\includegraphics[width=3cm]{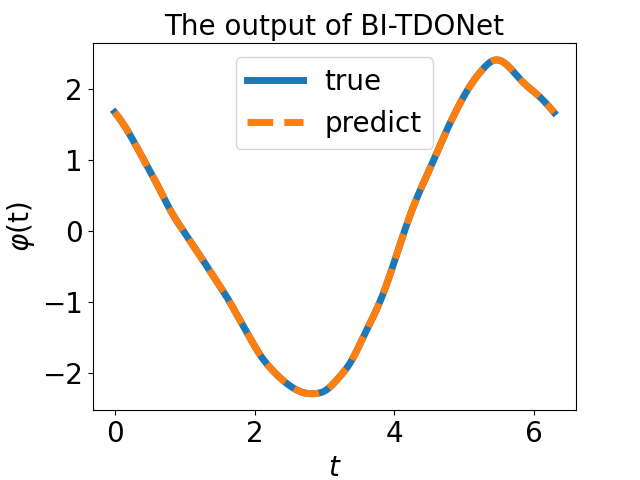}}
			&
			&\raisebox{-.9\height}{\includegraphics[width=3cm]{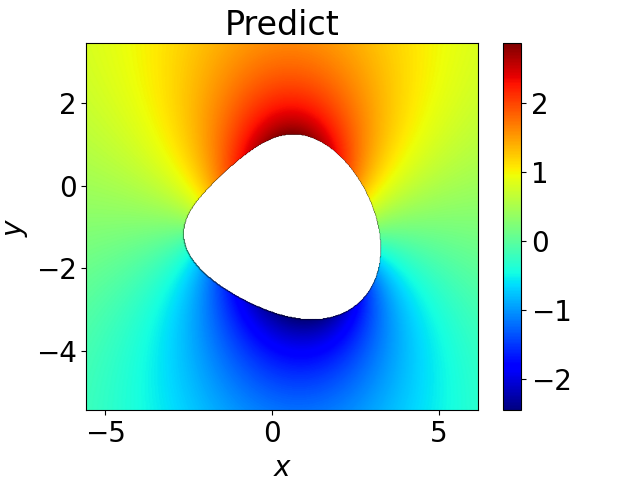}}
			&\raisebox{-.9\height}{\includegraphics[width=3cm]{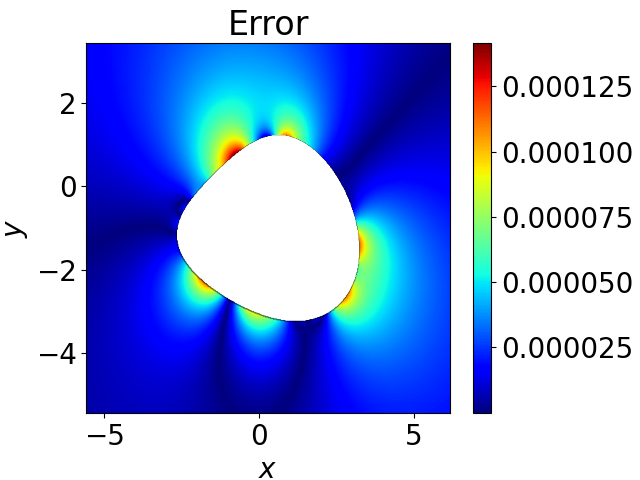}}\\
			\hline
			\raisebox{-6\height}{\tiny BI-DeepONet}
			&\raisebox{-.9\height}{\includegraphics[width=3cm]{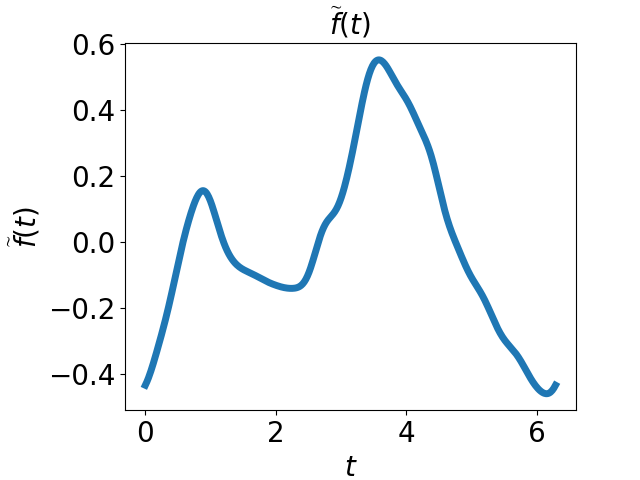}} &\raisebox{-.9\height}{\includegraphics[width=3cm]{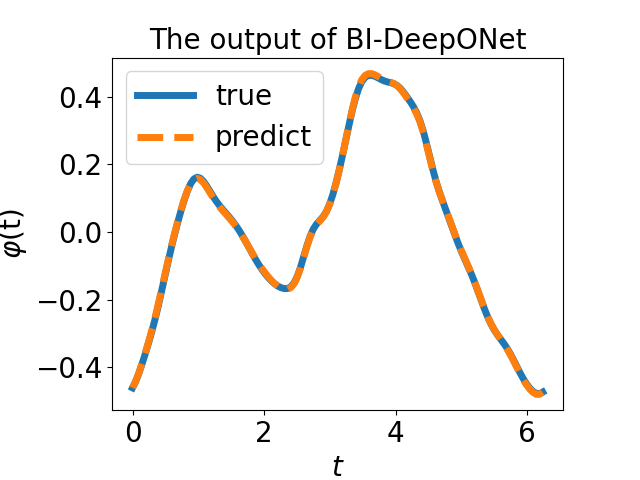}}
			&\multirow{2}*{\tabincell{c}{\raisebox{-1.2\height}{\includegraphics[width=4cm]{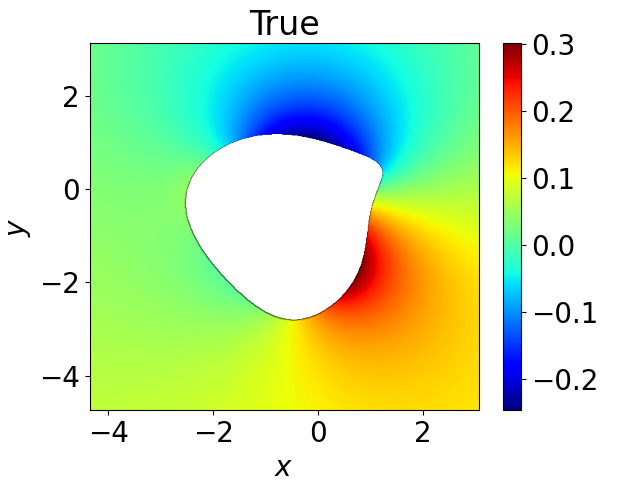}}}} &\raisebox{-.9\height}{\includegraphics[width=3cm]{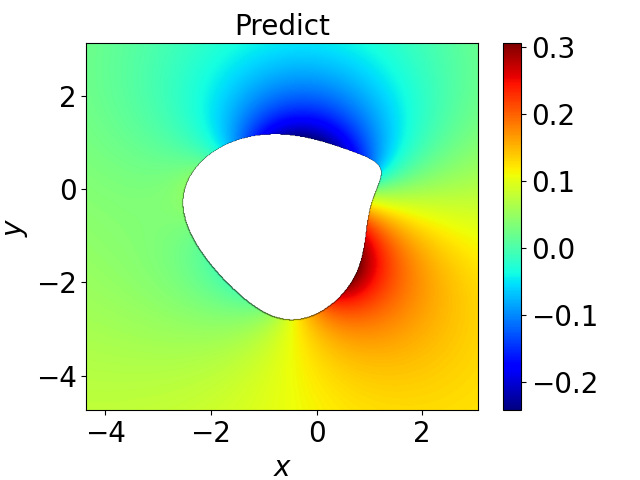}} &\raisebox{-.9\height}{\includegraphics[width=3cm]{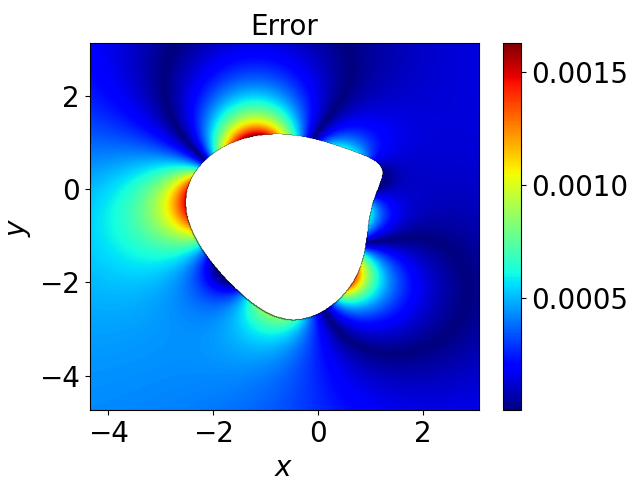}}\\
			\cline{1-1}\cline{3-3}\cline{5-6}
			\raisebox{-6\height}{\tiny BI-TDONet}
			&\raisebox{-.9\height}{\includegraphics[width=3cm]{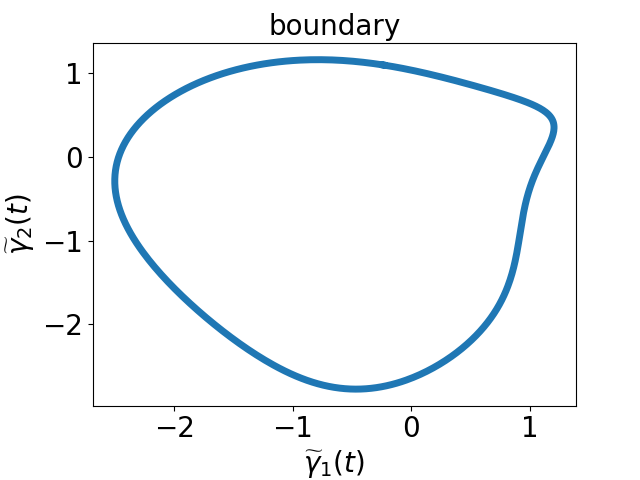}}
			&\raisebox{-.9\height}{\includegraphics[width=3cm]{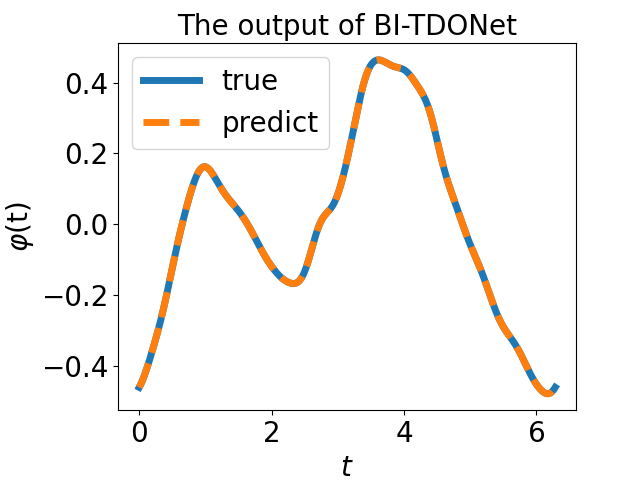}}
			&
			&\raisebox{-.9\height}{\includegraphics[width=3cm]{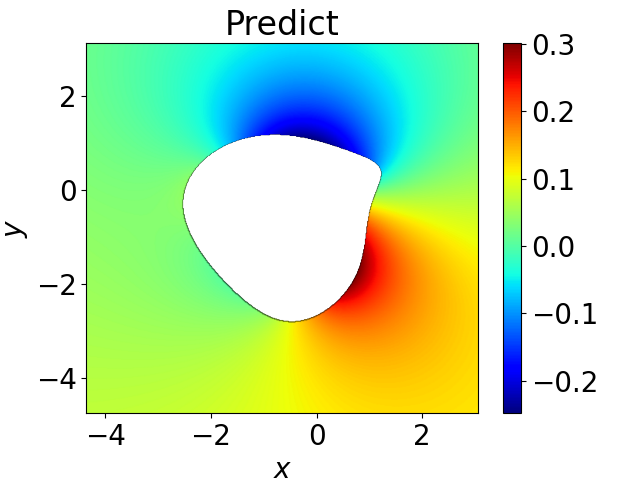}}
			&\raisebox{-.9\height}{\includegraphics[width=3cm]{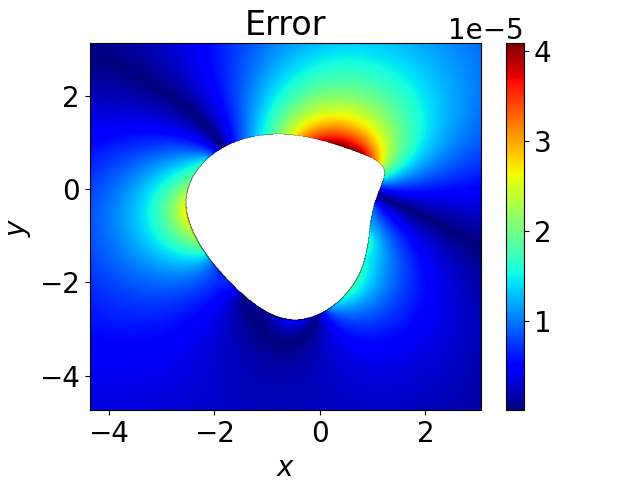}}
			\\ \hline
		\end{tabular}}
\caption{The solutions for two randomly selected samples of ENPs as processed by BI-DeepONet and BI-TDONet}
		\label{ENP result}
\end{table}

\begin{table}[ht]
	\vspace{-0.5cm}
	\centering
	\resizebox{\textwidth}{!}{
		\begin{tabular}{|c|c|c|c|c|c|} \hline
			Model
			&Model Input
			&Model Output
			&True solution
			&Predict solution
			&Error \\
			\hline
			\raisebox{-6\height}{\tiny BI-DeepOnet}
			&\raisebox{-.9\height}{\includegraphics[width=3cm]{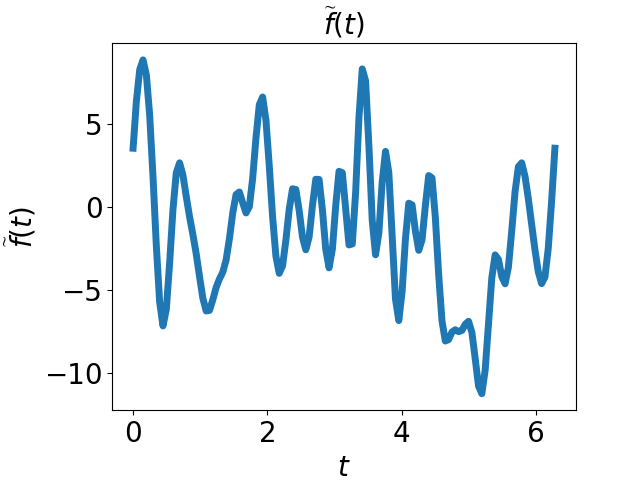}}
			&\raisebox{-.9\height}{\includegraphics[width=3cm]{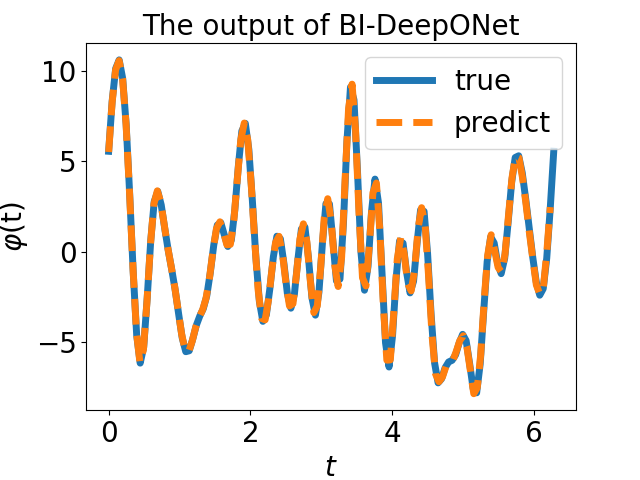}}
			&\multirow{2}*{\tabincell{c}{\raisebox{-1.15\height}{\includegraphics[width=4cm]{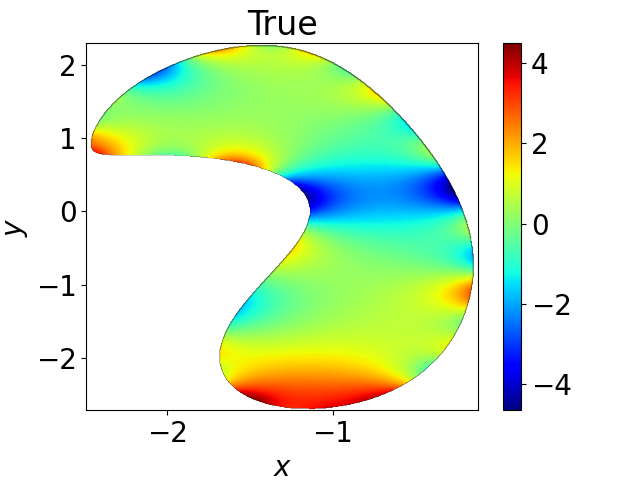}}
			}}
			&\raisebox{-.9\height}{\includegraphics[width=3cm]{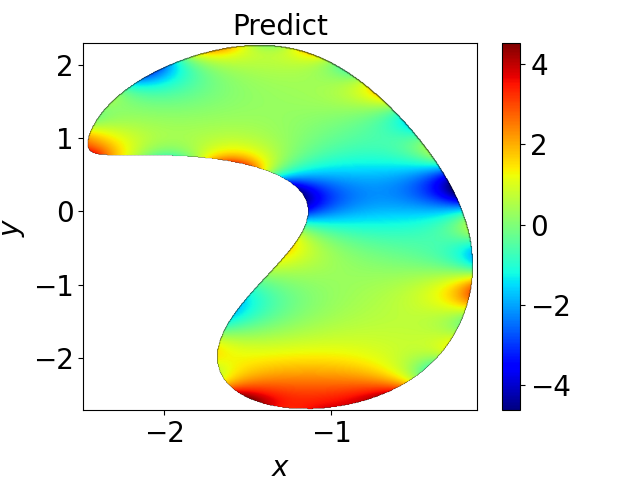}}
			&\raisebox{-.9\height}{\includegraphics[width=3cm]{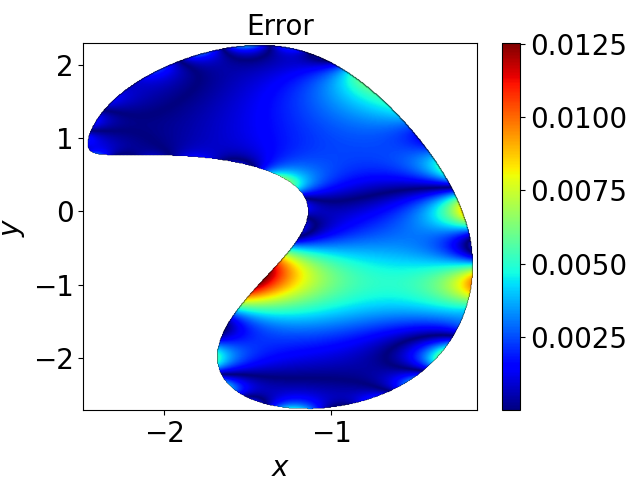}}\\
			\cline{1-1}\cline{3-3}\cline{5-6}
			\raisebox{-6\height}{\tiny{BI-TDONet }}
			&\raisebox{-.9\height}{\includegraphics[width=3cm]{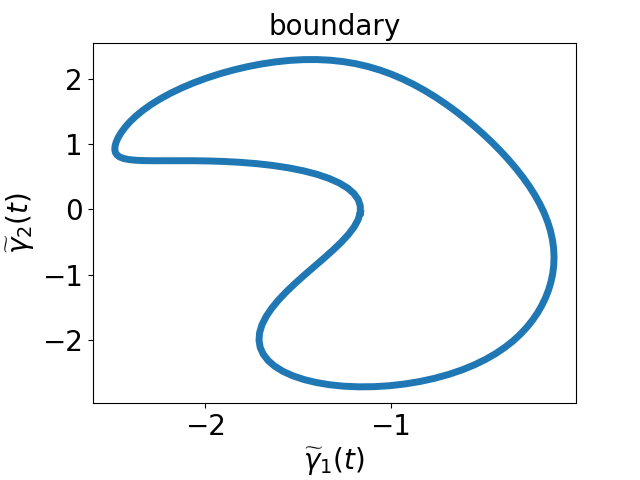}}
			&\raisebox{-.9\height}{\includegraphics[width=3cm]{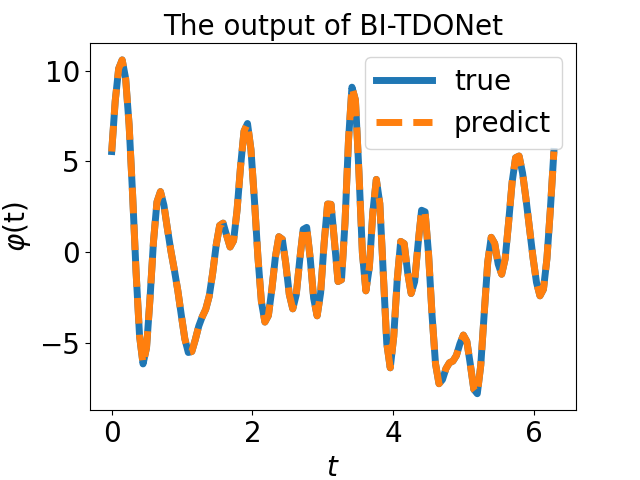}}
			&
			&\raisebox{-.9\height}{\includegraphics[width=3cm]{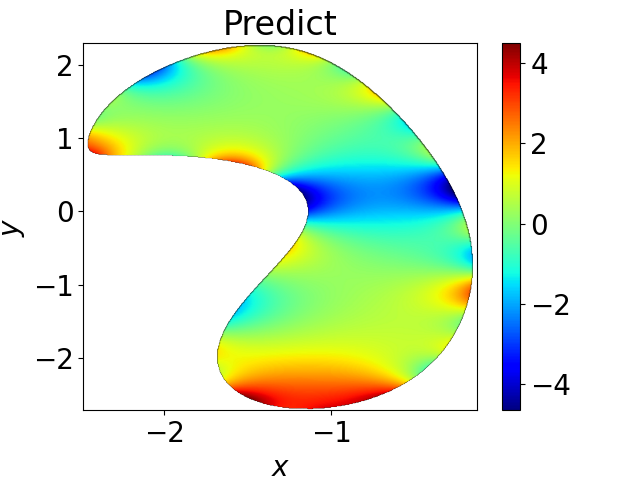}}
			&\raisebox{-.9\height}{\includegraphics[width=3cm]{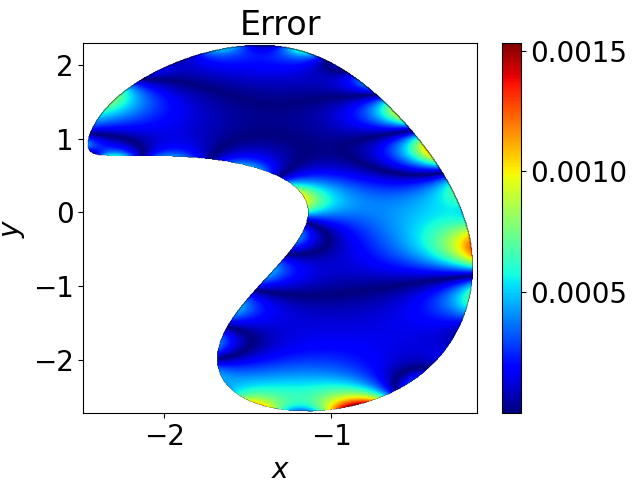}}\\
			\hline
			\raisebox{-6\height}{\tiny BI-DeepONet}
			&\raisebox{-.9\height}{\includegraphics[width=3cm]{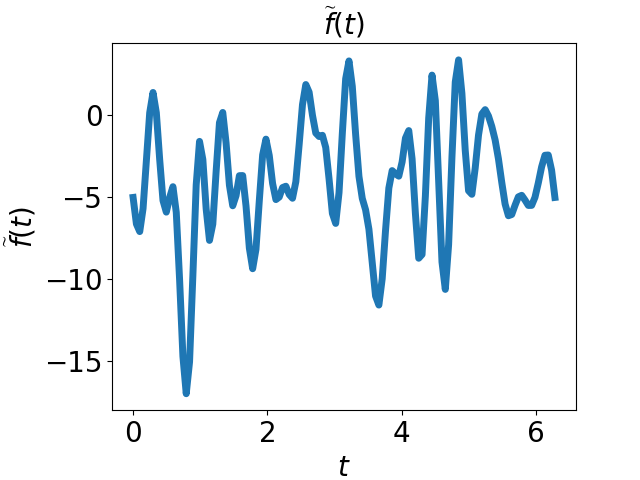}} &\raisebox{-.9\height}{\includegraphics[width=3cm]{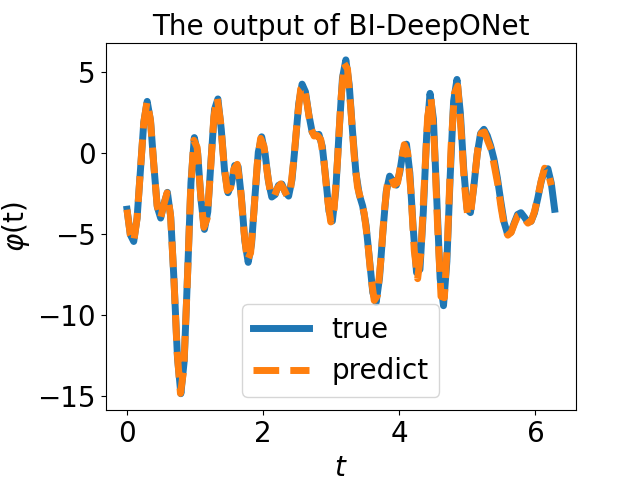}}
			&\multirow{2}*{\tabincell{c}{\raisebox{-1.2\height}{\includegraphics[width=4cm]{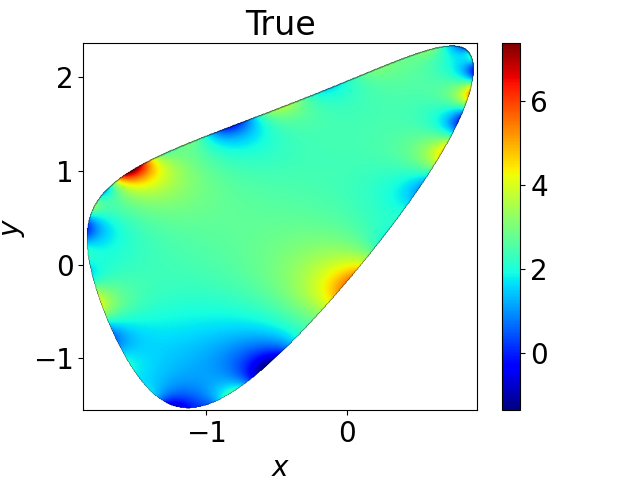}}}} &\raisebox{-.9\height}{\includegraphics[width=3cm]{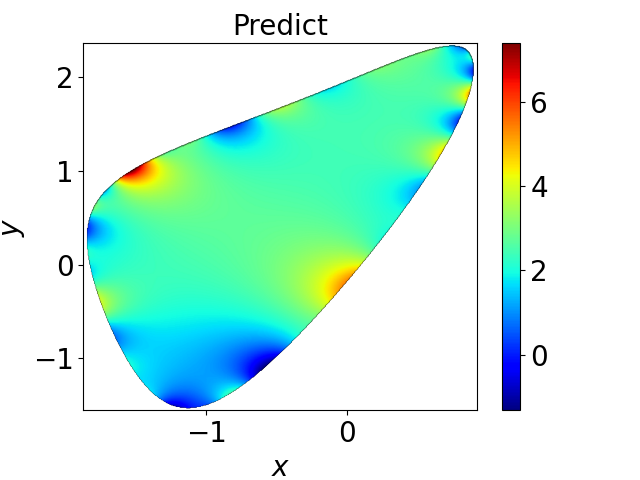}}
			&\raisebox{-.9\height}{\includegraphics[width=3cm]{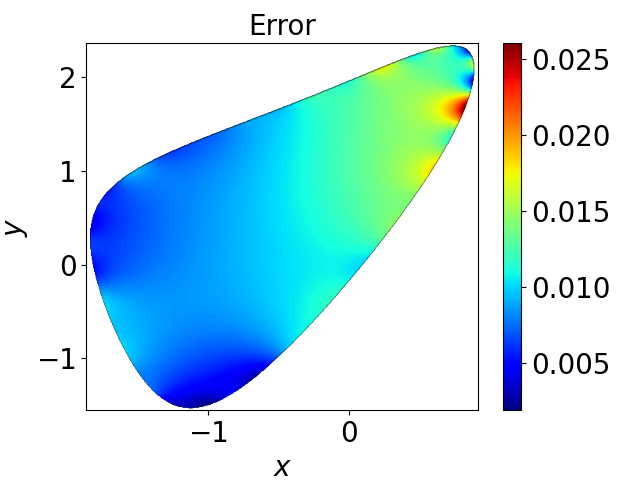}}\\
			\cline{1-1}\cline{3-3}\cline{5-6}
			\raisebox{-6\height}{\tiny BI-TDONet}
			&\raisebox{-.9\height}{\includegraphics[width=3cm]{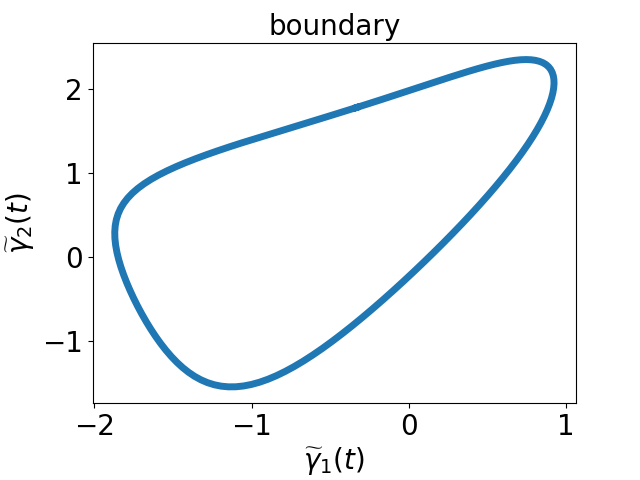}}
			&\raisebox{-.9\height}{\includegraphics[width=3cm]{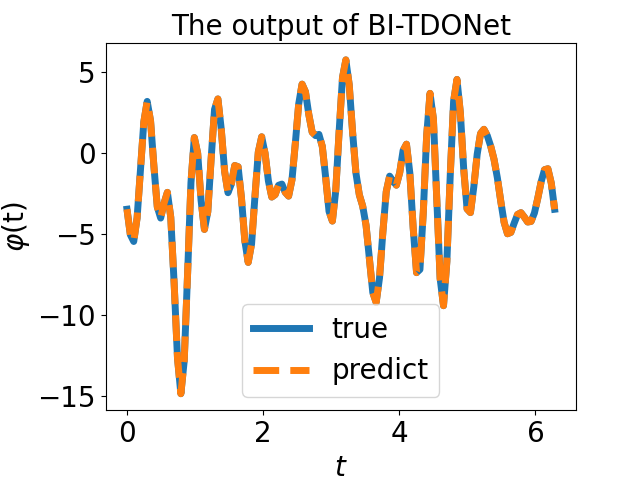}}
			&
			&\raisebox{-.9\height}{\includegraphics[width=3cm]{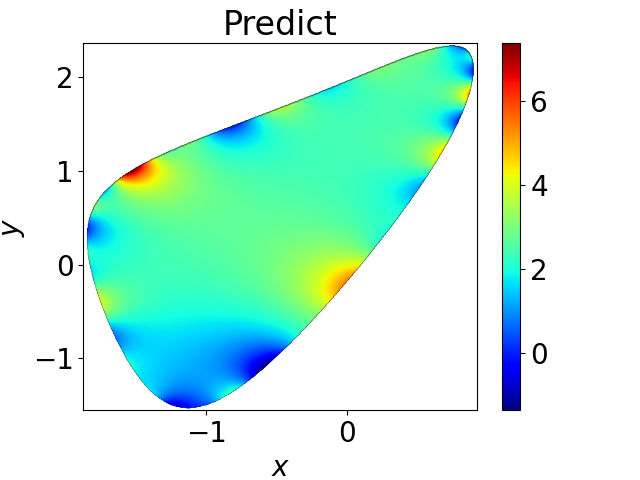}}
			&\raisebox{-.9\height}{\includegraphics[width=3cm]{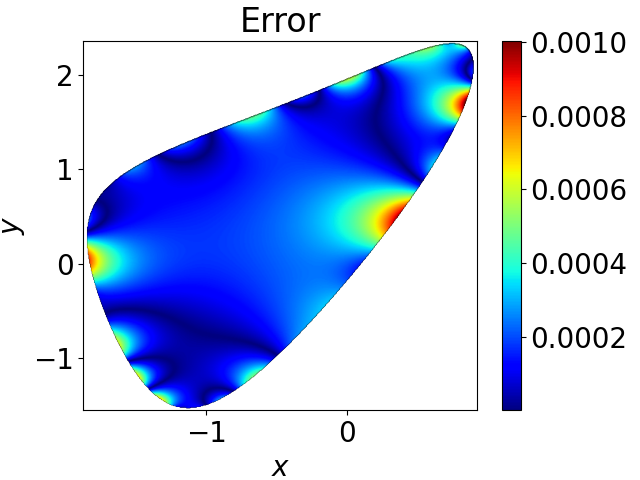}}\\
			\hline
		\end{tabular}}
\caption{The solutions for two randomly selected samples of HFIDPs as processed by BI-DeepONet and BI-TDONet}
		\label{osc result}
\end{table}

These visual results indicate that BI-TDONet significantly outperforms BI-DeepONet in solving LBVPs defined across various 2D domains. Moreover, both operator learning frameworks based on BIEs have been proven effective in addressing LBVPs across different 2D domains.

\subsection{Potential flow}

In this subsection, we explore the example of a two-dimensional potential flow that transitions uniformly from the negative to the positive half of the $x$-axis at a velocity of $v_0$, while circumventing an obstacle. In the context of potential flow problems, the velocity field $\boldsymbol v$ is derived as the gradient of the velocity potential function $u$, represented by $\boldsymbol v: = \nabla u$. Given that potential flow is incompressible, it follows that $\nabla \cdot \boldsymbol{v} = 0$. This condition ensures that the velocity potential function must satisfy the Laplace equation (see \cite{hess1967calculation})
\begin{equation*}
	\left\{
	\begin{aligned}
		\Delta u(\boldsymbol{x})=0\qquad \boldsymbol{x}&\in D,\\
		\frac{\partial u(\boldsymbol{x})}{\partial \nu(\boldsymbol{x}) }=0 \qquad \boldsymbol{x}&\in\partial  D.
	\end{aligned}
	\right.
\end{equation*}

In addressing potential flow problems, the velocity potential $u$ is typically split into two components: $u_1$, which corresponds to the uniform flow with a velocity field $\boldsymbol{v} = (v_0, 0)$, and $u_2$, a perturbation velocity potential that decays at infinity. This decomposition transforms the problem into solving an exterior Neumann Boundary Value Problem (NBVP) as follows:
\begin{equation}\label{potential}
\left\{
\begin{aligned}
\Delta u_2(\boldsymbol{x}) &= 0 &\quad \boldsymbol{x} &\in D,\\
\frac{\partial u_2(\boldsymbol{x})}{\partial \nu(\boldsymbol{x})} &= -v_0n_1 &\quad \boldsymbol{x} &\in\partial D,
\end{aligned}
\right.
\end{equation}
where $n_1$ represents the first component of the boundary's outward unit normal vector. This setup ensures that the perturbed flow component, $u_2$, meets the necessary conditions at the boundary and in the domain to account for the presence of obstacles.

Although models for the ENP were trained as detailed in Subsection 4.1, their effectiveness is limited by the amount of training data, which hampers the achievement of high precision. Consequently, in this subsection, we aim to enhance the training process by incorporating datasets that meet the criteria outlined in \eqref{potential}, followed by retraining the models. It is important to note that the boundary conditions specified in \eqref{potential} depend solely on the initial velocity $v_0$ and the boundary’s outward unit normal vector. For velocity $v_0$, which is uniformly sampled between $1$ and $10$, $200$ values are generated. Together with the $5,998$ boundaries created in Subsection 3.3, this approach results in a total of $1,199,600$ samples. These samples are thoroughly integrated with those from the ENP to compose the training and testing datasets for this subsection, resulting in a grand total of $2,399,200$ samples. Of these, $80\%$ are designated for training, while the remaining $20\%$ are set aside for testing. Given that half of the training dataset is derived from ENP, and the other half is generated according to \eqref{potential}, the dataset exhibits richer features. Consequently, in this subsection, we consider expanding the network scale to better capture this wealth of information. The network structure for BI-DeepONet is set as $[[256, 500, 500, 500, 500]$, $[128, 500, 500, 500, 500]$, $[1, 500, 500, 500, 500]]$. The structure of BI-TDONet is defined as $[[123, 500, 500, 500, 500, 41]$, $[82, 500, 500, 500, 500, 82]$, $[123, 500, 500, 500, 500, 41]]$.

\begin{table}[htb]
	\centering
	\begin{tabular}{c|c|c|c|c|c}
		\midrule
		Model
		&MNE
		&MRE
		&variance-MNE
		&variance-MRE
		&Mean-Time/ms\\
		\hline
		BI-DeepONet
		&$1.7753\times10^{-2}$
		&$8.1782\times10^{-3}$
		&$4.8971\times10^{-4}$
		&$1.9093\times10^{-4}$
		&$3.1752\times10^{-4}$\\
		\hline
		BI-TDONet
		&$1.2241\times10^{-2}$
		&$8.2359\times10^{-4}$
		&$1.8848\times10^{-4}$
		&$2.7980\times10^{-6}$
		&$3.1614\times10^{-4}$\\
		\midrule
	\end{tabular}	
\caption{Errors of BI-DeepONet and BI-TDONet for the potential flow problems on test set}
	\label{potential error}
\end{table}

During the training process of BI-DeepONet, we set the batch size to $8,192$ and the number of iterations to $3,000,000$. With the training dataset's sample size doubled, the number of epochs is effectively halved given the same number of iterations, resulting in approximately $12,500$ epochs. The initial learning rate was set at $0.001$, utilizing the Adam optimization algorithm. The learning rate decay was implemented on an inverse-time schedule, with a decay cycle every $1/100$ of the iterations and a decay rate of $0.5$.

For the training process of BI-TDONet, we also set the batch size to $8,192$ and targeted a total of $5,000$ epochs. The initial learning rate was established at $0.001$, again using the Adam optimization algorithm. The learning rate decay was dynamically managed by monitoring performance; if the loss does not decrease over $1/100$ of the epochs, the learning rate is halved. This approach ensures timely adjustments to maintain training efficiency.

Table \ref{potential error} displays the  MNEs and MREs for BI-DeepONet and BI-TDONet on the test set, including the variances of these metrics. The data reveal that the average of MREs for BI-DeepONet is approximately ten times that for BI-TDONet, yet BI-DeepONet still achieves the average of MREs of $8.1782\times10^{-3}$, demonstrating the robust generalization capabilities of both models on the test set. The final column in Table \ref{potential error} provides the average inference time per sample, illustrating that both BI-DeepONet and BI-TDONet not only deliver high accuracy but also process data swiftly. This performance underscores the effectiveness of operator learning methods in computational efficiency and accuracy.

\begin{figure}[htbp]
	\vspace{-0.4cm}
	\centering
	\begin{subfigure}{.32\linewidth}
		\includegraphics[width=\textwidth]{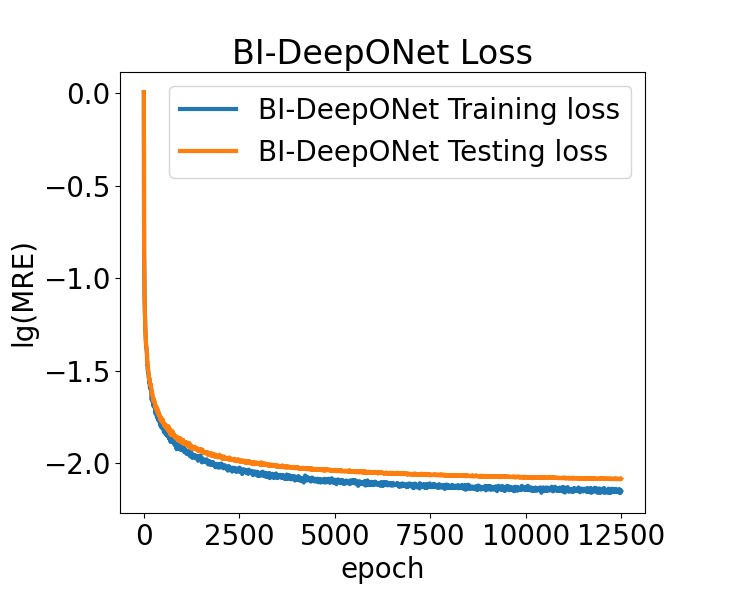}
		\subcaption{Training and testing loss of BI-DeepONet}
	\end{subfigure}
	\begin{subfigure}{.32\linewidth}
		\includegraphics[width=\textwidth]{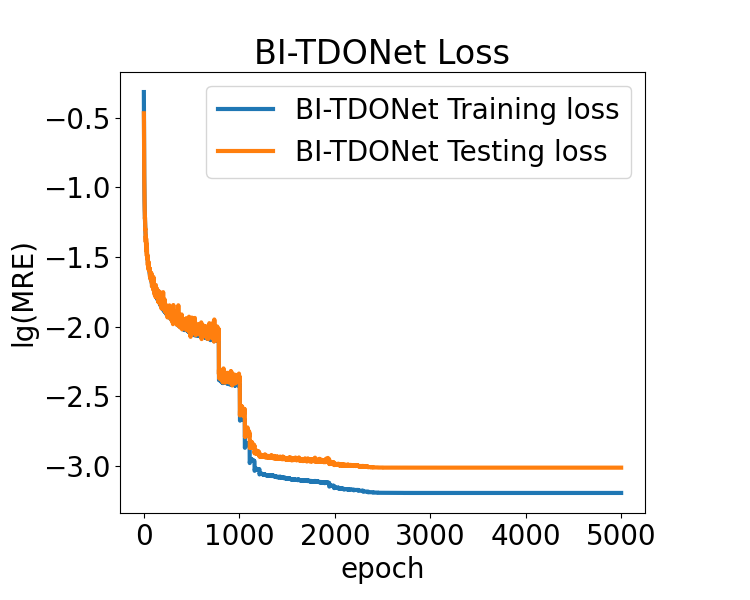}
		\subcaption{Training and testing loss of BI-TDONet}
	\end{subfigure}
	\caption{The logarithmic loss trajectories for BI-DeepONet and BI-TDONet in the problem of potential flow around obstacles.}
	\label{potential_loss}
\end{figure}

Figure \ref{potential_loss} clarifies the logarithmic loss trajectories of the models throughout their entire training and testing process. Observations from this figure reveal that BI-TDONet exhibits a faster convergence rate and achieves lower training and testing losses compared to BI-DeepONet. These findings indicate that BI-TDONet not only optimizes more efficiently but also excels in minimizing errors during both the training and testing phases, demonstrating its superior performance.

\begin{figure}[htbp]
		\vspace{-0.8cm}
	\centering
	\begin{subfigure}{.32\linewidth}
		\includegraphics[width=\textwidth]{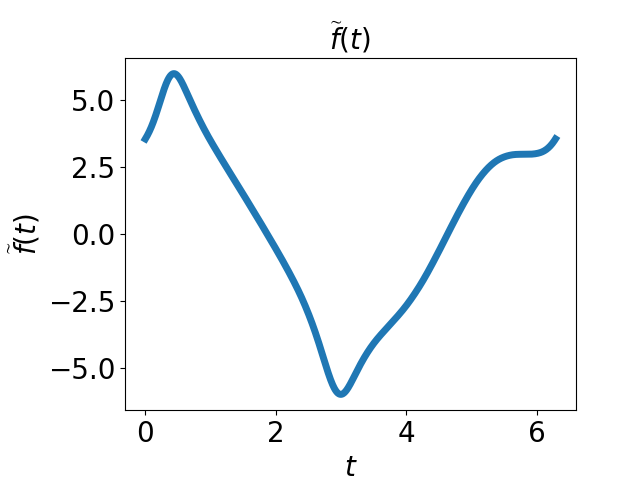}
		\subcaption{The right-hand side function $\widetilde{f}$ of the BIE}
	\end{subfigure}
	\begin{subfigure}{.32\linewidth}
		\includegraphics[width=\textwidth]{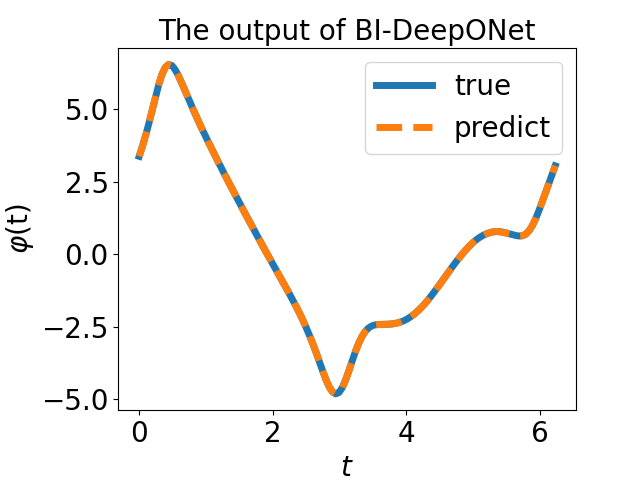}
		\subcaption{The predicted solution of the BI-DeepONet.}
	\end{subfigure}
	\begin{subfigure}{.32\linewidth}
		\includegraphics[width=\textwidth]{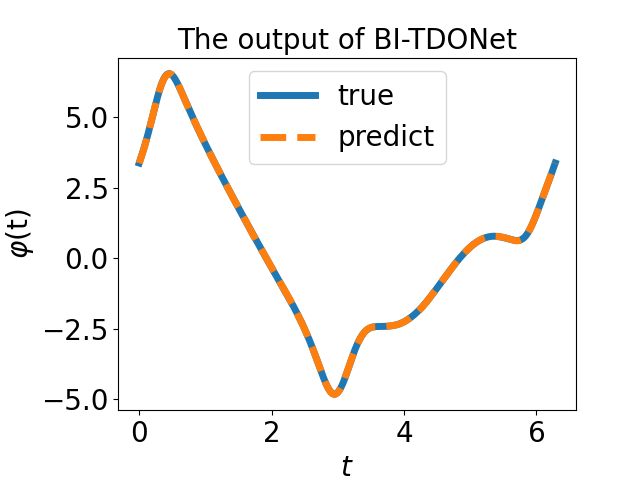}
		\subcaption{The predicted solution of the BI-TDONet.}
	\end{subfigure}
	\begin{subfigure}{.32\linewidth}
		\includegraphics[width=\textwidth]{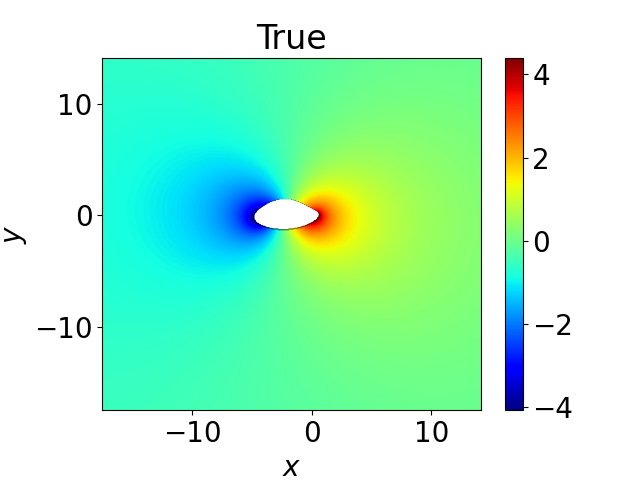}
		\subcaption{The true solution of the velocity potential function $ u_2$}
	\end{subfigure}
	\begin{subfigure}{0.32\textwidth}
		\centering
		\includegraphics[width=\textwidth]{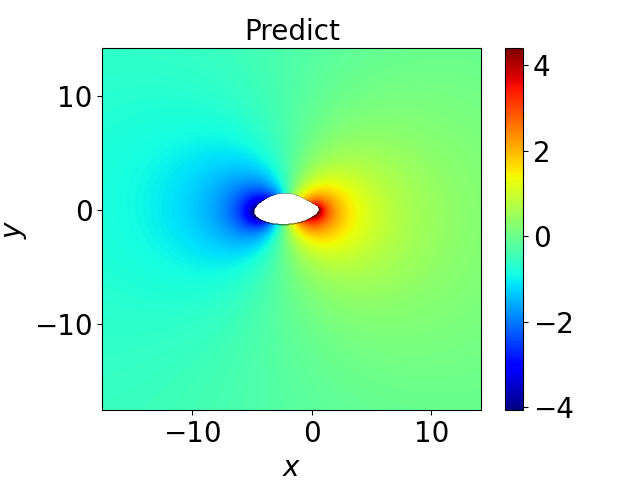}
		\subcaption{The predicted solution of the velocity potential function $ u_2$ in BI-DeepONet}
	\end{subfigure}
	\begin{subfigure}{0.32\textwidth}
		\centering
		\includegraphics[width=\textwidth]{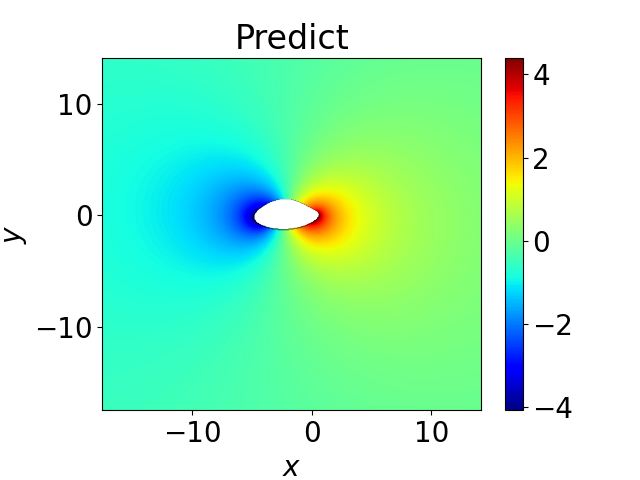}
		\subcaption{The predicted solution of the velocity potential function $ u_2$ in BI-TDONet}
	\end{subfigure}
	\begin{subfigure}{0.32\textwidth}
		\centering
		\includegraphics[width=\textwidth]{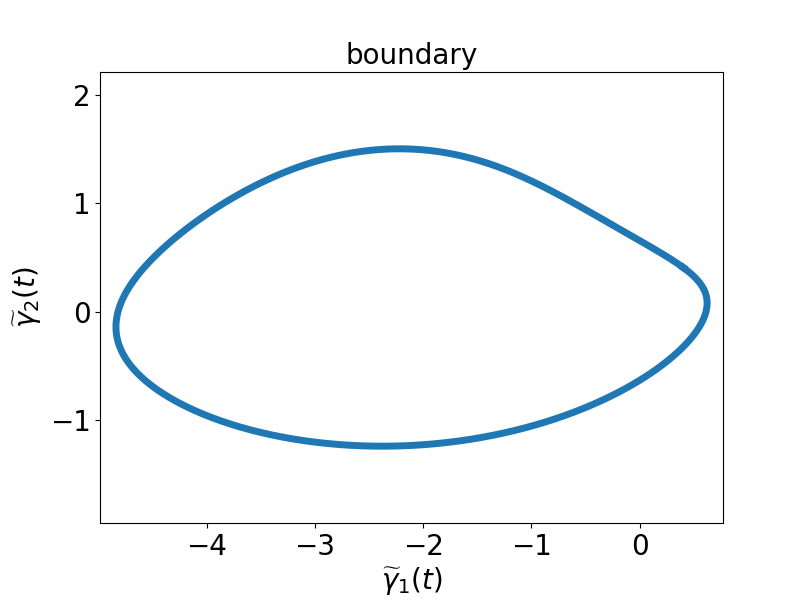}
		\subcaption{The geometric of the two-dimensional potential flow around an obstacle problem}
	\end{subfigure}
	\begin{subfigure}{0.32\textwidth}
		\centering
		\includegraphics[width=\textwidth]{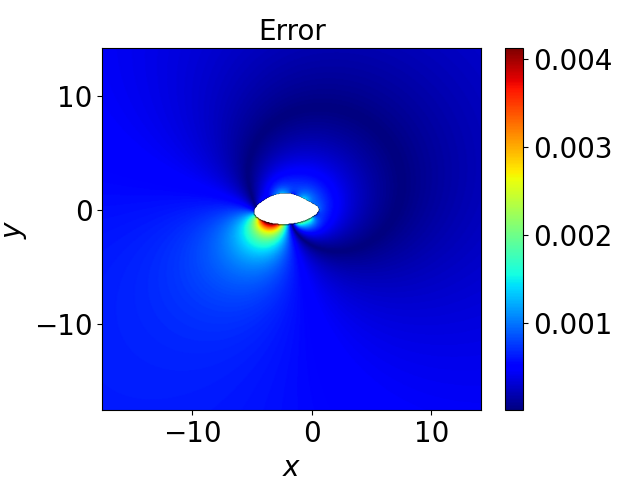}
		\subcaption{	The error of the velocity potential function $ u_2 $ in BI-DeepONet}
	\end{subfigure}
	\begin{subfigure}{0.32\textwidth}
		\centering
		\includegraphics[width=\textwidth]{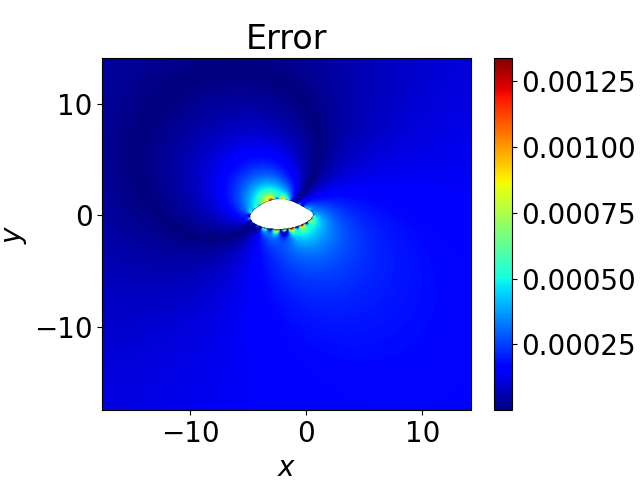}
		\subcaption{	The error of the velocity potential function $ u_2$ in BI-TDONet}
	\end{subfigure}
	\begin{subfigure}{.32\linewidth}
		\includegraphics[width=\textwidth]{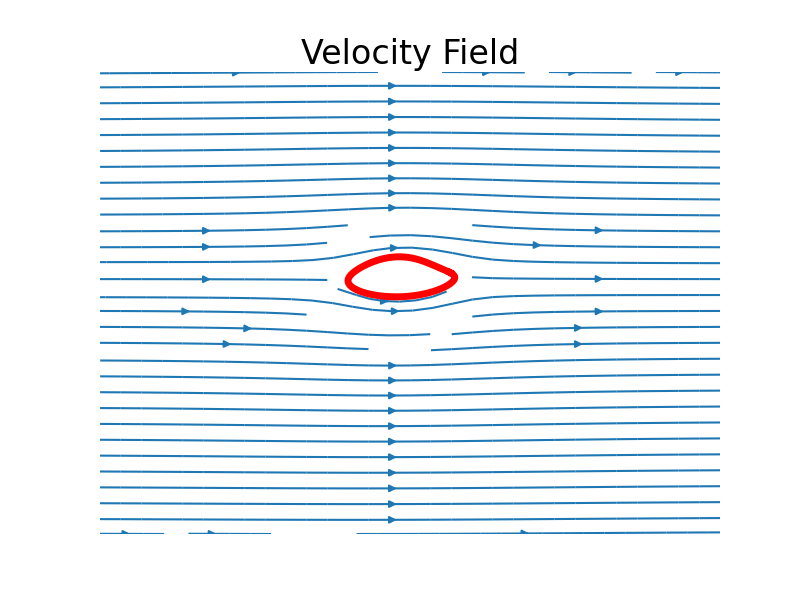}
		\subcaption{The true solution of the velocity field $ \boldsymbol{v} $}
	\end{subfigure}
	\begin{subfigure}{0.32\textwidth}
		\centering
		\includegraphics[width=\textwidth]{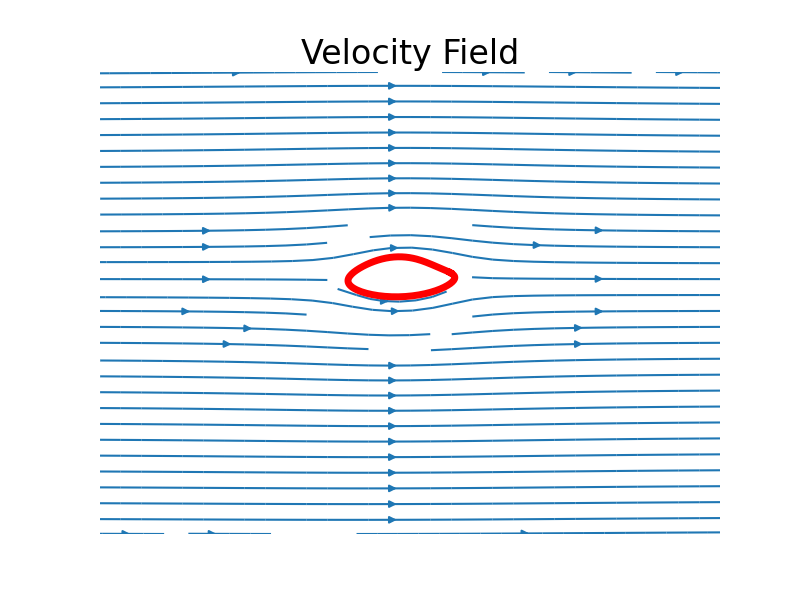}
		\subcaption{The predicted velocity field in BI-DeepONet}
	\end{subfigure}
	\begin{subfigure}{0.32\textwidth}
		\centering
		\includegraphics[width=\textwidth]{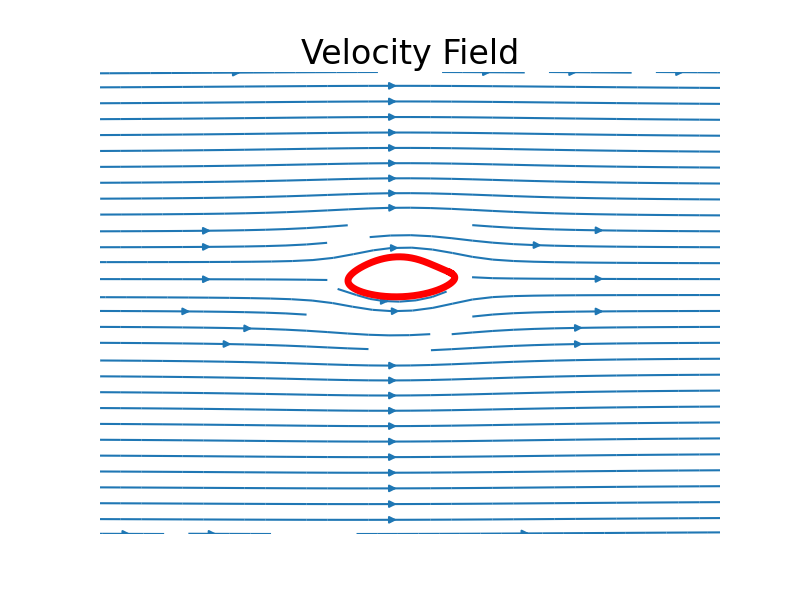}
		\subcaption{The predicted velocity field in BI-TDONet}
	\end{subfigure}
	\caption{The performance of BI-DeepONet and BI-TDONet for the first example of the two-dimensional potential flow around an obstacle problem.}
	\label{potential_result}
\end{figure}

\begin{figure}[htbp]
		\vspace{-0.8cm}
		\centering
		\begin{subfigure}{.32\linewidth}
			\includegraphics[width=\textwidth]{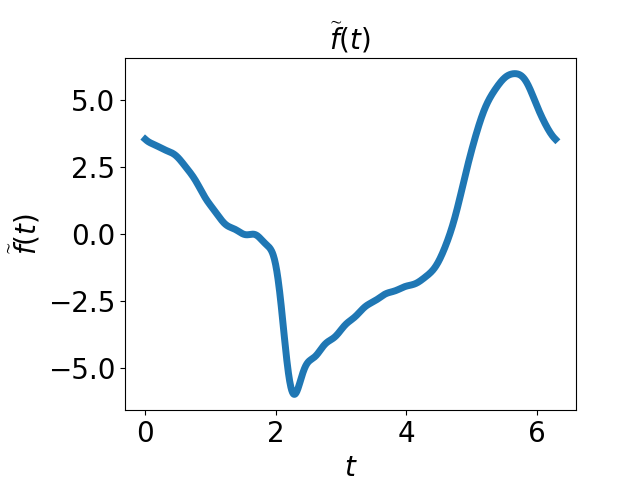}
			\subcaption{The right-hand side function $\widetilde{f}$ of the BIE}
		\end{subfigure}
		\begin{subfigure}{.32\linewidth}
			\includegraphics[width=\textwidth]{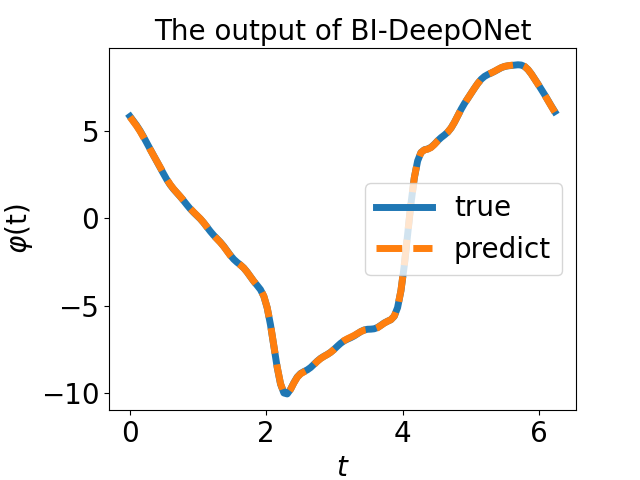}
			\subcaption{The predicted solution of the BI-DeepONet.}
		\end{subfigure}
		\begin{subfigure}{.32\linewidth}
			\includegraphics[width=\textwidth]{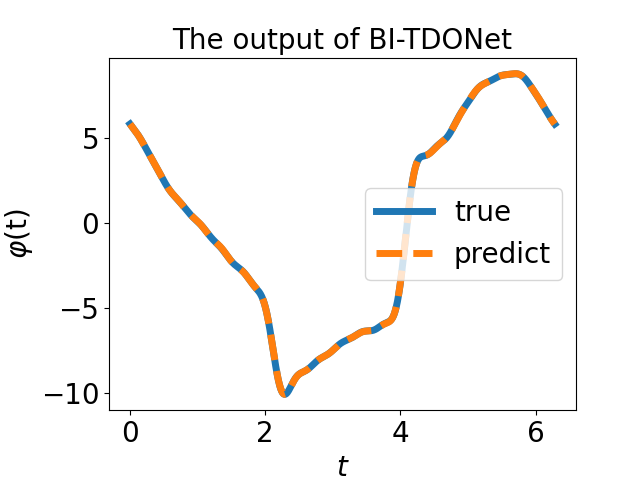}
			\subcaption{The predicted solution of the BI-TDONet.}
		\end{subfigure}
		\begin{subfigure}{.32\linewidth}
			\includegraphics[width=\textwidth]{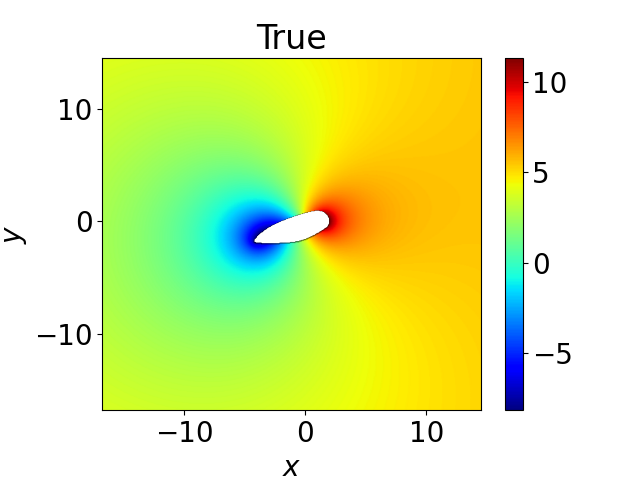}
			\subcaption{The true solution of the velocity potential function $ u_2$}
		\end{subfigure}
		\begin{subfigure}{0.32\textwidth}
			\centering
			\includegraphics[width=\textwidth]{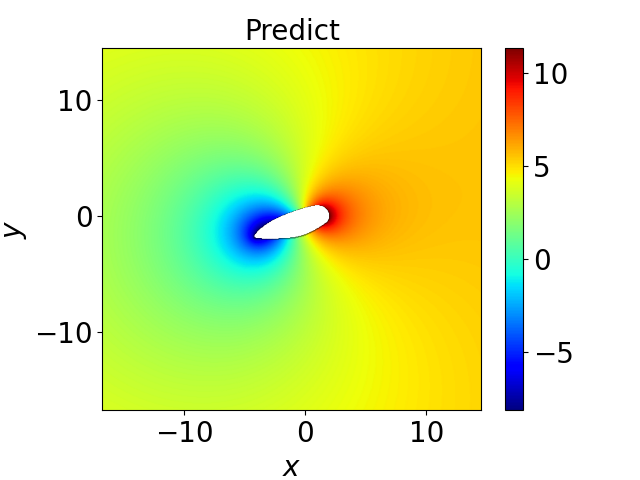}
			\subcaption{The predicted solution of the velocity potential function $ u_2$ in BI-DeepONet}
		\end{subfigure}
		\begin{subfigure}{0.32\textwidth}
			\centering
			\includegraphics[width=\textwidth]{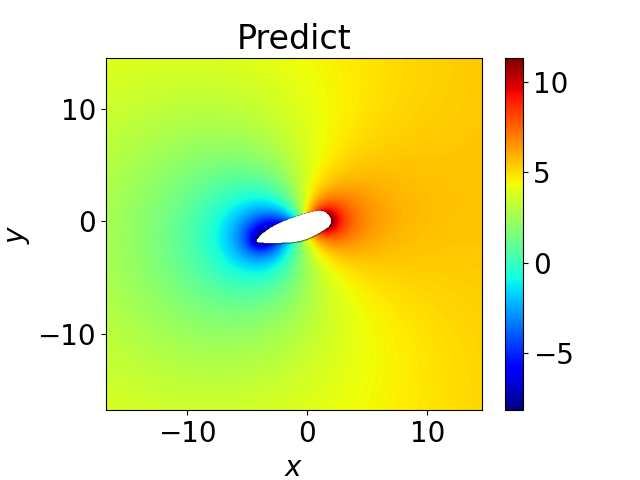}
			\subcaption{The predicted solution of the velocity potential function $ u_2 $ in BI-TDONet}
		\end{subfigure}
		\begin{subfigure}{0.32\textwidth}
			\centering
			\includegraphics[width=\textwidth]{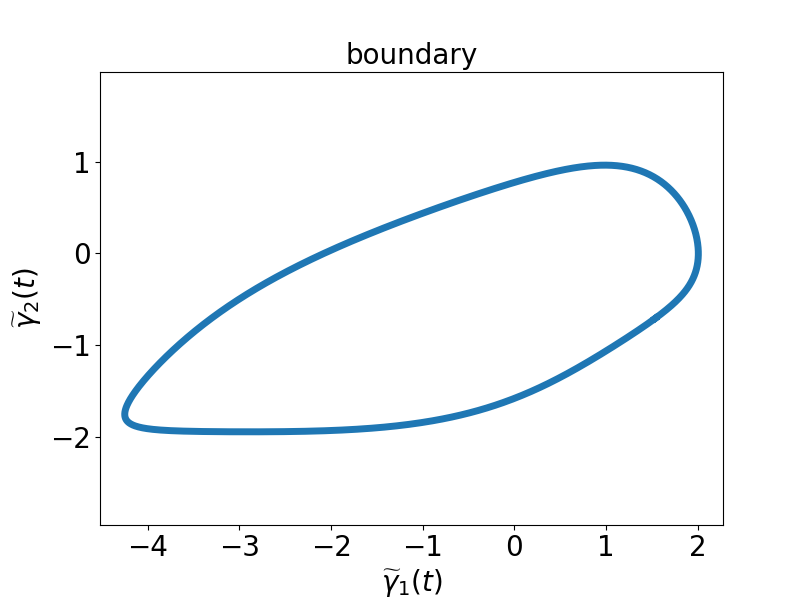}
			\subcaption{	The geometric of the two-dimensional potential flow around an obstacle problem}
		\end{subfigure}
		\begin{subfigure}{0.32\textwidth}
			\centering
			\includegraphics[width=\textwidth]{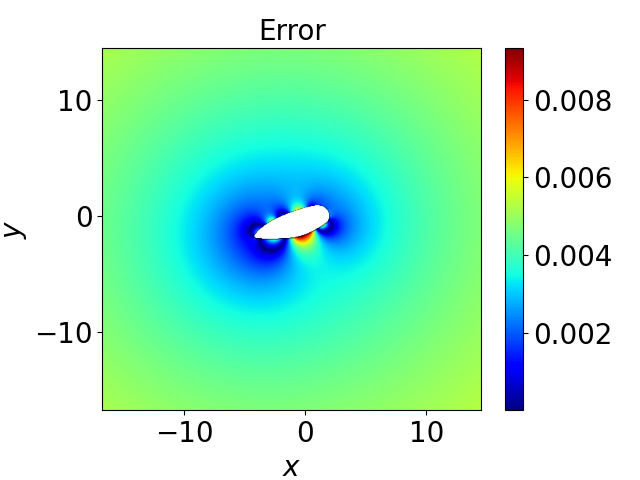}
			\subcaption{	The error of the velocity potential function $ u_2 $ in BI-DeepONet}
		\end{subfigure}
		\begin{subfigure}{0.32\textwidth}
			\centering
			\includegraphics[width=\textwidth]{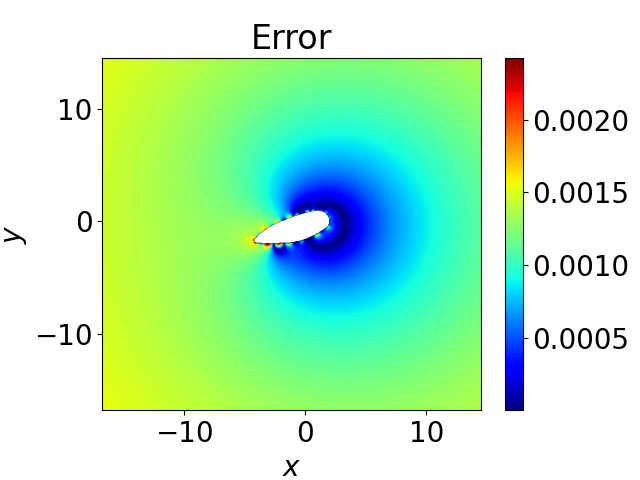}
			\subcaption{	The error of the velocity potential function $ u_2 $ in BI-TDONet}
		\end{subfigure}
		\begin{subfigure}{.32\linewidth}
			\includegraphics[width=\textwidth]{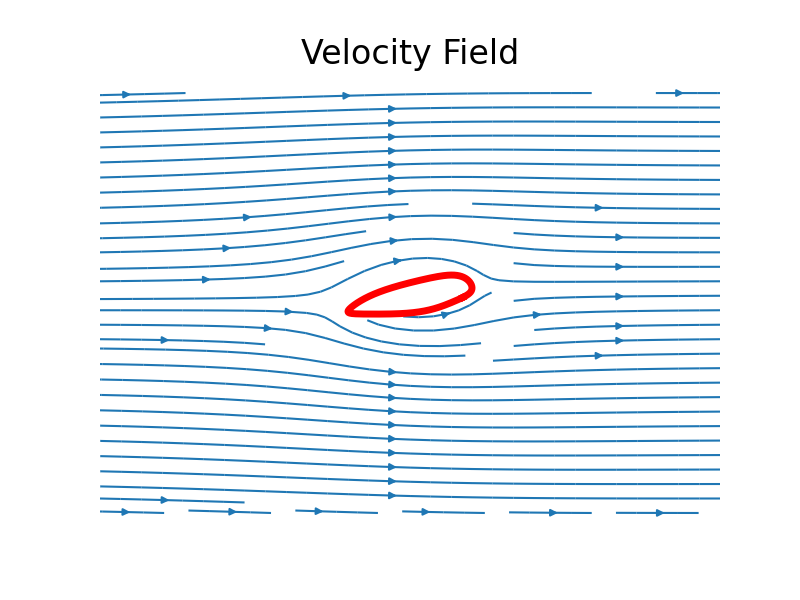}
			\subcaption{The true solution of the velocity field $ \boldsymbol{v} $}
		\end{subfigure}
		\begin{subfigure}{0.32\textwidth}
			\centering
			\includegraphics[width=\textwidth]{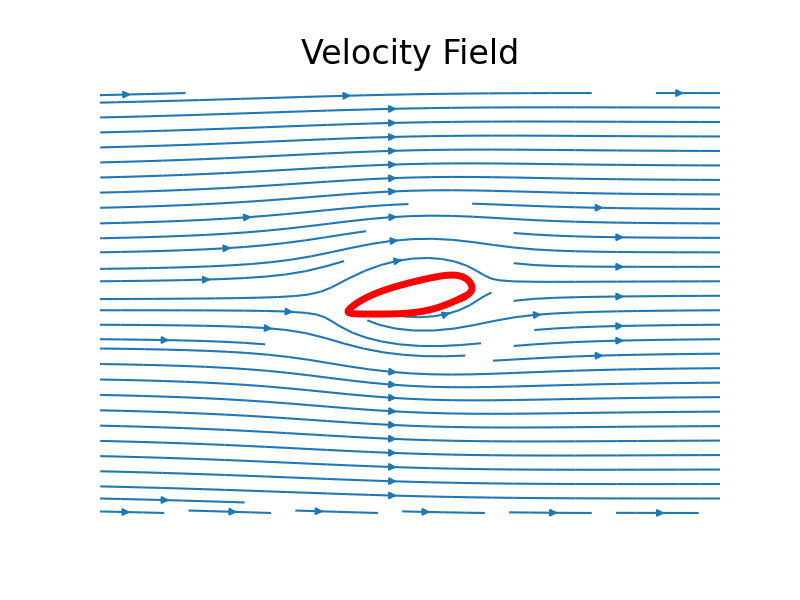}
			\subcaption{The predicted velocity field in BI-DeepONet}
		\end{subfigure}
		\begin{subfigure}{0.32\textwidth}
			\centering
			\includegraphics[width=\textwidth]{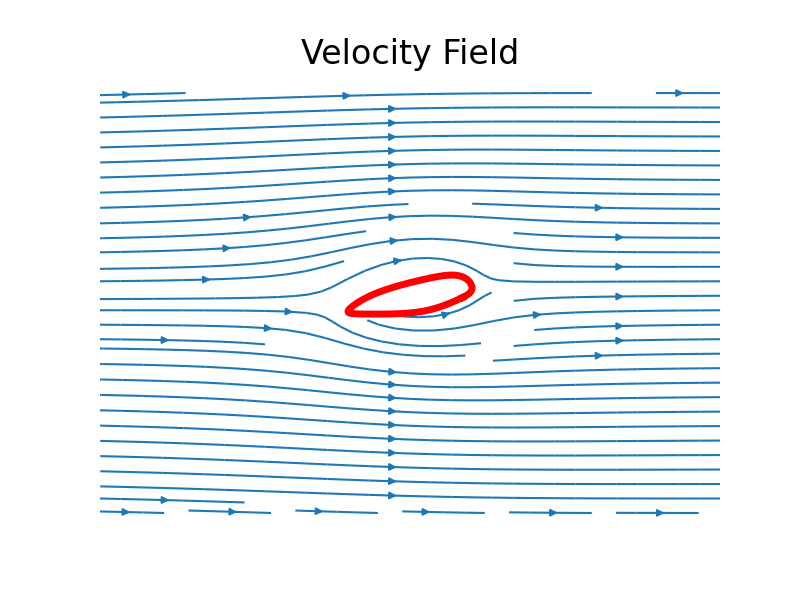}
			\subcaption{The predicted velocity field in BI-TDONet}
		\end{subfigure}
		\caption{The performance of BI-DeepONet and BI-TDONet for the second example of the two-dimensional potential flow around an obstacle problem.}
		\label{potential_result1}
\end{figure}

To further assess the performances of BI-DeepONet and BI-TDONet, we set $v_0 = 3$ and randomly selected two boundaries from the dataset. The results are illustrated in Figures \ref{potential_result} and \ref{potential_result1}. Figures \ref{potential_result} and \ref{potential_result1} (a) display the right-hand side of the BIE, while parts (b) of both figures visualize the true solution of the BIE alongside the predicted solution by BI-DeepONet. Parts (c) show the comparison between the true solution of the BIE and the predicted solution by BI-TDONet.

Figures \ref{potential_result} and \ref{potential_result1} (d) illustrate the true solution of the velocity potential $u_2(\boldsymbol{x})$. Parts (e) and (f) in each figure respectively depict the predicted solutions of $u_2(\boldsymbol{x})$ by BI-DeepONet and BI-TDONet. Part (g) visualizes the boundary configurations.

Parts (h) and (i) in both figures detail the absolute errors between the predicted solutions and the true solutions of the velocity potential $u_2(\boldsymbol{x})$ for BI-DeepONet and BI-TDONet, respectively. Finally, parts (j), (k), and (l) present the true solution of the velocity field along with the predicted solutions by BI-DeepONet and BI-TDONet, highlighting the accuracy and differences between the models.

Although BI-DeepONet exhibits significantly larger errors in predicting the velocity potential $u_2(\boldsymbol{x})$ compared to BI-TDONet, the visualizations of the predicted velocity fields in Figures \ref{potential_result} and \ref{potential_result1} indicate that both BI-DeepONet and BI-TDONet effectively predict the velocity fields. These visual outcomes convincingly demonstrate that the operator learning network models based on BIEs can proficiently address potential flow problems.

\subsection{Elastostatic problems}
In this subsection, we consider the elastostatics problem concerning a plate with smooth boundaries, focusing on cases without body forces. The governing equations are the Navier equations, mathematically expressed as
\begin{equation}\label{elastostatic}
	\left\{
	\begin{aligned}
	\Delta\boldsymbol{u}(\boldsymbol{x})+\frac{1}{1-2\nu}\nabla(\nabla\cdot\boldsymbol{u}(\boldsymbol{x}))=0\qquad \boldsymbol{x}&\in  D,\\
	\boldsymbol{u}(\boldsymbol{x})={\boldsymbol{v}}(\boldsymbol{x})\qquad \boldsymbol{x}&\in  \partial D,
	\end{aligned}
	\right.
\end{equation}
where $\boldsymbol{u}$ represents the displacement field and $\nu$ is Poisson's ratio. To avoid confusion in terminology, particularly since $t$ typically denotes stress in elastostatics, we will use $t$ to denote stress and $\theta$ to denote the parametric domain only within this subsection.  As established in prior works \cite{Aliabadi2003, brebbia2012boundary}, the components $u_1$ and $u_2$ of the displacement field $\boldsymbol{u}$ within domain $D$ can be mathematically represented as
\begin{equation*}
	u_{\alpha}(\boldsymbol{x}) = \int_{\partial D} u_{\alpha\beta}^*(\boldsymbol{x}, \boldsymbol{y})t_{\beta}(\boldsymbol{y})ds(\boldsymbol{y}) - \int_{\partial D} t_{\alpha\beta}^*(\boldsymbol{x}, \boldsymbol{y})v_{\beta}(\boldsymbol{y})ds(\boldsymbol{y}), \quad \boldsymbol{x} \in  D, \, \boldsymbol{y} \in \partial D, \, \alpha, \beta = 1, 2,
\end{equation*}
where the components $t_1$ and $t_2$ of the unknown stress filed $\boldsymbol{t}$, denoted by $\boldsymbol{t}:=[t_1, t_2]$, satisfy the BIEs
\begin{equation}\label{elastostatic BIE}
	\int_{\partial D} u_{\alpha\beta}^{*}(\boldsymbol{x}, \boldsymbol{y})t_{\beta}(\boldsymbol{y})ds(\boldsymbol{y}) = C_{\alpha\beta}(\boldsymbol{x})v_{\beta}(\boldsymbol{x}) + \int_{\partial D} t_{\alpha\beta}^*(\boldsymbol{x}, \boldsymbol{y})v_{\beta}(\boldsymbol{y})ds(\boldsymbol{y}), \quad \boldsymbol{x}, \boldsymbol{y} \in \partial D, \quad \alpha,\beta = 1,2,
\end{equation}
where $C_{\alpha\beta} := \frac{\delta_{\alpha\beta}}{2}$, and $\delta_{\alpha\beta}$ is the Kronecker delta. The fundamental solutions of the Navier equation \eqref{elastostatic}, represented by $t_{\alpha\beta}^*$ and $u_{\alpha\beta}^*$, are defined as
\begin{equation*}
	\begin{aligned}
		t_{\alpha\beta}^*(\boldsymbol{x}, \boldsymbol{y}) &= - \frac{1}{4\pi (1 - \nu)r} \left[ \frac{\partial r}{\partial n} \left( (1 - 2\nu)\delta_{\alpha\beta} + 2r'_{\alpha}r'_{\beta} \right) - (1 - 2\nu) (r'_{\alpha}n_{\beta} - r'_{\beta}n_{\alpha}) \right],\\
		u_{\alpha\beta}^*(\boldsymbol{x}, \boldsymbol{y}) &= \frac{1}{8\pi G(1 - \nu)} \left[ (3 - 4\nu) \ln \left( \frac{1}{r} \right) \delta_{\alpha\beta} + r'_{\alpha}r'_{\beta} \right],
	\end{aligned}
\end{equation*}
where $r := \left\|\boldsymbol{x} - \boldsymbol{y}\right\|$, $r'_{i}$ indicates the derivative of $r$ with respect to the $i$-th component of $\boldsymbol x$, and $ G $ is the shear modulus.

We reformulate \eqref{elastostatic BIE} as an operator equation by defining
\begin{equation*}
	\begin{aligned}
		\mathcal{U}=\begin{bmatrix}
			U_{11}&U_{12}\\
			U_{21}&U_{22}
		\end{bmatrix},\\
		\mathcal{T}=\begin{bmatrix}
			T_{11}&T_{12}\\
			T_{21}&T_{22}
		\end{bmatrix},
	\end{aligned}
\end{equation*}
where $\left( U_{\alpha\beta}t_{\beta}\right) (\boldsymbol{x}):=\int_{\partial D} u_{\alpha\beta}^*(\boldsymbol{x}, \boldsymbol{y})t_{\beta}(\boldsymbol{y})ds(\boldsymbol{y}),\left( T_{\alpha\beta}v_{\beta}\right) (\boldsymbol{x}):=\int_{\partial D}t_{\alpha\beta}^*(\boldsymbol{x}, \boldsymbol{y})v_{\beta}(\boldsymbol{y})ds(\boldsymbol{y}),\alpha, \beta = 1, 2$. Consequently, the operator equation derived from \eqref{elastostatic BIE} is expressed as
\begin{equation}\label{ela BIO equation}
	\mathcal{U}\boldsymbol{t}= \left( \frac{1}{2}\mathcal{I}-\mathcal{T}\right) \boldsymbol{v}.
\end{equation}
Assuming that $\mathcal{U}^{-1}$ exists, for a fixed boundary $\boldsymbol{\gamma}$, both $\frac{1}{2}\mathcal{I}-\mathcal{T}$ and $\mathcal{U}^{-1}$ are bounded linear operators. The nonlinearity of the operator $\mathcal{U}^{-1}\left( \frac{1}{2}\mathcal{I}-\mathcal{T}\right) $ comes from $\boldsymbol{\gamma}$. Therefore, we can utilize BI-TDONet and BI-DeepONet to approximate this operator, leveraging its capability to encapsulate and represent the complex relationships dictated by the boundary $\boldsymbol{\gamma}$.

When $\boldsymbol{v}(\boldsymbol{x})$ is specified, the right-hand side of the BIE \eqref{ela BIO equation} is determined, thereby classifying \eqref{ela BIO equation} as a typical first kind BIE. The forthcoming experiments will illustrate that both BI-DeepONet and BI-TDONet excel in approximating first kind BIOs, showcasing their outstanding performance.

It is important to note that the kernel functions $t_{\alpha\beta}^*$ and $u_{\alpha\beta}^*$ have singularities. To avoid the calculation of singular integrals during data generation, we derive the training and testing data for this subsection by specifying the displacements at the boundary and computing the corresponding boundary stresses using the strain-stress relationship and Hooke’s law. Given the known boundary displacement $\boldsymbol{v}(\boldsymbol{x})$, the boundary stress $\boldsymbol{t}(\boldsymbol{x})$ is computed as
\begin{equation}\label{stress-strain}
	\begin{aligned}
		\boldsymbol{\epsilon}&=\frac{1}{2}\left(\nabla\boldsymbol{v}+(\nabla\boldsymbol{v})^\top\right),\\
		\boldsymbol{\sigma}&=2G\boldsymbol{\epsilon}+\frac{2G\nu}{1-2\nu}tr(\boldsymbol{\epsilon})\mathbf{I},\\
		\boldsymbol{t}&=\boldsymbol{\sigma}\cdot\boldsymbol{n}.
	\end{aligned}
\end{equation}

For generating training and testing data, we let $v_1(\boldsymbol{x})=a_1x_1+b_1x_2$ and $v_2(\boldsymbol{x})=a_2x_1+b_2x_2$, and compute the stress using \eqref{stress-strain}. By randomly generating $a_1$, $a_2$, $b_1$ and $b_2$ from a standard normal distribution, we created $1,000$ instances of $\boldsymbol{v}$ and $\boldsymbol{t}$. Along with the $5,998$ boundaries generated in Subsection 3.3, this approach yielded a total of $5,998,000$ samples, with $80\%$ allocated for training and $20\%$ for testing.

In the BI-DeepONet, we configured the number of points $p=128$ and defined the inputs to the trunk net as uniformly distributed points along the interval $I$, with the inputs to the branch net being boundary parameter functions $ \boldsymbol{\gamma} $ and displacement fields $ u_1, u_2$ evaluated at these points. Given that the output of BI-DeepONet represents the evaluation of functions at the trunk net and considering that when the output function is vector-valued, the components of this vector-valued function produced by BI-DeepONet will be identical, BI-DeepONet faces challenges in effectively learning operators where the output function is vector-valued. A practical solution to enable BI-DeepONet to learn operators with vector-valued functions effectively is to treat each component of the vector-valued function as a separate output of different BI-DeepONet instances. Let $e_1 := [1,0]^\top$ and $e_2 := [0,1]^\top$. Specifically, we utilize BI-DeepONet-1 to learn the operator $e_1^\top\mathcal{U}^{-1}\left( \frac{1}{2}\mathcal{I}-\mathcal{T}\right)$ in operator equation
\begin{equation*}
	e_1^\top\boldsymbol{t} = e_1^\top\mathcal{U}^{-1}\left( \frac{1}{2}\mathcal{I}-\mathcal{T}\right)\boldsymbol{v},
\end{equation*}
and employ  BI-DeepONet-2 to learn the operator $e_2^\top\mathcal{U}^{-1}\left( \frac{1}{2}\mathcal{I}-\mathcal{T}\right)$ as expressed in the following operator equation
\begin{equation*}
	e_2^\top\boldsymbol{t} = e_2^\top\mathcal{U}^{-1}\left( \frac{1}{2}\mathcal{I}-\mathcal{T}\right)\boldsymbol{v}.
\end{equation*}
By dividing the learning tasks in this way, each instance of BI-DeepONet can specialize in a particular component of the output, enhancing the network’s ability to model complex, multi-dimensional operator relationships effectively.

The network architecture for both BI-DeepONet-1 and BI-DeepONet-2 is defined as $$[[256, 600, 600, 600,600], [128, 600, 600, 600,600], [1, 600, 600, 600,600]].$$
For BI-TDONet, with $n=20$, the network structure can be represented as
$$
[[164, 600, 600, 600,600, 82], [82, 600, 600, 600,600, 82], [164, 600, 600, 600,600, 82]].
$$

During the training of BI-DeepONet-1 and BI-DeepONet-2, we set the batch size to $3,200$ and the number of iterations to $3,000,000$, which corresponds to approximately $20,000$ epochs. The initial learning rate was set at $0.001$, using the Adam optimization algorithm. The learning rate decay was implemented on an inverse-time schedule, with a decay cycle every $1/100$ of the iterations and a decay rate of $0.5$.

In the training process of BI-TDONet, the batch size was set to $8,192$ and the number of epochs to $5,000$. The initial learning rate was also established at $0.001$, utilizing the Adam optimization algorithm. Learning rate decay was managed by monitoring; if the loss does not decrease within $1/100$ of the epochs, the learning rate is halved to ensure effective adaptation to training progress.

\begin{table}[htb]
	\vspace{-0.2cm}
	\centering

	\begin{tabular}{c|c|c|c|c|c}
		\midrule
		Model
		&MNE
		&MRE
		&variance-MNE
		&variance-MRE
		&Mean-Time/ms\\
		\hline
		
		BI-DeepONet-1
		&$2.4632\times10^{-2}$
		&$2.1423\times10^{-2}$
		&$4.8275\times10^{-4}$
		&$2.5635\times10^{-4}$
		&$2.2588\times10^{-4}$\\
		\hline
		BI-DeepONet-2
		&$5.4307\times10^{-2}$
		&$4.8494\times10^{-2}$
		&$8.4837\times10^{-3}$
		&$3.4136\times10^{-3}$
		&$2.2684\times10^{-4}$\\
		\hline
		BI-TDONet
		&$1.5833\times10^{-2}$
		&$8.7361\times10^{-3}$
		&$1.0292\times10^{-4}$
		&$1.9158\times10^{-5}$
		&$2.4619\times10^{-4}$\\
		
		\midrule
	\end{tabular}	
    \caption{Error table on the elastostatic problem test set.}
	\label{elastostatic error}
\end{table}

Table \ref{potential error} presents the MNE and MRE, along with the variances of these metrics, for BI-DeepONet-1, BI-DeepONet-2, and BI-TDONet on the test set. In the context of elastostatic problems, the data from Table \ref{potential error} indicate that the MNE and MRE for BI-DeepONet-1, BI-DeepONet-2, and BI-TDONet are relatively similar, marking a departure from the trends observed in other types of problems. This suggests that utilizing multiple BI-DeepONet instances to independently learn different components of vector-valued functions can be an effective strategy.

However, a significant drawback of this approach for BI-DeepONet is scalability: the more components the vector-valued function has, the more BI-DeepONet models are needed for training. This escalation can significantly increase the complexity and resource demands of the training process, potentially limiting the practicality of this approach in more complex applications.

\begin{figure}[htb]
	\vspace{-0.5cm}
	\centering
	\begin{subfigure}{.32\linewidth}
		\includegraphics[width=\textwidth]{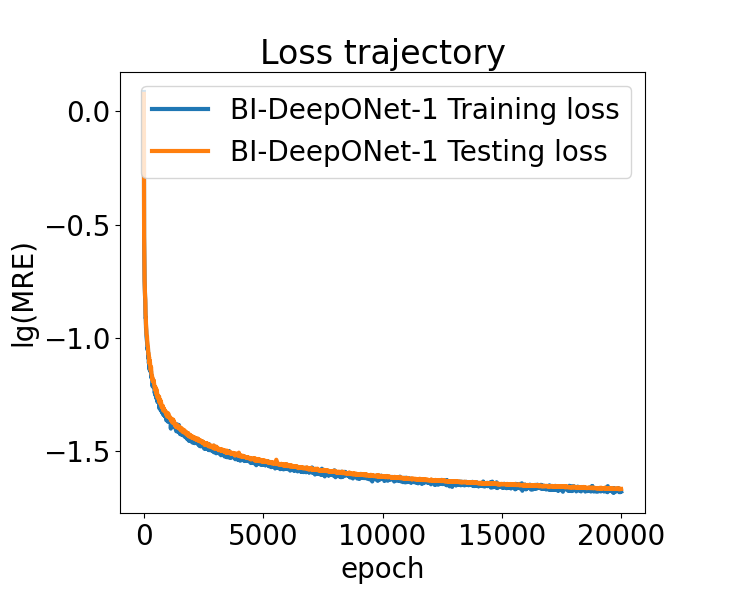}
		\subcaption{Training and testing loss of BI-DeepONet-1}
	\end{subfigure}
	\begin{subfigure}{.32\linewidth}
		\includegraphics[width=\textwidth]{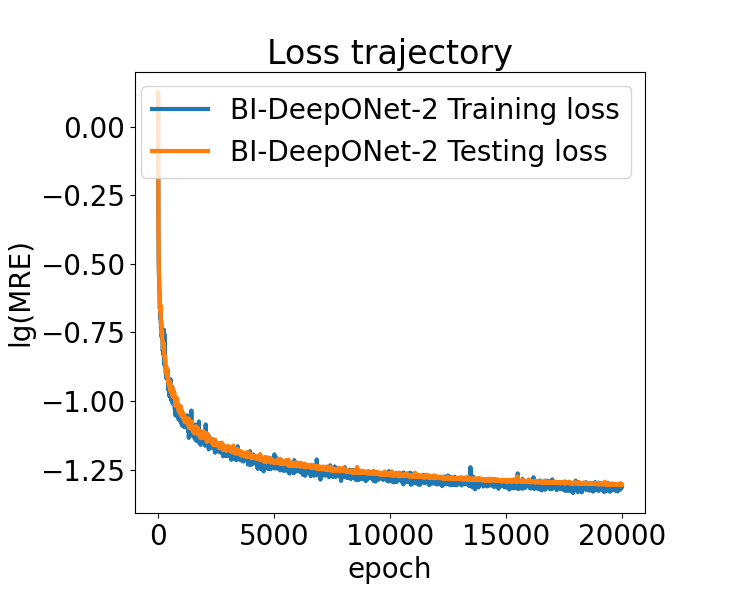}
		\subcaption{Training and testing loss of BI-DeepONet-2}
	\end{subfigure}
	\begin{subfigure}{.32\linewidth}
		\includegraphics[width=\textwidth]{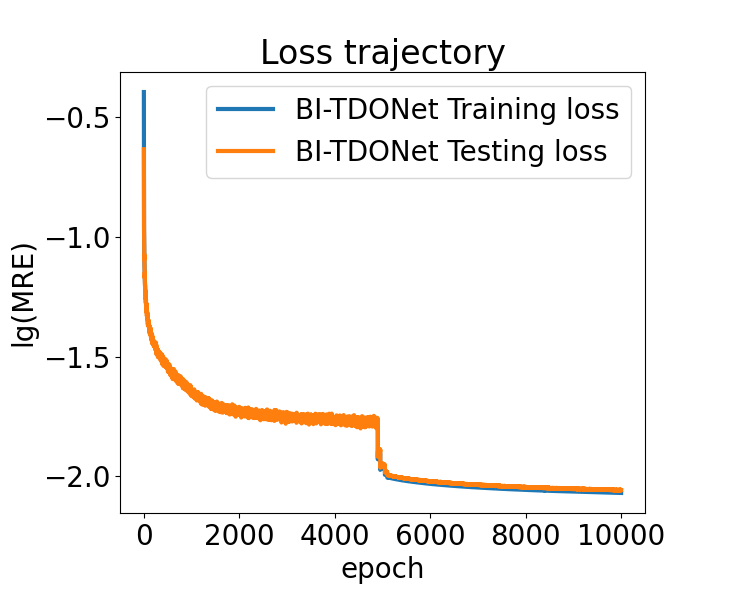}
		\subcaption{Training and testing loss of BI-TDONet}
	\end{subfigure}
	\caption{The logarithmic loss trajectories for BI-DeepONet and BI-TDONet in the elastostatic problem.}
	\label{ela_loss}
\end{figure}

Figure \ref{ela_loss} displays the logarithmic loss trajectories for MRE during the training and testing phases. BI-TDONet shows a more rapid decline in loss compared to BI-DeepONet when the epoch count is less than $1,000$. As the learning rate for BI-TDONet decreases, it reaches convergence by epoch $3,000$, while BI-DeepONet achieves convergence later, around epoch $20,000$. This demonstrates that BI-TDONet has a more efficient initial learning phase and achieves faster convergence under similar training conditions, highlighting its greater effectiveness in loss reduction over BI-DeepONet.

\begin{figure}[htb]
	\centering
	\begin{subfigure}{0.24\textwidth}
		\centering
		\includegraphics[width=\textwidth]{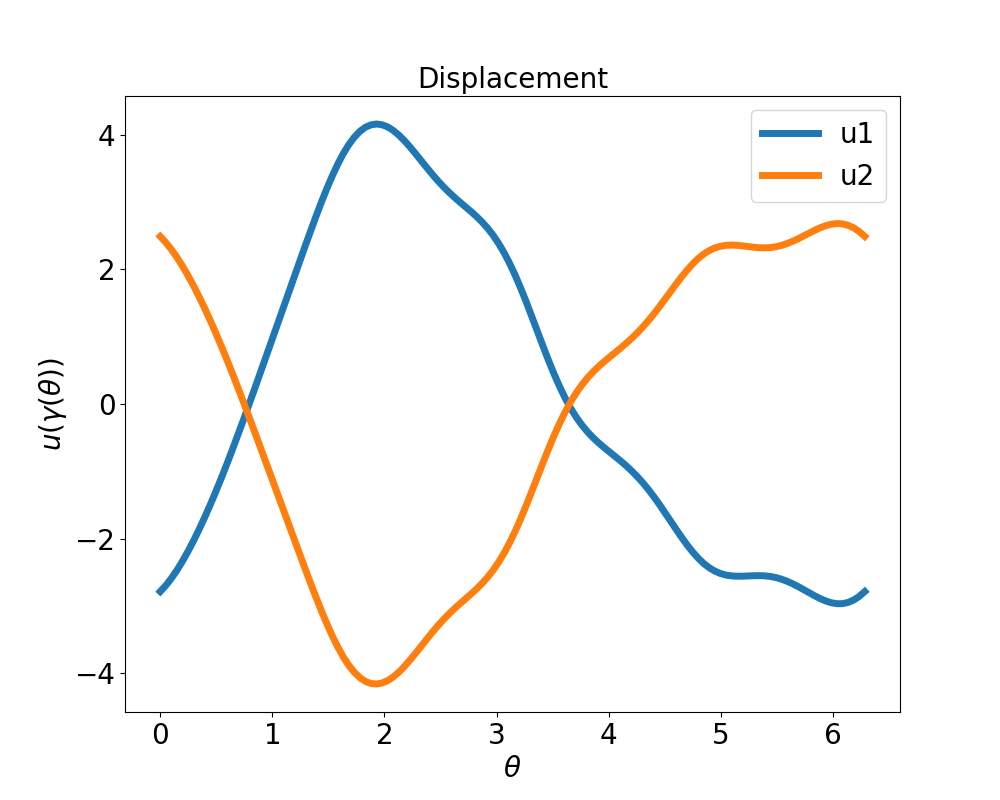}
		\subcaption{The displacement on the boundary}
	\end{subfigure}
	\hfill 
	\begin{subfigure}{0.24\textwidth}
		\centering
		\includegraphics[width=\textwidth]{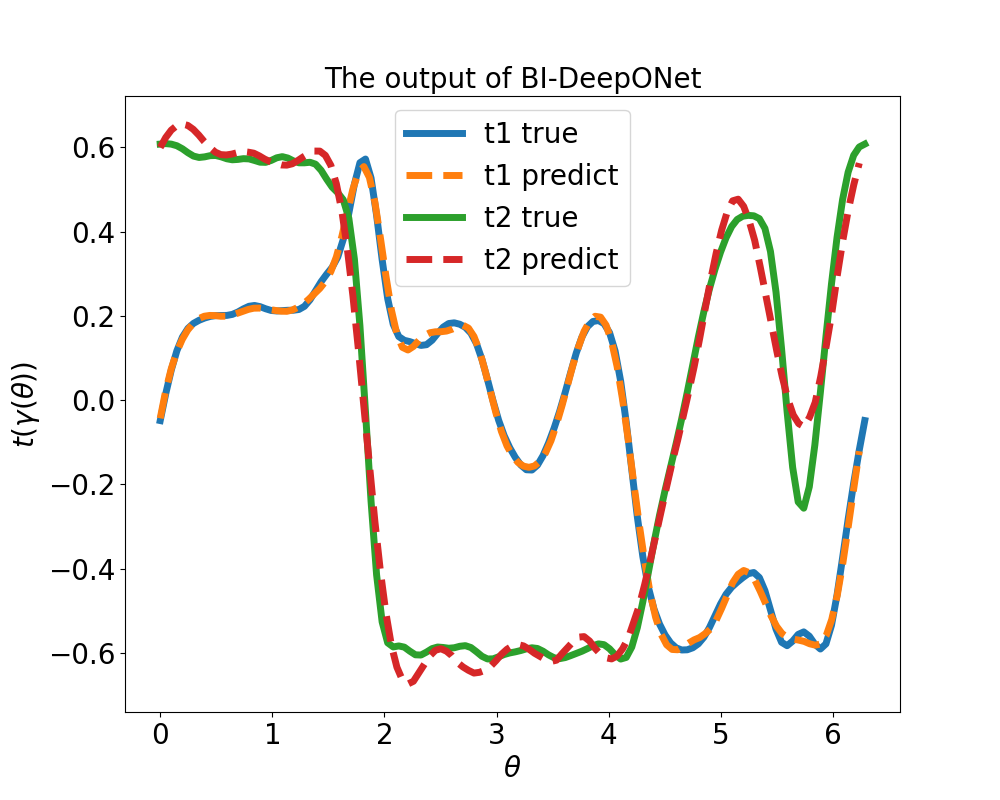}
		\subcaption{The true solution and predicted solution of stress on the boundary}
	\end{subfigure}
	\hfill 
	\begin{subfigure}{0.24\textwidth}
		\centering
		\includegraphics[width=\textwidth]{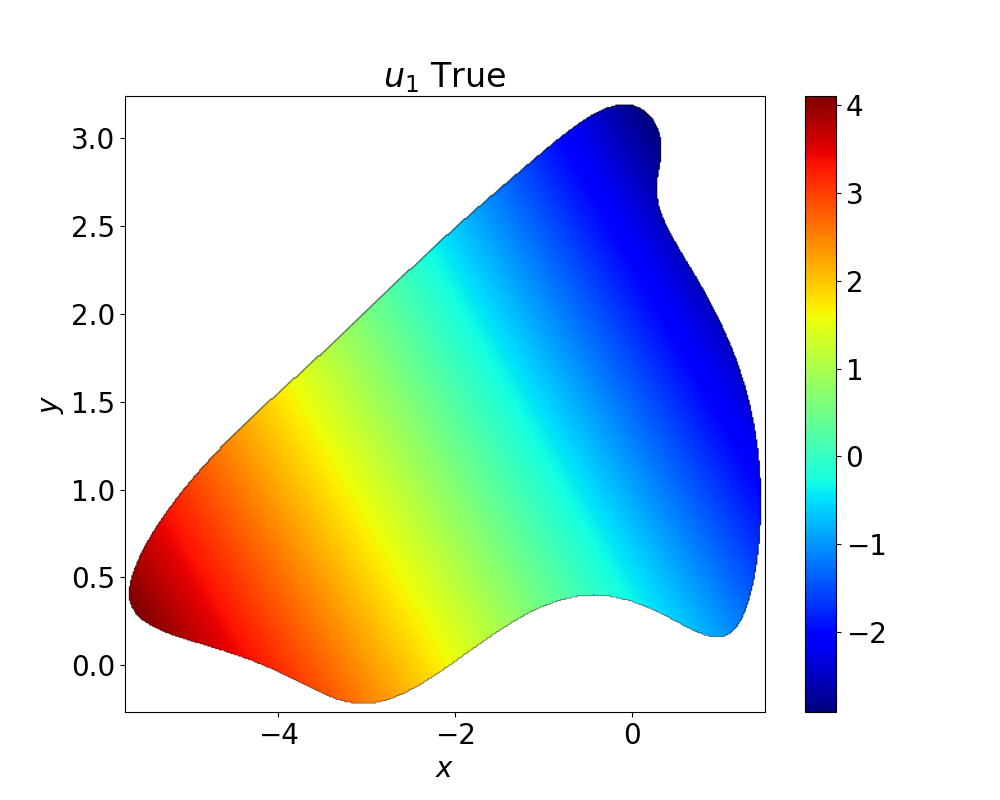}
		\subcaption{The true solution of the displacement field $u_1(\boldsymbol{x})$}
	\end{subfigure}
	\hfill 
	\begin{subfigure}{0.24\textwidth}
		\centering
		\includegraphics[width=\textwidth]{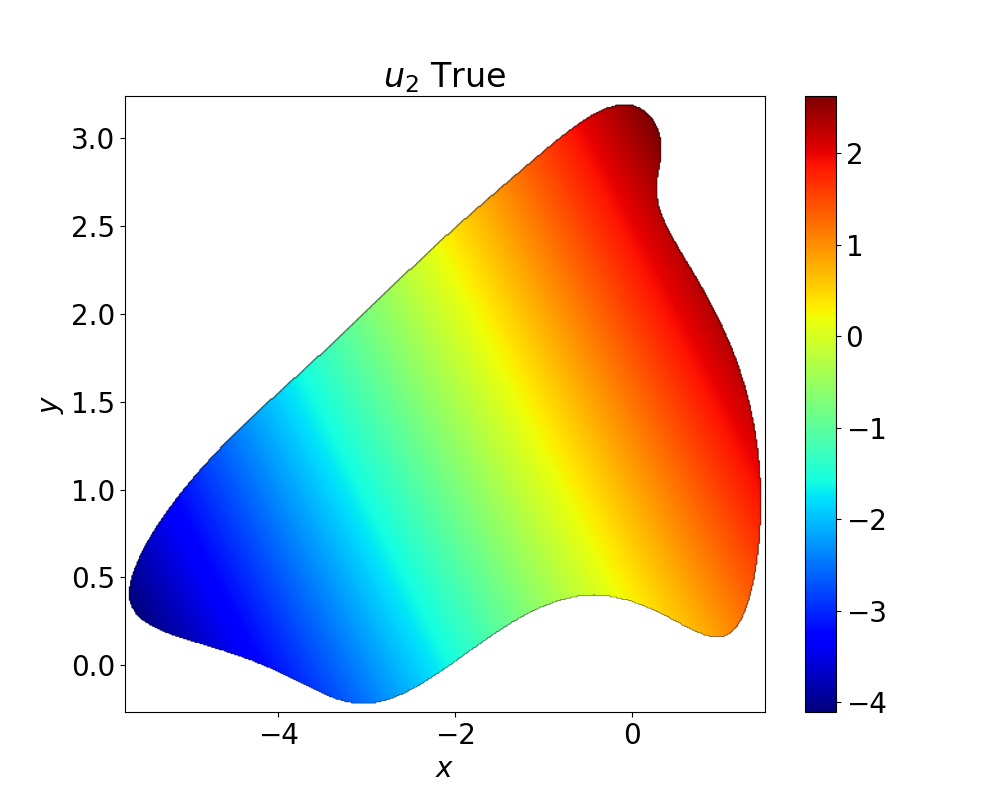}
		\subcaption{The true solution of the displacement field $u_2(\boldsymbol{x})$}
	\end{subfigure}
	
	\begin{subfigure}{0.24\textwidth}
		\centering
		\includegraphics[width=\textwidth]{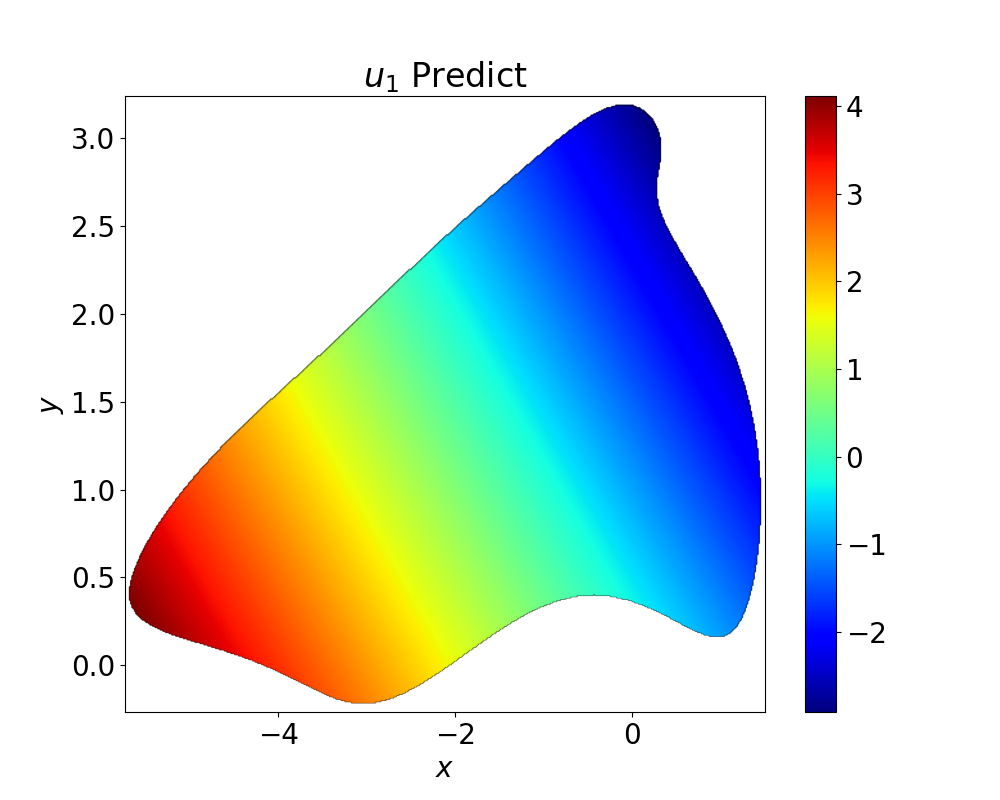}
		\subcaption{The predicted solution of the displacement field $u_1(\boldsymbol{x})$}
	\end{subfigure}
	\hfill 
	\begin{subfigure}{0.24\textwidth}
		\centering
		\includegraphics[width=\textwidth]{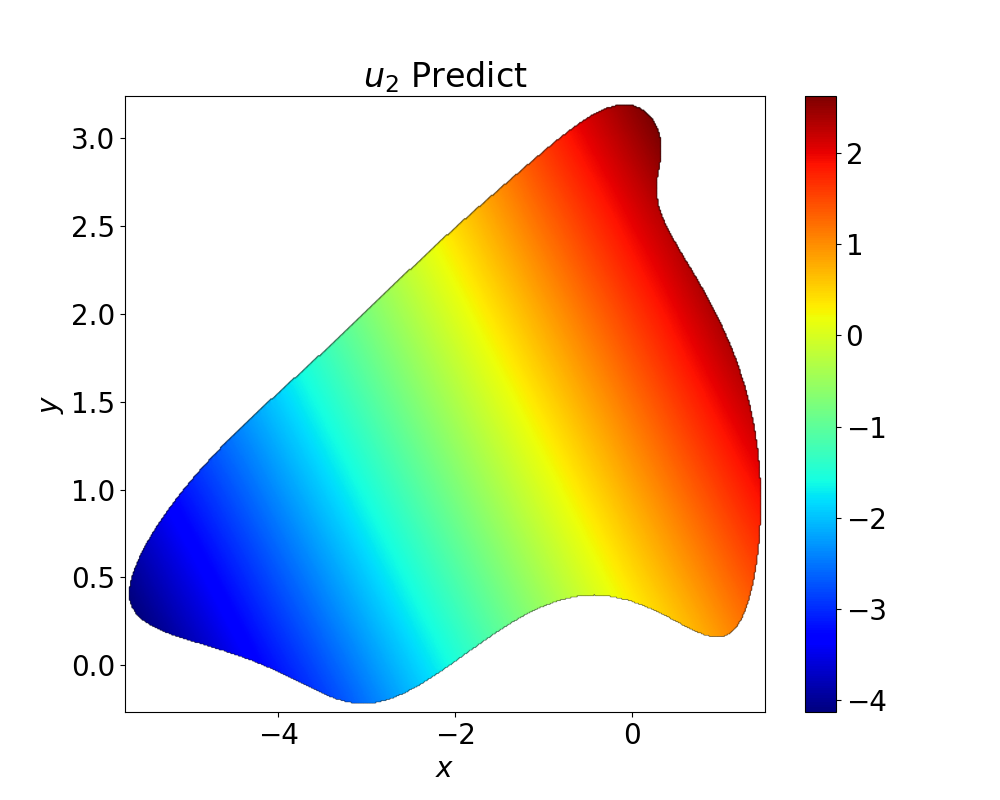}
		\subcaption{The predicted solution of the displacement field $u_2(\boldsymbol{x})$}
	\end{subfigure}
	\hfill 
	\begin{subfigure}{0.24\textwidth}
		\centering
		\includegraphics[width=\textwidth]{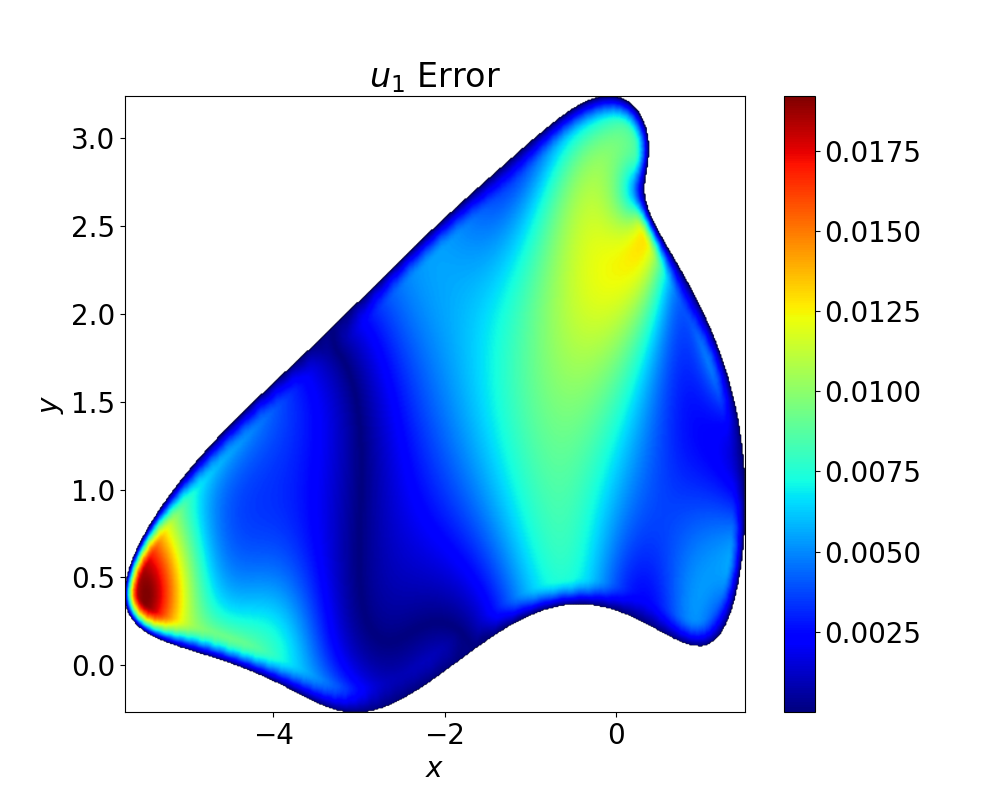}
		\subcaption{The error of the displacement field $u_1(\boldsymbol{x})$}
	\end{subfigure}
	\hfill 
	\begin{subfigure}{0.24\textwidth}
		\centering
		\includegraphics[width=\textwidth]{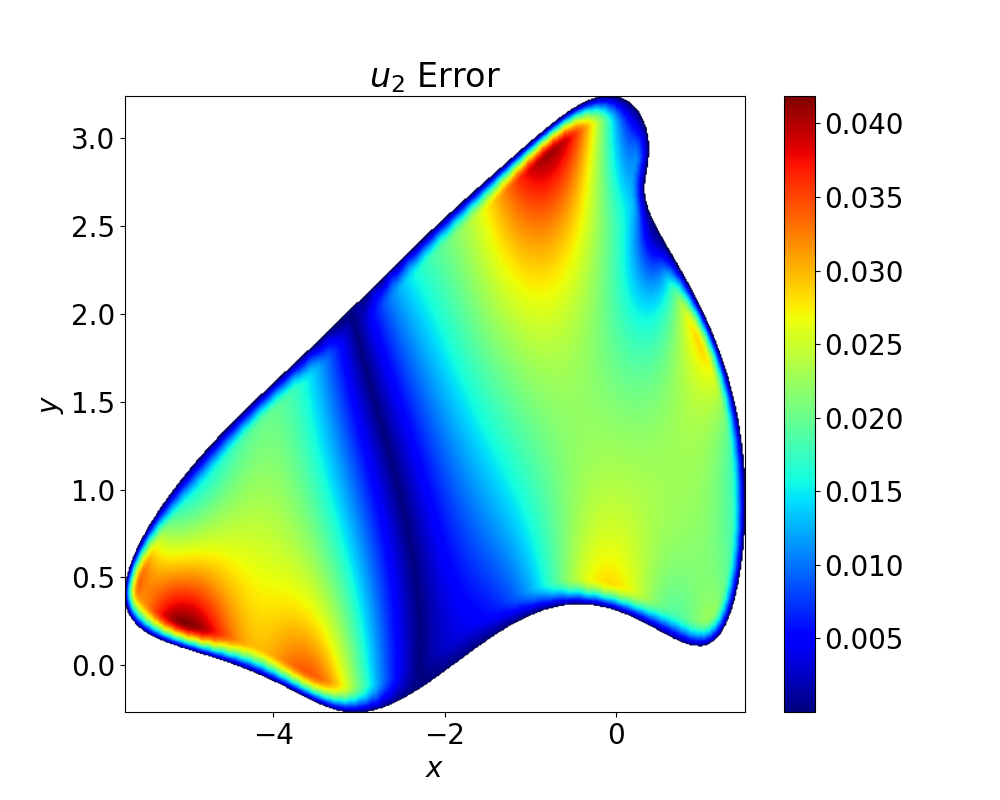}
		\subcaption{The error of the displacement field $u_2(\boldsymbol{x})$}
	\end{subfigure}
	
	\caption{The performance of BI-DeepONet in the first example of elastostatics problems.}
	\label{ela_xde_result}
\end{figure}

\begin{figure}[htb]
	\centering
	\begin{subfigure}{0.24\textwidth}
		\centering
		\includegraphics[width=\textwidth]{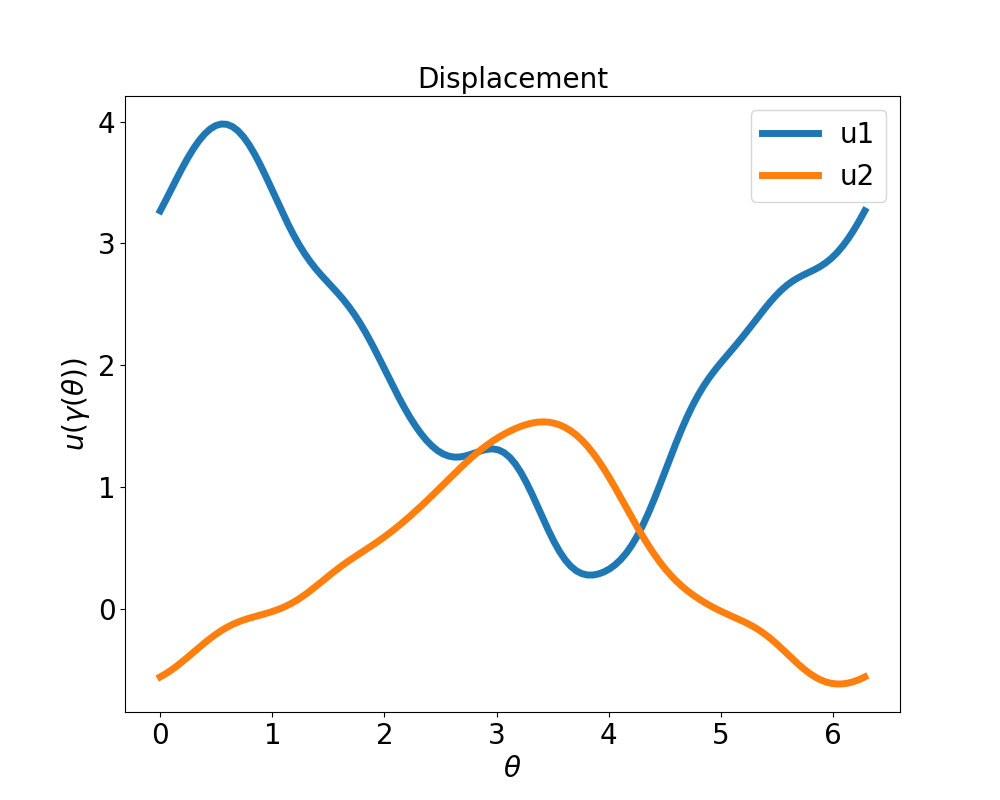}
		\subcaption{The displacement on the boundary}
	\end{subfigure}
	\hfill 
	\begin{subfigure}{0.24\textwidth}
		\centering
		\includegraphics[width=\textwidth]{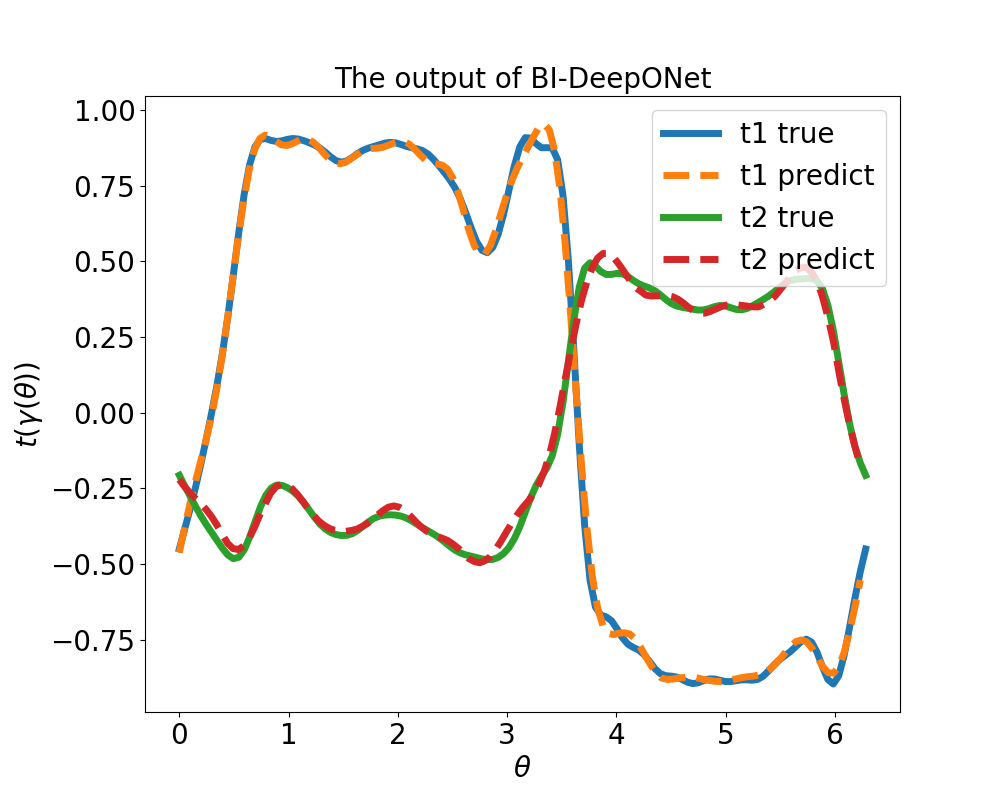}
		\subcaption{The true solution and predicted solution of stress on the boundary}
	\end{subfigure}
	\hfill 
	\begin{subfigure}{0.24\textwidth}
		\centering
		\includegraphics[width=\textwidth]{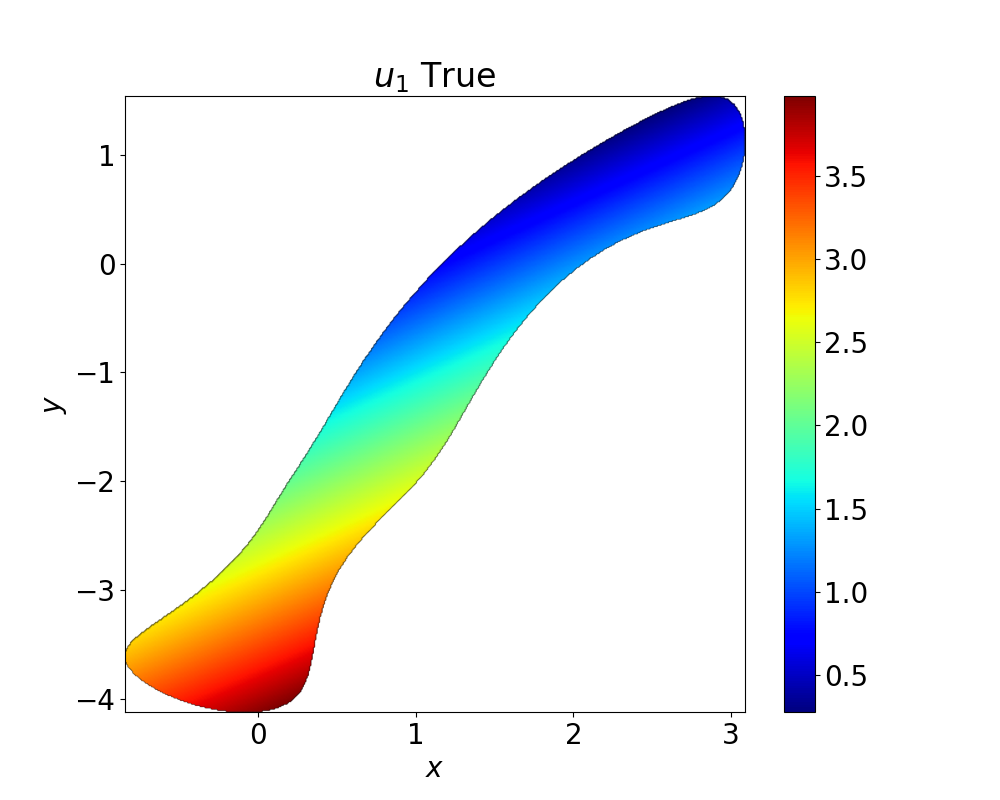}
		\subcaption{The true solution of the displacement field $u_1(\boldsymbol{x})$}
	\end{subfigure}
	\hfill 
	\begin{subfigure}{0.24\textwidth}
		\centering
		\includegraphics[width=\textwidth]{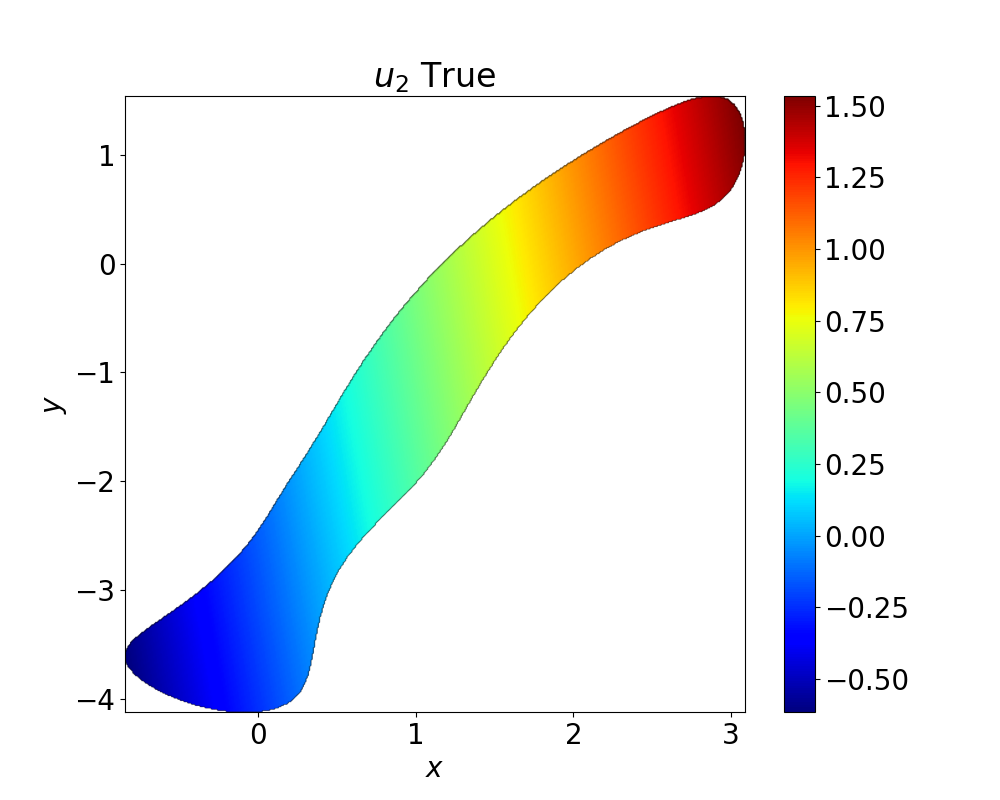}
		\subcaption{The true solution of the displacement field $u_2(\boldsymbol{x})$}
	\end{subfigure}
	
	\begin{subfigure}{0.24\textwidth}
		\centering
		\includegraphics[width=\textwidth]{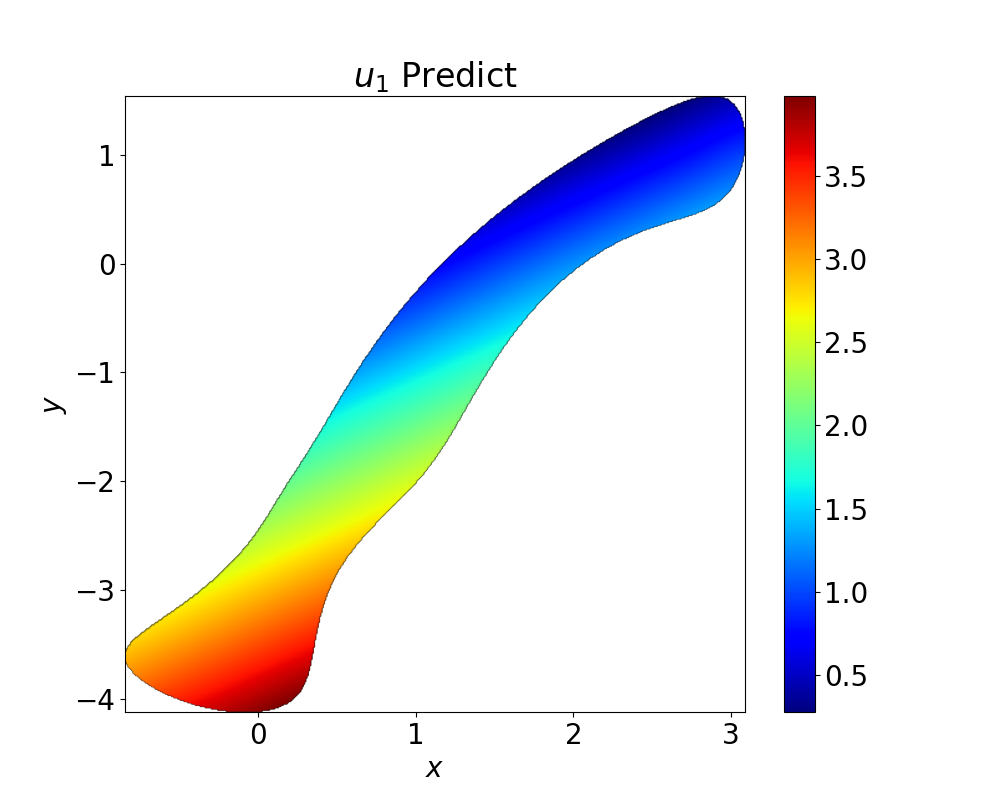}
		\subcaption{The predicted solution of the displacement field $u_1(\boldsymbol{x})$}
	\end{subfigure}
	\hfill 
	\begin{subfigure}{0.24\textwidth}
		\centering
		\includegraphics[width=\textwidth]{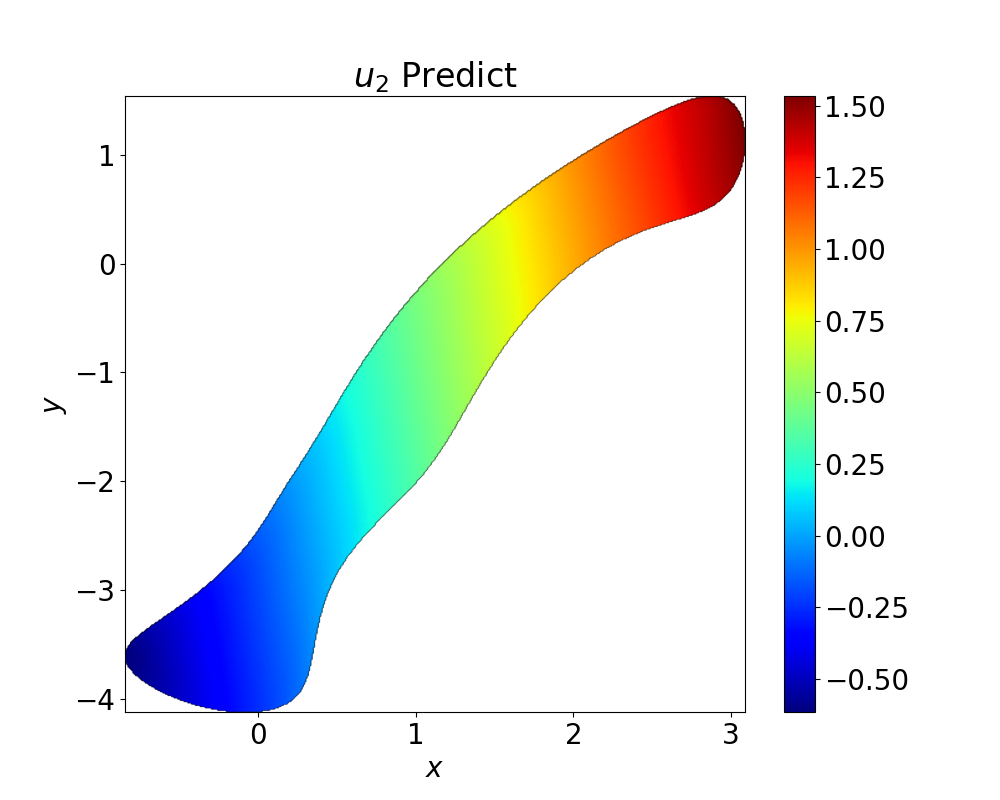}
		\subcaption{The predicted solution of the displacement field $u_2(\boldsymbol{x})$}
	\end{subfigure}
	\hfill 
	\begin{subfigure}{0.24\textwidth}
		\centering
		\includegraphics[width=\textwidth]{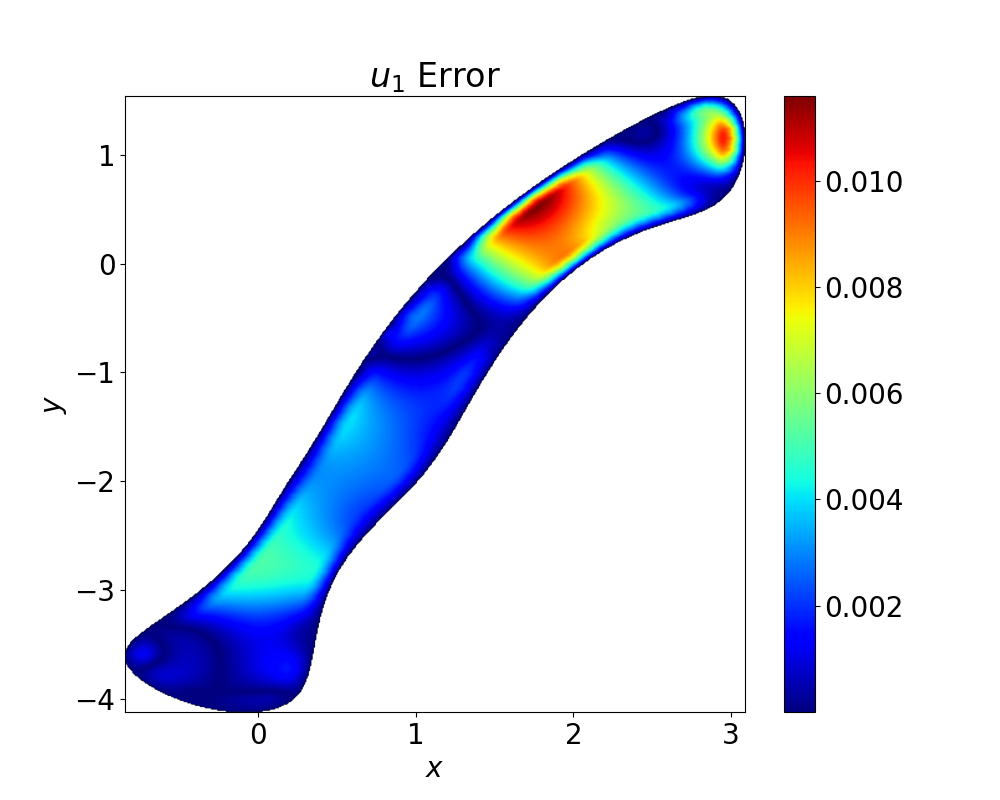}
		\subcaption{The error of the displacement field $u_1(\boldsymbol{x})$}
	\end{subfigure}
	\hfill 
	\begin{subfigure}{0.24\textwidth}
		\centering
		\includegraphics[width=\textwidth]{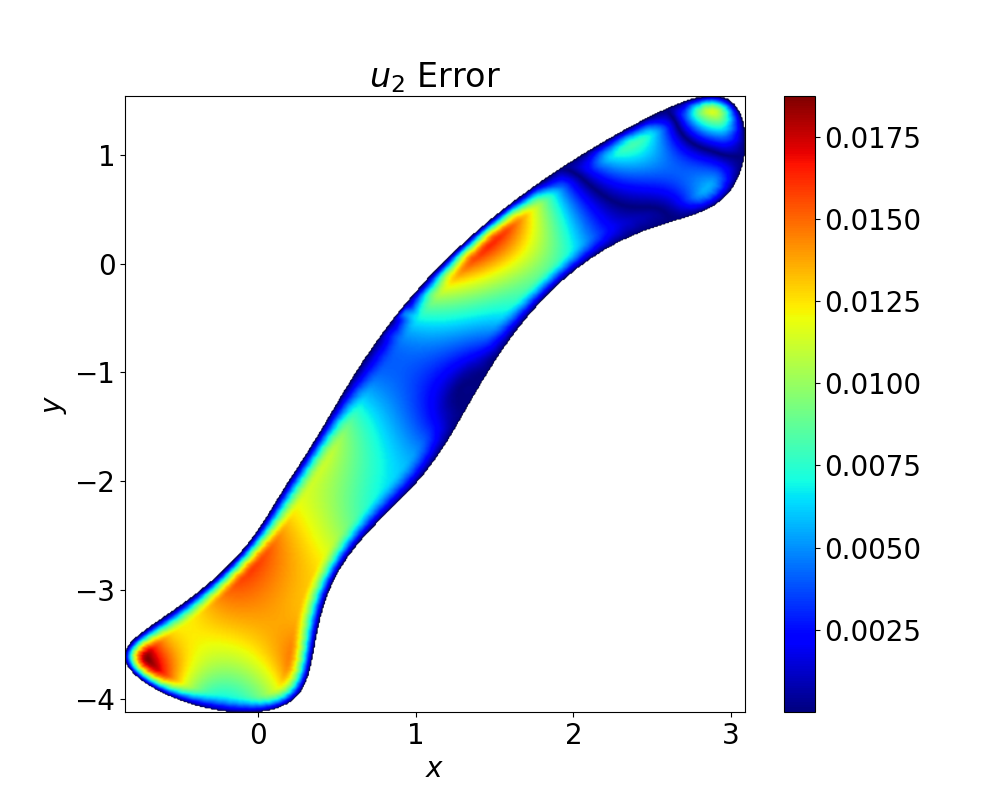}
		\subcaption{The error of the displacement field $u_2(\boldsymbol{x})$}
	\end{subfigure}
	
	\caption{The performance of BI-DeepONet in the second example of elastostatics problems.}
	\label{ela_xde_result1}
\end{figure}

\begin{figure}[htb]
	\centering
	\begin{subfigure}{0.24\textwidth}
		\centering
		\includegraphics[width=\textwidth]{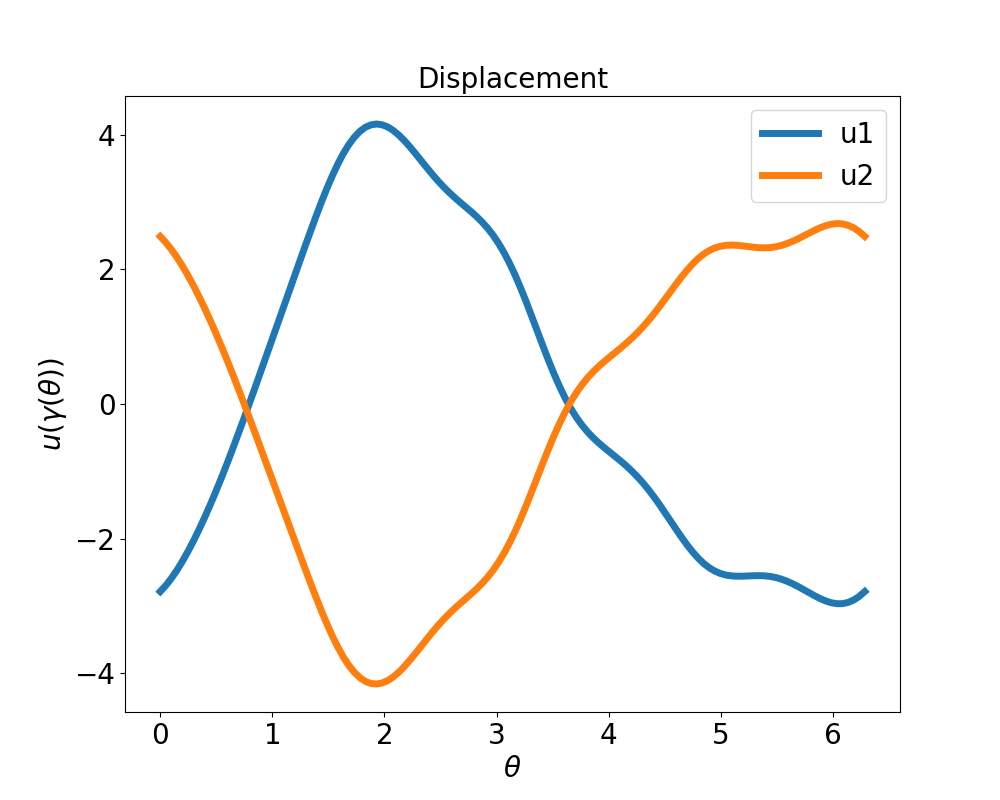}
		\subcaption{The displacement on the boundary}
	\end{subfigure}
	\hfill 
	\begin{subfigure}{0.24\textwidth}
		\centering
		\includegraphics[width=\textwidth]{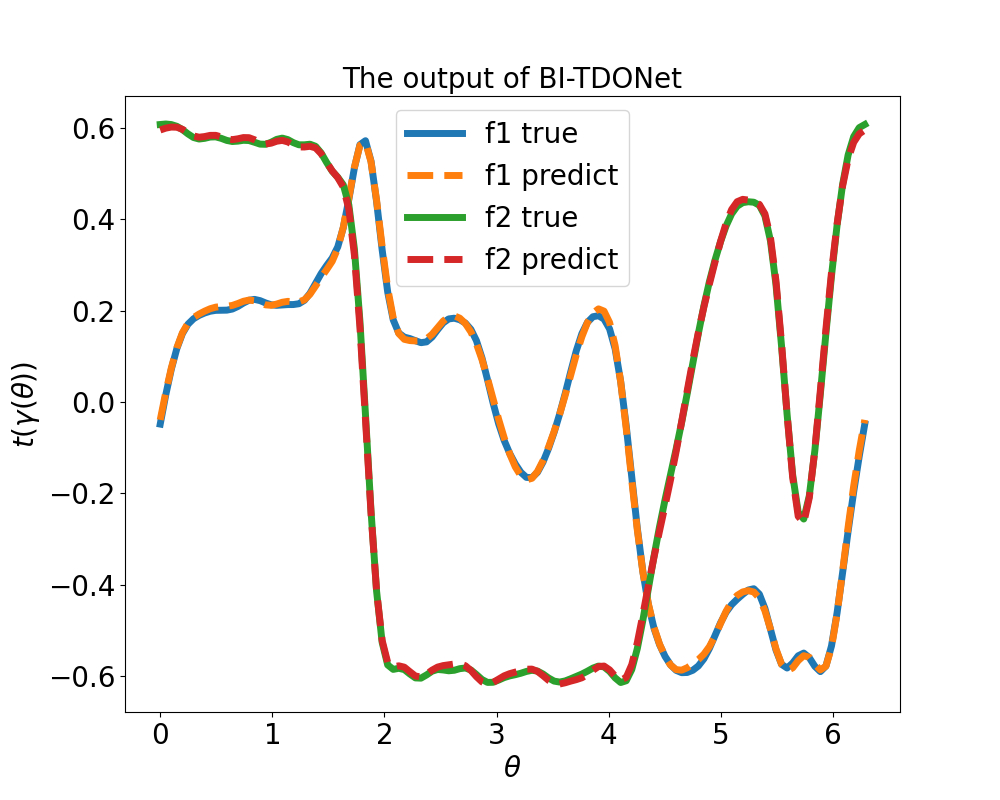}
		\subcaption{The true solution and predicted solution of stress on the boundary}
	\end{subfigure}
	\hfill 
	\begin{subfigure}{0.24\textwidth}
		\centering
		\includegraphics[width=\textwidth]{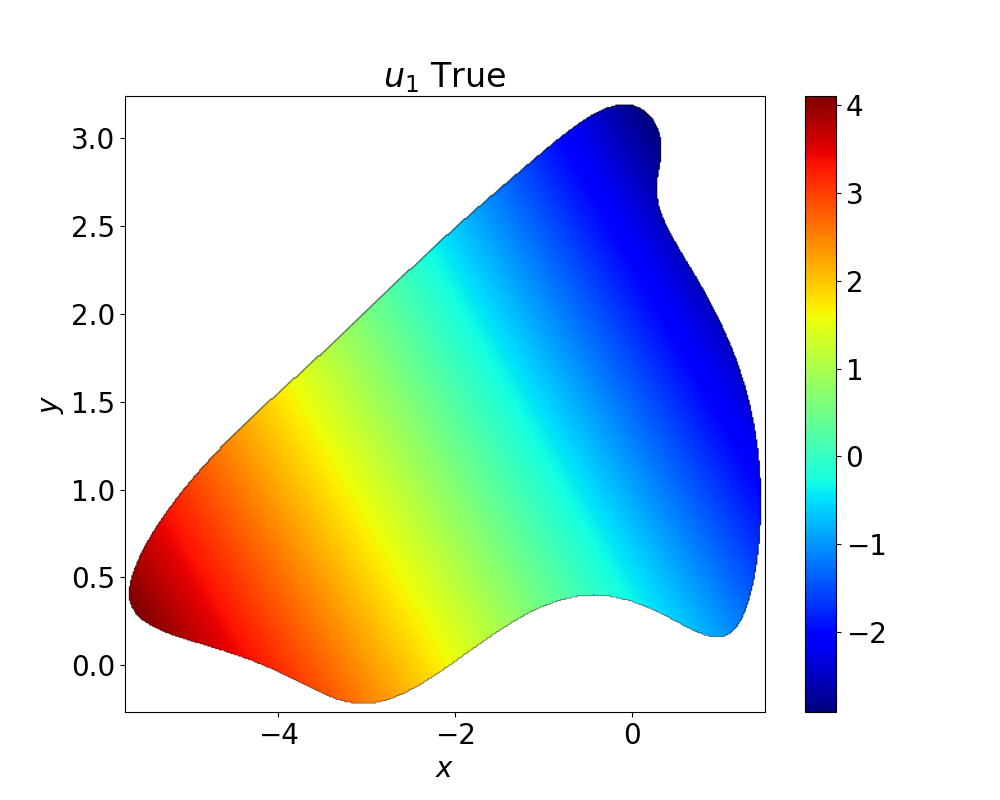}
		\subcaption{The true solution of the displacement field $u_1(\boldsymbol{x})$}
	\end{subfigure}
	\hfill 
	\begin{subfigure}{0.24\textwidth}
		\centering
		\includegraphics[width=\textwidth]{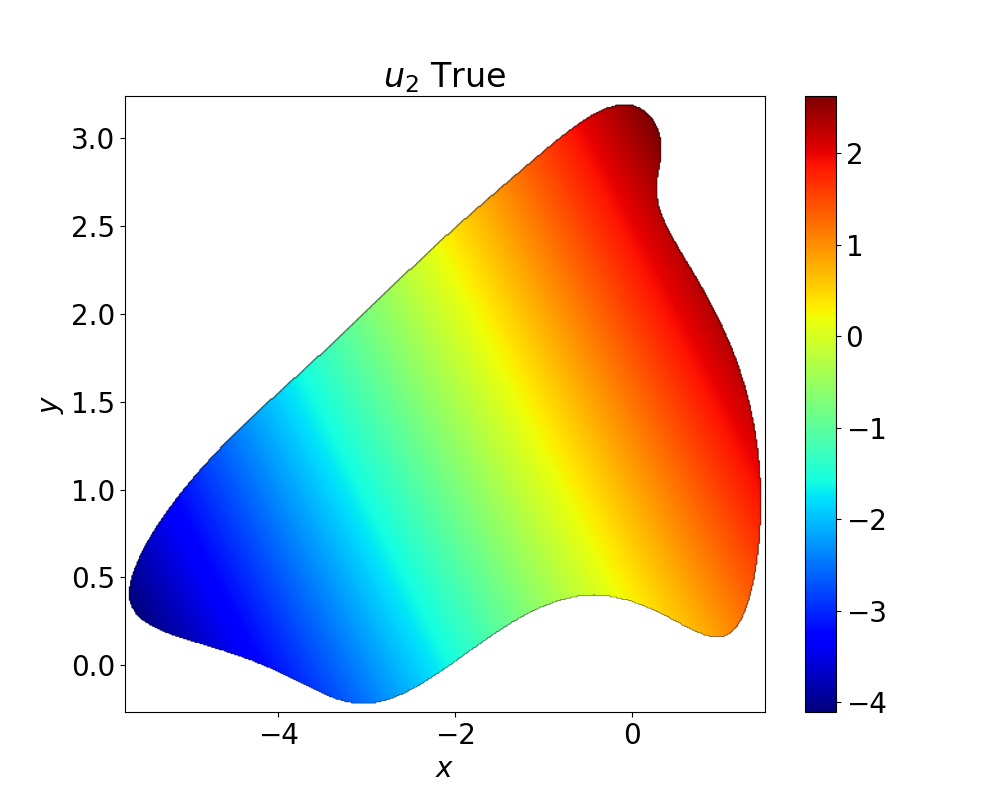}
		\subcaption{The true solution of the displacement field $u_2(\boldsymbol{x})$}
	\end{subfigure}
	
	\begin{subfigure}{0.24\textwidth}
		\centering
		\includegraphics[width=\textwidth]{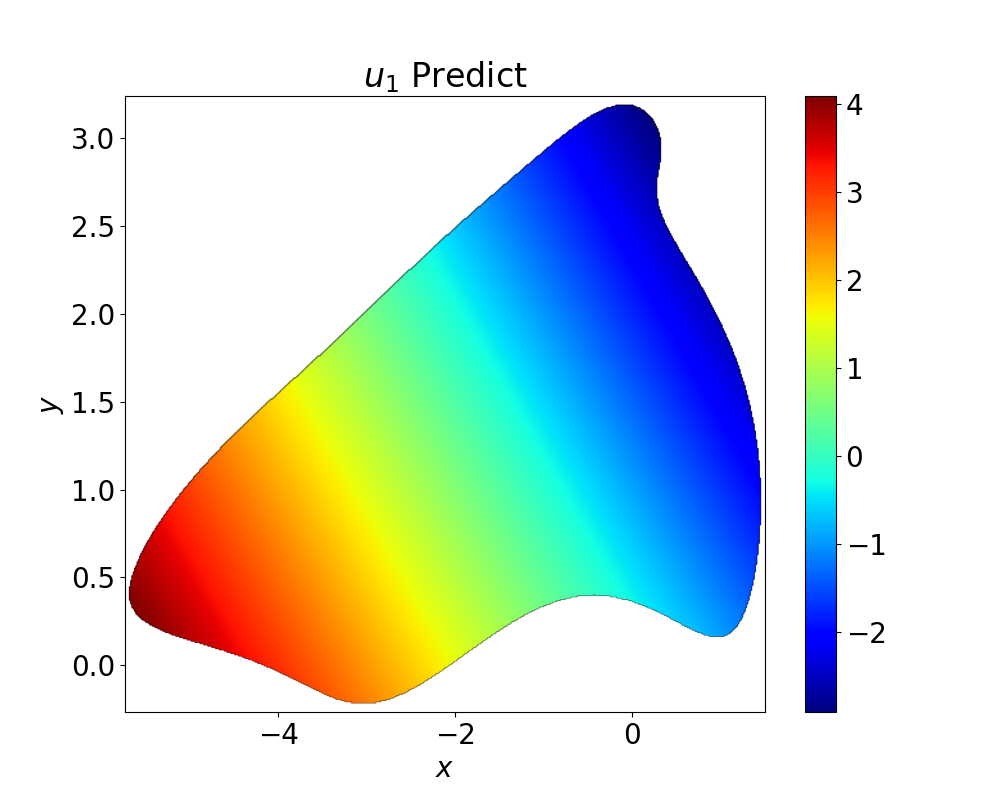}
		\subcaption{The predicted solution of the displacement field $u_1(\boldsymbol{x})$}
	\end{subfigure}
	\hfill 
	\begin{subfigure}{0.24\textwidth}
		\centering
		\includegraphics[width=\textwidth]{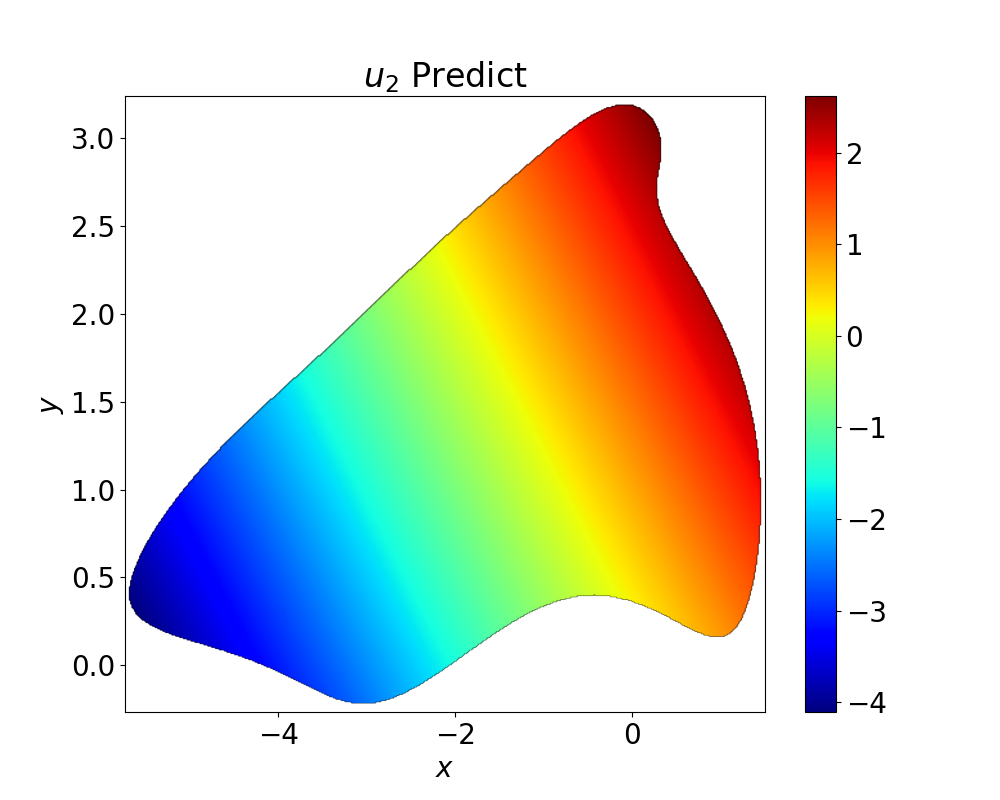}
		\subcaption{The predicted solution of the displacement field $u_2(\boldsymbol{x})$}
	\end{subfigure}
	\hfill 
	\begin{subfigure}{0.24\textwidth}
		\centering
		\includegraphics[width=\textwidth]{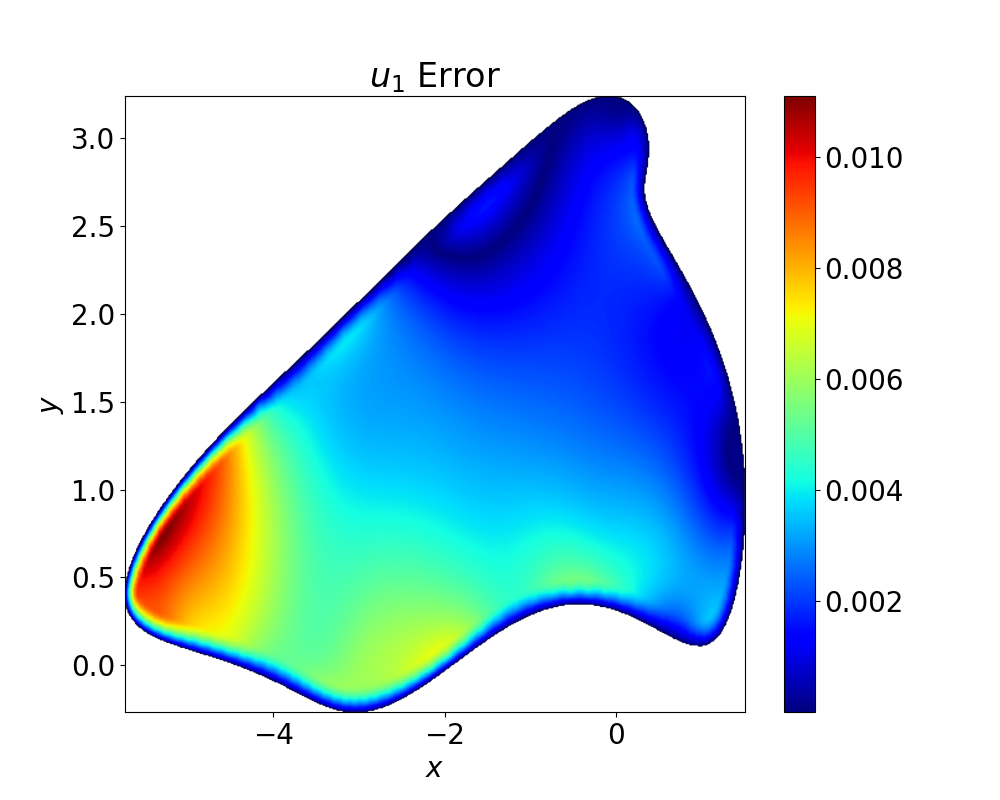}
		\subcaption{The error of the displacement field $u_1(\boldsymbol{x})$}
	\end{subfigure}
	\hfill 
	\begin{subfigure}{0.24\textwidth}
		\centering
		\includegraphics[width=\textwidth]{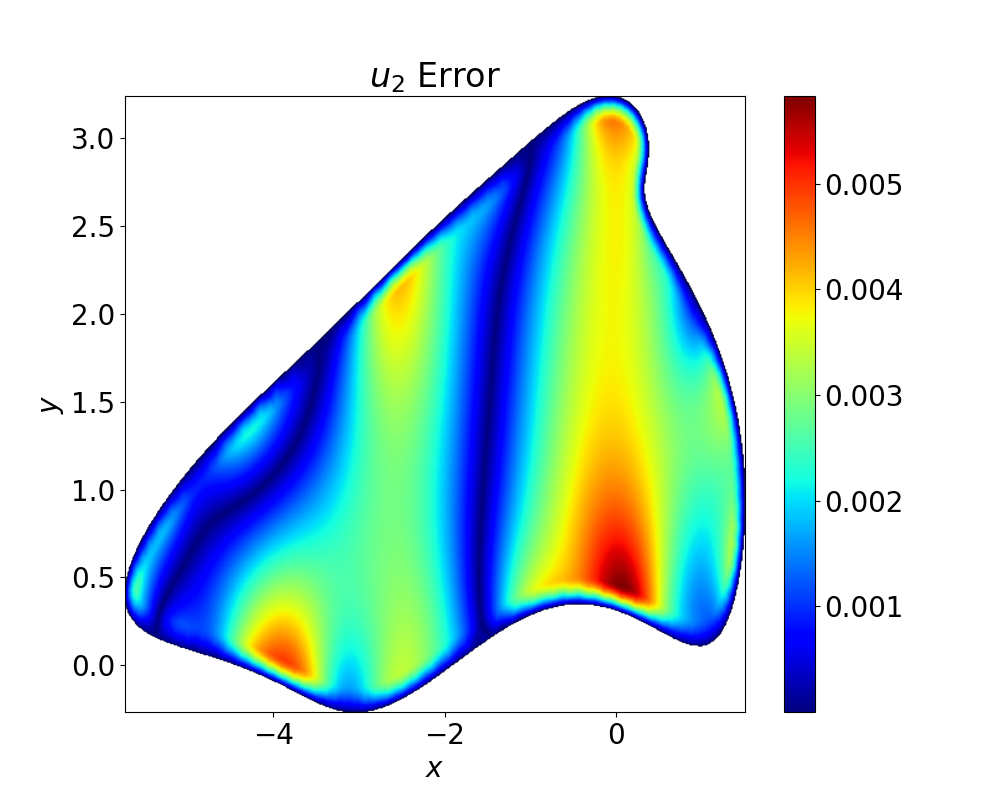}
		\subcaption{The error of the displacement field $u_2(\boldsymbol{x})$}
	\end{subfigure}
	
	\caption{The performance of BI-TDONet in the first example of elastostatics problems.}
	\label{ela_result}
\end{figure}

\begin{figure}[htb]
	\centering
	\begin{subfigure}{0.24\textwidth}
		\centering
		\includegraphics[width=\textwidth]{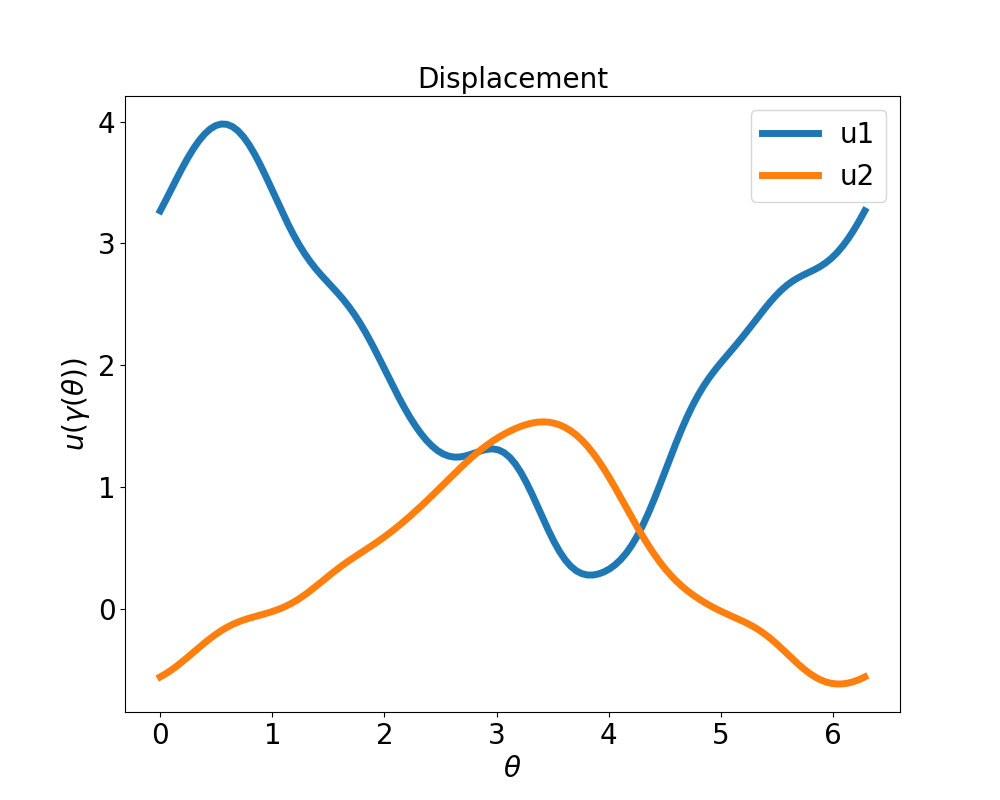}
		\subcaption{The displacement on the boundary}
	\end{subfigure}
	\hfill 
	\begin{subfigure}{0.24\textwidth}
		\centering
		\includegraphics[width=\textwidth]{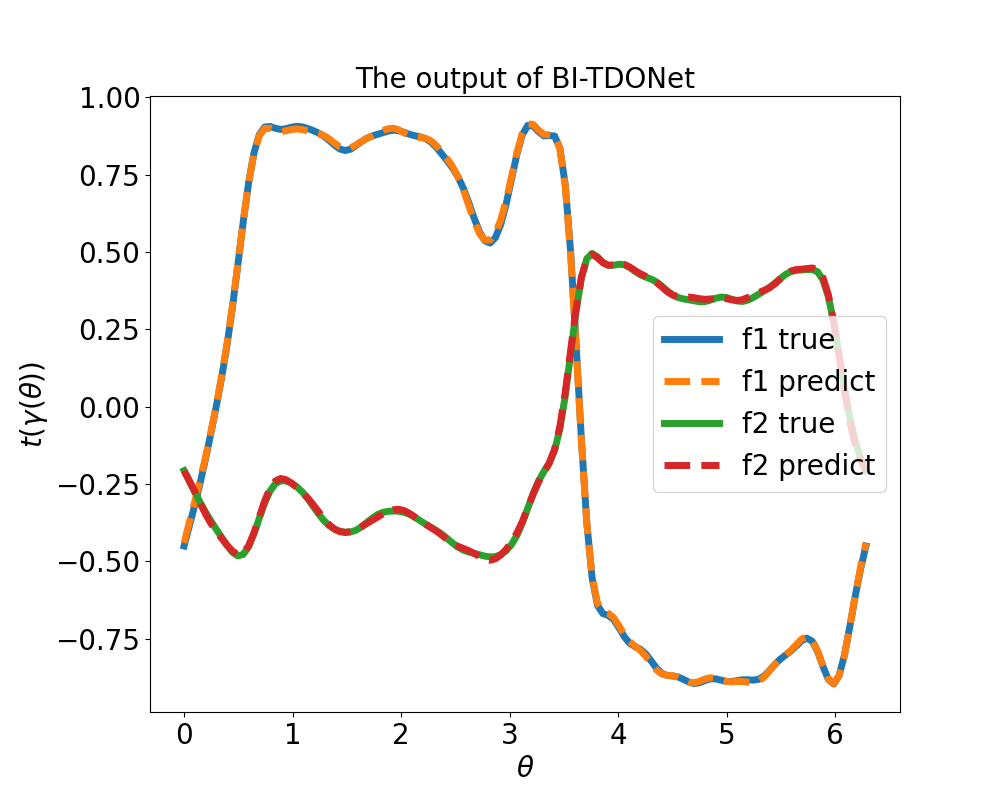}
		\subcaption{The true solution and predicted solution of stress on the boundary}
	\end{subfigure}
	\hfill 
	\begin{subfigure}{0.24\textwidth}
		\centering
		\includegraphics[width=\textwidth]{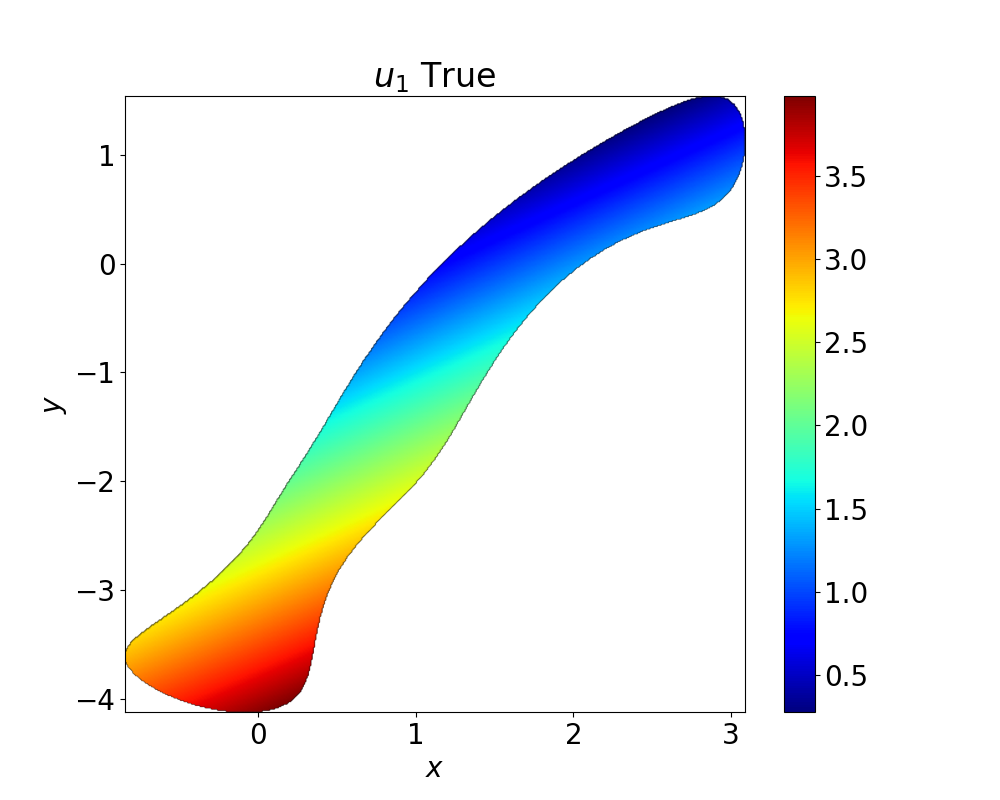}
		\subcaption{The true solution of the displacement field $u_1(\boldsymbol{x})$}
	\end{subfigure}
	\hfill 
	\begin{subfigure}{0.24\textwidth}
		\centering
		\includegraphics[width=\textwidth]{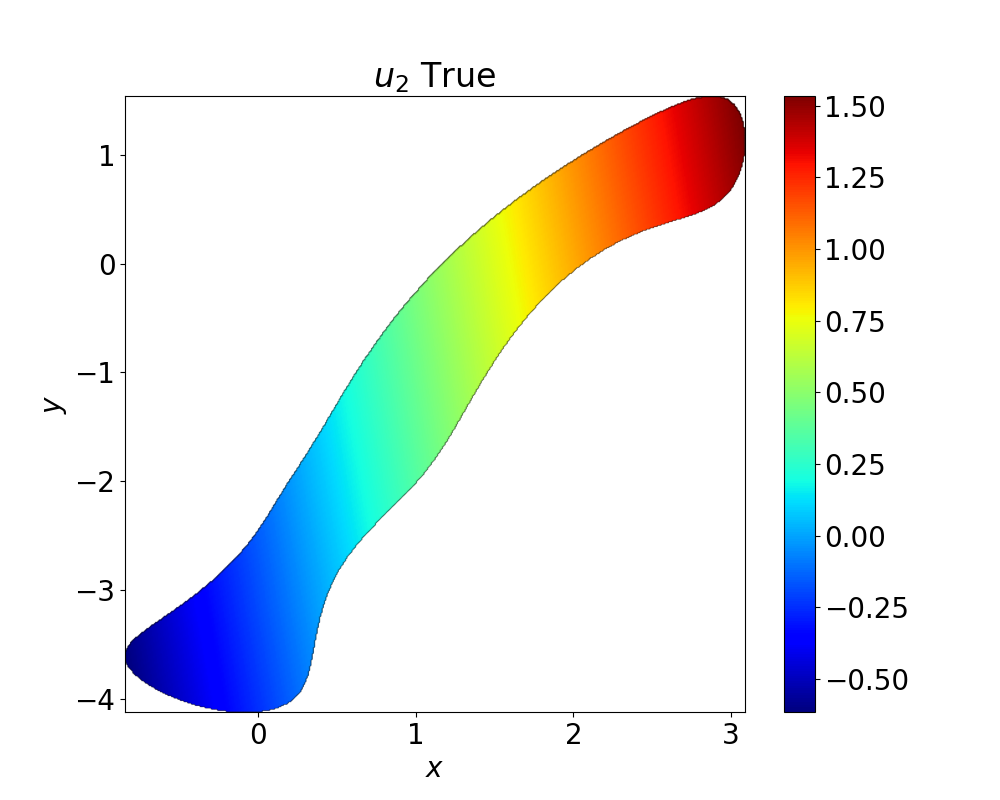}
		\subcaption{The true solution of the displacement field $u_2(\boldsymbol{x})$}
	\end{subfigure}
	
	\begin{subfigure}{0.24\textwidth}
		\centering
		\includegraphics[width=\textwidth]{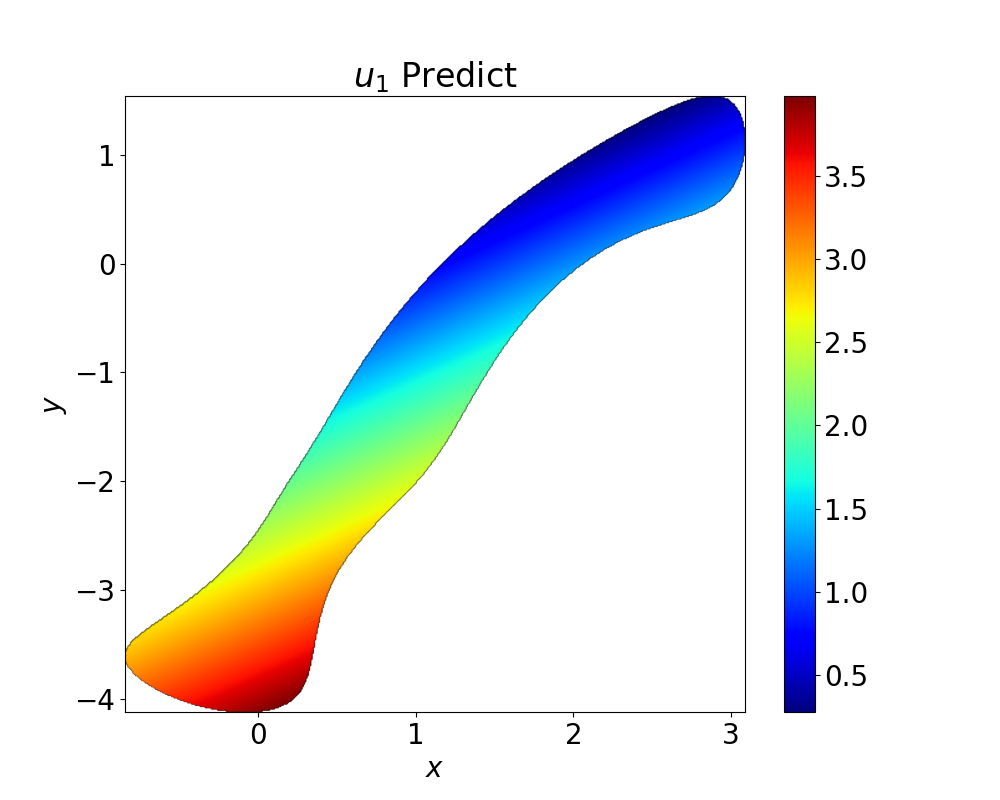}
		\subcaption{The predicted solution of the displacement field $u_1(\boldsymbol{x})$}
	\end{subfigure}
	\hfill 
	\begin{subfigure}{0.24\textwidth}
		\centering
		\includegraphics[width=\textwidth]{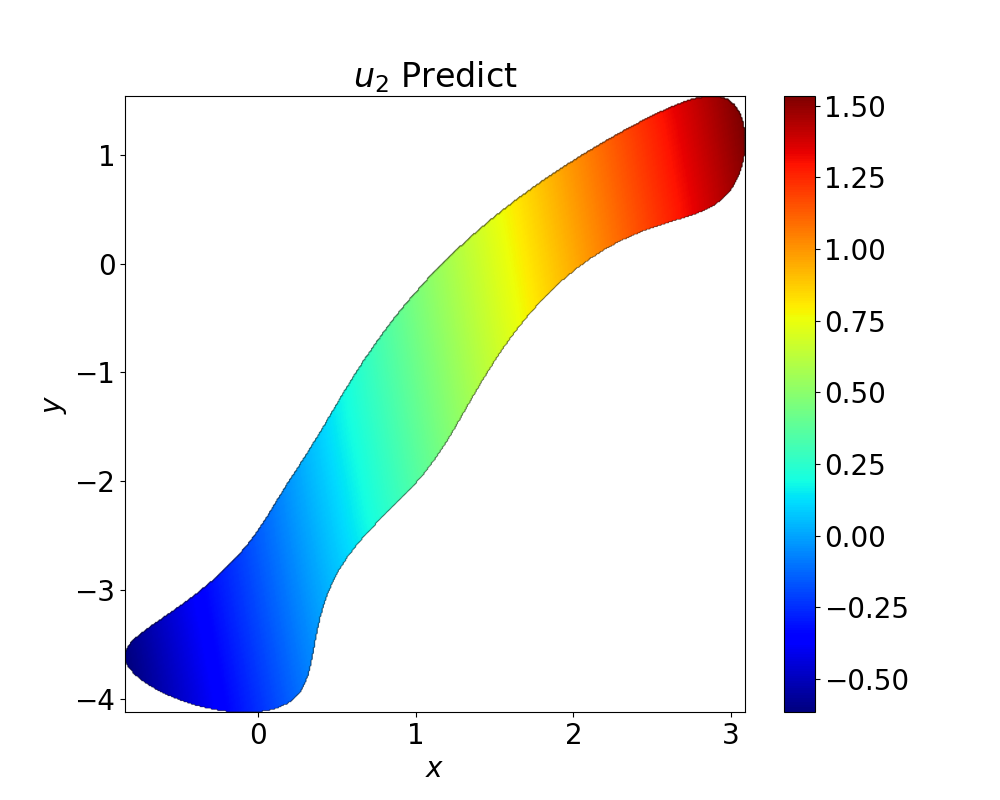}
		\subcaption{The predicted solution of the displacement field $u_2(\boldsymbol{x})$}
	\end{subfigure}
	\hfill 
	\begin{subfigure}{0.24\textwidth}
		\centering
		\includegraphics[width=\textwidth]{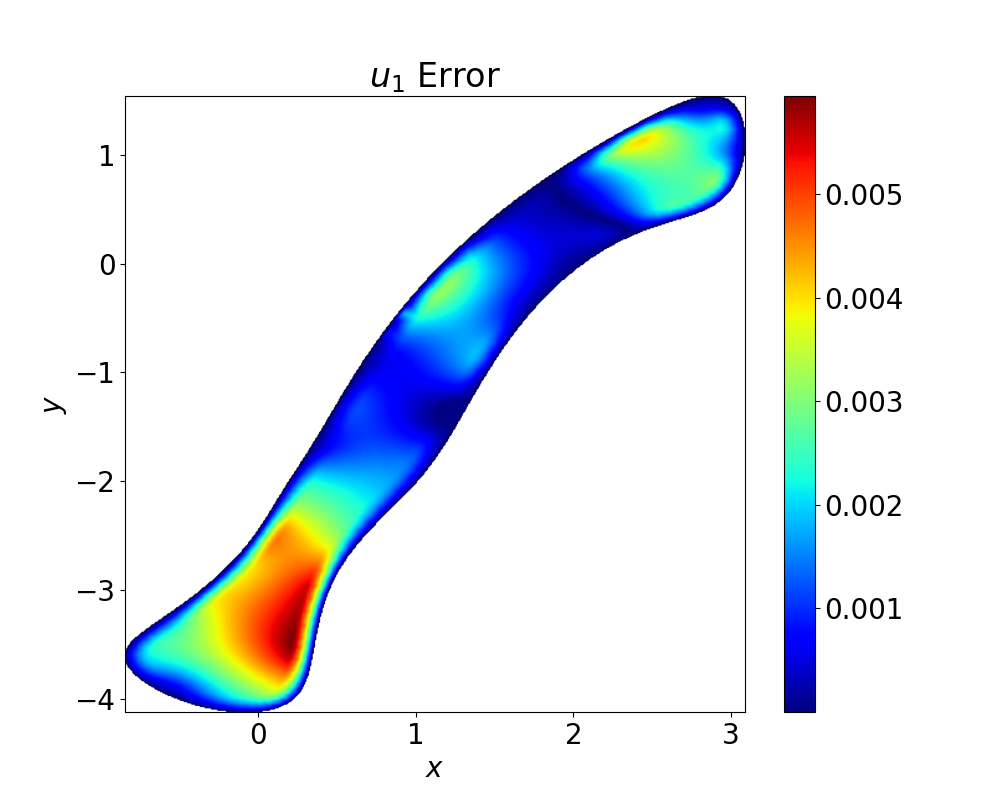}
		\subcaption{The error of the displacement field $u_1(\boldsymbol{x})$}
	\end{subfigure}
	\hfill 
	\begin{subfigure}{0.24\textwidth}
		\centering
		\includegraphics[width=\textwidth]{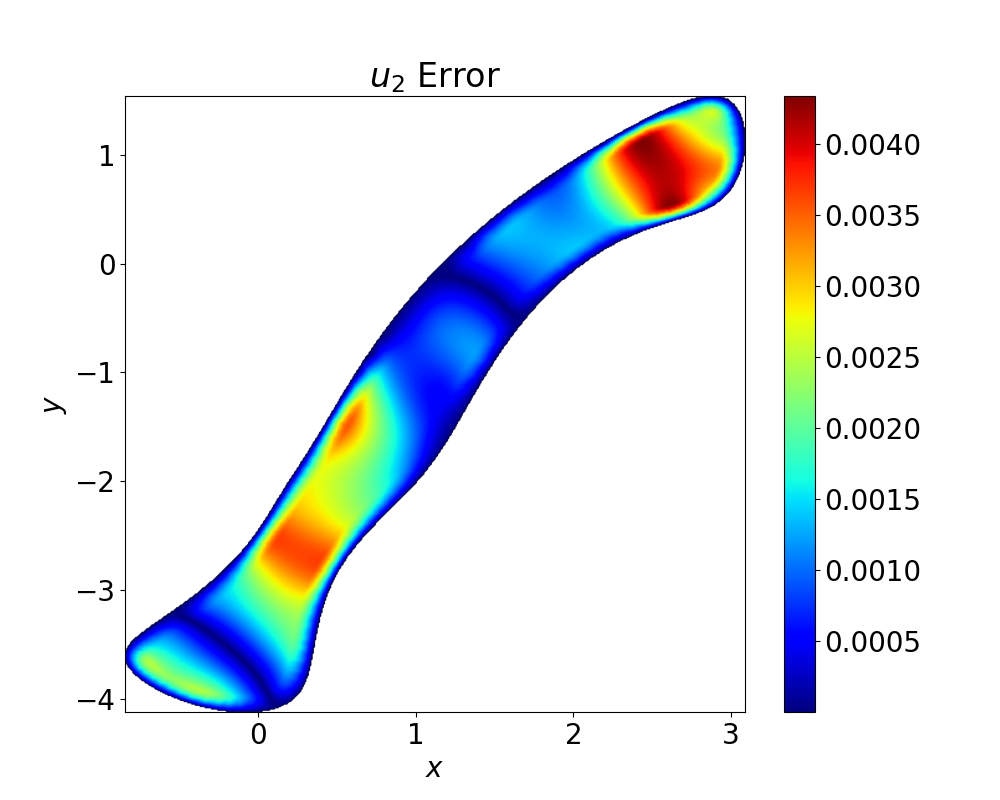}
		\subcaption{The error of the displacement field $u_2(\boldsymbol{x})$}
	\end{subfigure}	
	\caption{The performance of BI-TDONet in the second example of elastostatics problems.}
	\label{ela_result1}
\end{figure}

We continue to utilize two randomly selected samples from the test set to demonstrate the effectiveness of BI-DeepONet and BI-TDONet in addressing elastostatics problems. The visualization results for both models are displayed in Figures \ref{ela_xde_result} - \ref{ela_result1}, showcasing their capabilities in this context.

Figures \ref{ela_xde_result} - \ref{ela_result1} (a) display visualizations of the displacement at the boundary. Figures \ref{ela_xde_result} - \ref{ela_result1} (b) present the true and predicted solutions of the operator equation, with predictions for $t_1(\boldsymbol{x})$ and $t_2(\boldsymbol{x})$ in Figures \ref{ela_xde_result} and \ref{ela_xde_result1} from BI-DeepONet-1 and BI-DeepONet-2, respectively. Figures \ref{ela_result} and \ref{ela_result1} showcase predictions from BI-TDONet, indicating that both BI-DeepONet and BI-TDONet can accurately predict the stress from the displacement at the boundary.

Figures \ref{ela_xde_result} - \ref{ela_result1} (c) and (d) illustrate the true solutions for the displacement fields $u_1(\boldsymbol{x})$ and $u_2(\boldsymbol{x})$, while Figures \ref{ela_xde_result} - \ref{ela_result1} (e) and (f) display their predicted solutions. Figures \ref{ela_xde_result} - \ref{ela_result1} (g) and (h) depict the discrepancies between the predicted and true solutions for the displacement fields $u_1(\boldsymbol{x})$ and $u_2(\boldsymbol{x})$. Due to the near-singular integrals encountered during numerical integration, errors tend to concentrate near the boundary. We have chosen to perform interpolation of the displacement field at positions within $0.1$ from the boundary. This approach is aimed at significantly improving the accuracy of our computational results.

The visual results from Figures \ref{ela_xde_result} to \ref{ela_result1} demonstrate that both BI-DeepONet and BI-TDONet not only effectively predict the solutions to the operator equations but also achieve high accuracy in the resultant displacement fields derived from these predictions, highlighting their distinct advantages in addressing elastostatics problems. Although BI-DeepONet also achieves commendable results, its requirement for multiple training sessions to handle vector-valued functions significantly increases the training burden, a challenge that is especially pronounced when dealing with high-dimensional vector-valued function problems.

\subsection{Obstacle scattering problem for acoustic waves}
In this subsection, we examine the problem of two-dimensional harmonic acoustic wave scattering by a sound-soft obstacle. This scattering problem is governed by the Helmholtz equation as follows (see \cite{colton1998inverse})
\begin{equation*}
	\left\{
	\begin{aligned}
		\Delta u(\boldsymbol{x})+k^2u(\boldsymbol{x})=0\qquad \boldsymbol{x}&\in D,\\
		u(\boldsymbol{x}) =0 \qquad \boldsymbol{x}&\in\partial  D,
	\end{aligned}
	\right.
\end{equation*}
where the $k$ is wave number. Given a plane incident wave $ u^i(\boldsymbol{x}) = e^{ik\boldsymbol{x} \cdot \boldsymbol{d}} $, with $ \boldsymbol{d} $ representing the direction of the incident wave, the scattered field also satisfies the Helmholtz equation
\begin{equation}\label{helmholtz}
		\left\{
	\begin{aligned}
		\Delta u^s(\boldsymbol{x})+k^2u^s(\boldsymbol{x})=0\qquad \boldsymbol{x}&\in D,\\
		u^s(\boldsymbol{x}) =-e^{ik\boldsymbol{x} \cdot \boldsymbol{d}} \qquad \boldsymbol{x}&\in\partial  D.
	\end{aligned}
	\right.
\end{equation}
The fundamental solution to the two-dimensional Helmholtz equation is given by $ \Phi(\boldsymbol{x}, \boldsymbol{y}) = \frac{i}{4} H_0^{(1)}(k|\boldsymbol{x}-\boldsymbol{y}|) $ for $ \boldsymbol{x} \neq \boldsymbol{y} $. The scattered field $ u^s(\boldsymbol{x}) $ can then be expressed as
\begin{equation}\label{scatter}
	u^s(\boldsymbol{x})=\int_{\partial D}\left\lbrace \frac{\partial \Phi(\boldsymbol{x},\boldsymbol{y})}{\partial \nu(\boldsymbol{y})}-i\eta\Phi(\boldsymbol{x},\boldsymbol{y})\right\rbrace \varphi(\boldsymbol{y})ds(\boldsymbol{y}),\qquad \boldsymbol{x}\in \partial D,
\end{equation}
where $ \eta \neq 0 $ is a real coupling parameter, and set to $\eta = k$ when the wave number $k > \frac{1}{2}$, as per references \cite{kress1985,colton1998inverse}. The layer potential $\varphi$ satisfies the BIE, which  integrates the effects of the single-layer \eqref{single-layer potential} and double-layer potential \eqref{double-layer potential},
\begin{equation}\label{helmholtz BIE}
	\varphi + \mathcal{D}\varphi - i\eta \mathcal{S}\varphi = -2e^{ik\boldsymbol{x} \cdot \boldsymbol{d}}.
\end{equation}

In this subsection, we use BI-TDONet to learn the operator $(\mathcal{I}+\mathcal{D}-i\eta \mathcal{S})^{-1}$. Initially, to train and test BI-TDONet, we generate the required data. We set $\boldsymbol{d} = [1,0]^\top$ and uniformly sample the wave number $k$ from 40 to 50, yielding 100 plane wave incident functions. Due to the large wave number $k$, these plane wave functions exhibit significant oscillations, necessitating a larger $n$ in \eqref{trigonometic} to accurately represent the functions. Consequently, we set $n=300$ to adequately represent the incident wave functions.

\begin{figure}[htbp]
	\centering
	\begin{subfigure}{0.45\textwidth}
		\centering
		\includegraphics[width=\textwidth]{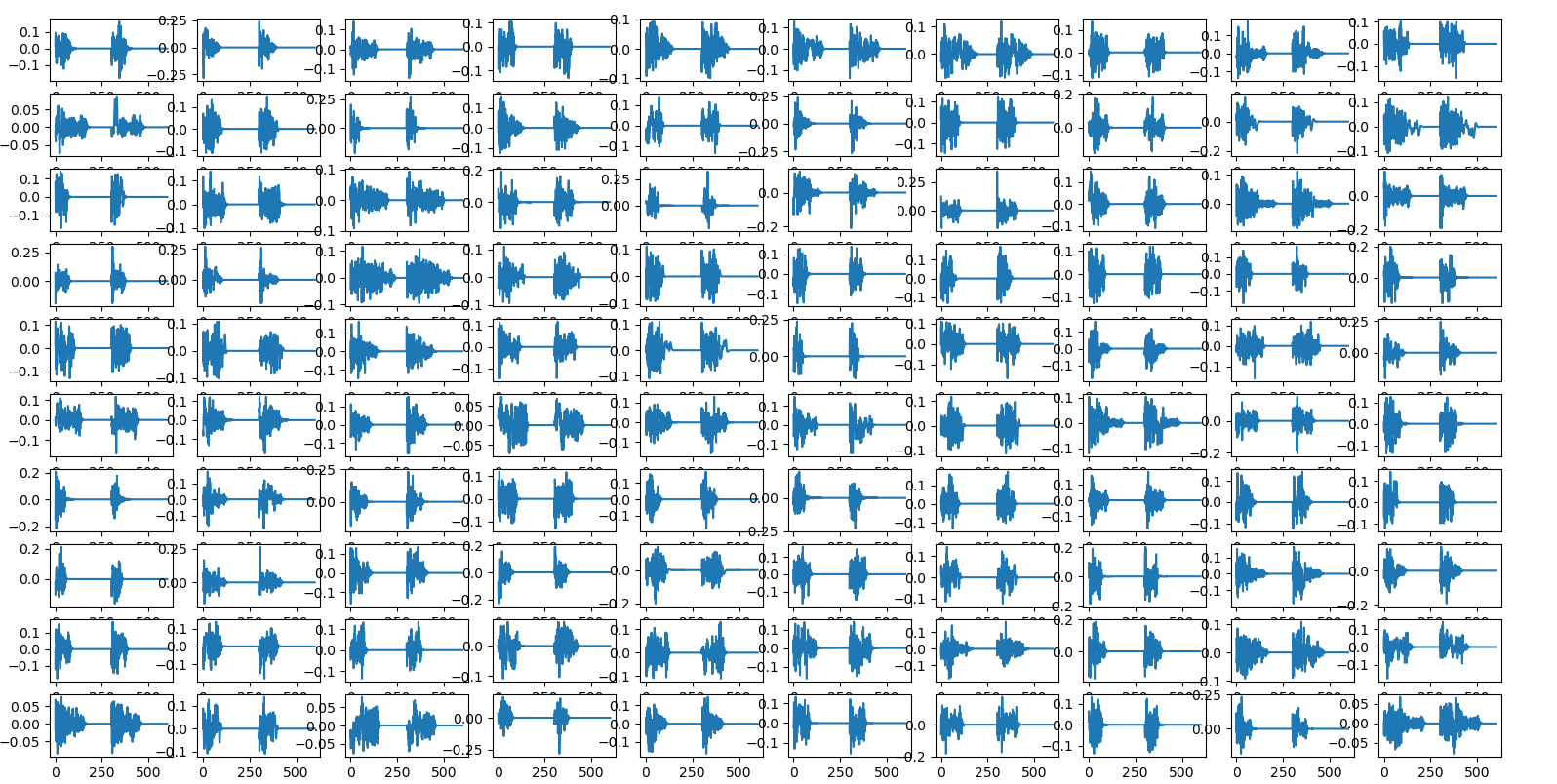}
		\subcaption{The trigonometric coefficients of the real part.}
	\end{subfigure}
	\begin{subfigure}{0.45\textwidth}
		\centering
		\includegraphics[width=\textwidth]{Fig/Helmholtz_coefficient_real.png}
		\subcaption{The trigonometric coefficients of the imaginary part. }
	\end{subfigure}
	\caption{(a) and (b) respectively display partial trigonometric coefficients of the real and imaginary parts of the true solution of the BIE. From the figures, we can observe that their values less than $0.3$.}
	\label{frequency}
\end{figure}

The dataset necessary for network training was generated by solving the \eqref{helmholtz BIE} using the Fourier-Galerkin method. As depicted in Figure $\eqref{frequency}$, the Fourier-Galerkin matrix is characterized as a high-dimensional dense matrix, resulting in significant computational costs for data generation. To facilitate the smooth execution of the experiments, we employed a comparatively smaller boundary dataset for training than in other experiments.

When generating boundaries, a large $n$ would lead to unnecessary storage usage and increase the training burden on the neural network. Following the method described in subsection $3.3$, we set $n=20$ and $\rho=0.3$ to generate $597$ boundaries. Tensor operations between the plane wave functions and the boundaries resulted in a dataset of $59,700$ samples, with $80\%$ allocated for training and $20\%$ for testing.

It is important to note that the operators $(\mathcal{I}+\mathcal{D}-i\eta \mathcal{S})^{-1}$ in the operator equations \eqref{helmholtz BIE} contain imaginary parts, and both the input and output functions are complex-valued. In BI-TDONet, we treat the real and imaginary parts as separate components of vector-valued functions. Thus, the network structure for BI-TDONet is configured as $[[1604, 1500, 1500], [802, 1500, 1500], [2302, 1500, 802]]$.  

When the input space contains high-frequency information, neural operators face significant challenges \cite{li2021fno}. In this subsection, the input space for BI-DeepONet is particularly rich in high-frequency details, which posed considerable challenges during network training and yielded unsatisfactory outcomes. As a result, we have opted not to present the experimental results for BI-DeepONet here. Zhu et al. \cite{zhu2023} explored the integration of Fourier features into DeepONet for full waveform inversion (FWI), and reported improved results, indicating that DeepONet enhanced with Fourier features can effectively manage inputs characterized by high-frequency features. Nevertheless, this article does not delve into these modifications further, concentrating solely on evaluating the performance of BI-DeepONet within the vanilla DeepONet framework.

During the training process of BI-TDONet, we set the batch size to $8,192$ and targeted
$80,000$ epochs for completion. The initial learning rate was set at
$0.001$, utilizing the Adam optimization algorithm for efficient gradient descent. We implemented adaptive learning rate decay by monitoring the loss: if there was no decrease in the loss within
$1/100$ of the epochs, the learning rate was reduced to $0.1$ times its initial value to fine-tune the network's convergence.

\begin{figure}[htbp]
	\centering
	\includegraphics[width=0.8\linewidth]{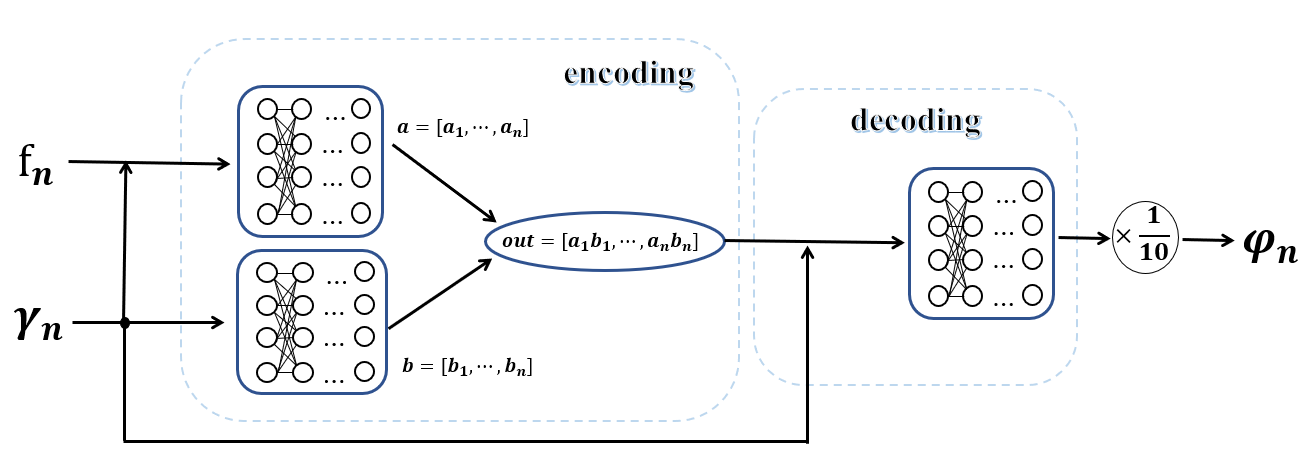}
	\caption{The network architecture of BI-TDONet for the obstacle scattering problem}
	\label{BITDONet structure2}
\end{figure}

As illustrated in Figure \ref{frequency}, the frequency of the true solution of the BIE is relatively low. To enhance the neural network's ability to capture these lower frequency details during training, we integrated a scaling layer before the output layer of BI-TDONet. The scaling factor was set to $1/10$, optimizing the network's sensitivity to subtle variations, as detailed in Figure \ref{BITDONet structure2}.


\begin{table}[htb]
	\centering
	
	\begin{tabular}{c|c|c|c|c|c}
		\midrule
		Model
		&MNE
		&MRE
		&variance-MNE
		&variance-MRE
		&Mean-Time/ms\\
		\hline
		BI-TDONet
		&$1.6443\times10^{-6}$
		&$3.5122\times10^{-2}$
		&$3.4386\times10^{-11}$
		&$7.8963\times10^{-4}$
		&$3.7378\times10^{-2}$\\
		
		\midrule
	\end{tabular}	
    \caption{Error table on the obstacle scattering problem test set.}
	\label{helmholtz error}	
\end{table}

Table \ref{helmholtz error} presents the MNE and MRE for BI-TDONet on the test set, along with the variances of these metrics. The data from Table \ref{helmholtz error} reveal that the MRE for BI-TDONet on the test set is $3.5122\%$, signifying that BI-TDONet is highly effective at learning the sound-soft obstacle scattering problem for plane wave acoustic incidents. This underscores BI-TDONet's robust capability to manage complex acoustic scattering simulations with high accuracy.

\begin{figure}[htb]
	\centering
	\includegraphics[width=0.5\textwidth]{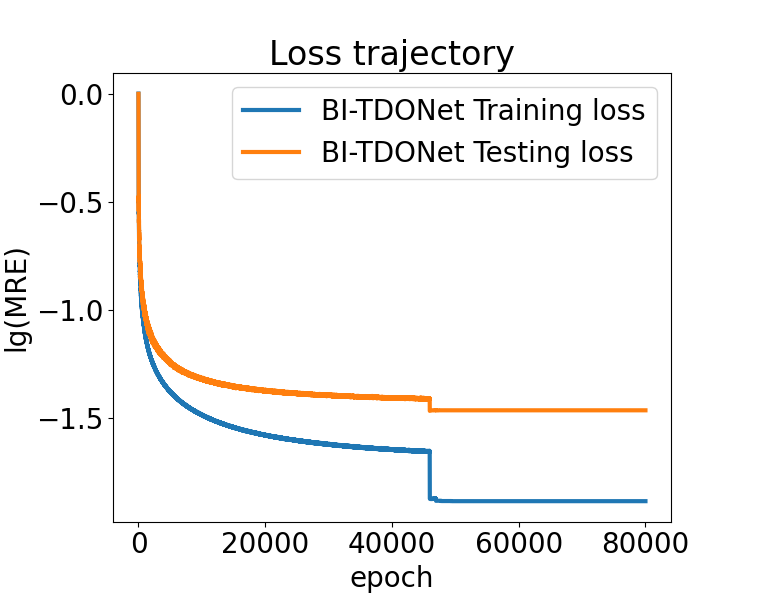}
	\caption{The logarithmic loss trajectories for BI-TDONet in the obstacle scattering problem.}
	\label{helmholtz_loss}
\end{figure}

Figure \ref{helmholtz_loss} shows the logarithmic MRE trajectories during the training and testing phases. BI-TDONet reaches convergence at epoch $50,000$, demonstrating the model's ability to stabilize and refine its performance effectively over the course of training. This indicates strong adaptability and efficiency in the model's learning process, enhancing its practical utility in acoustic modeling scenarios.

\begin{figure}[htb]
	\vspace{-1.cm}
	\centering
	\begin{subfigure}{.3\linewidth}
		\includegraphics[width=\textwidth]{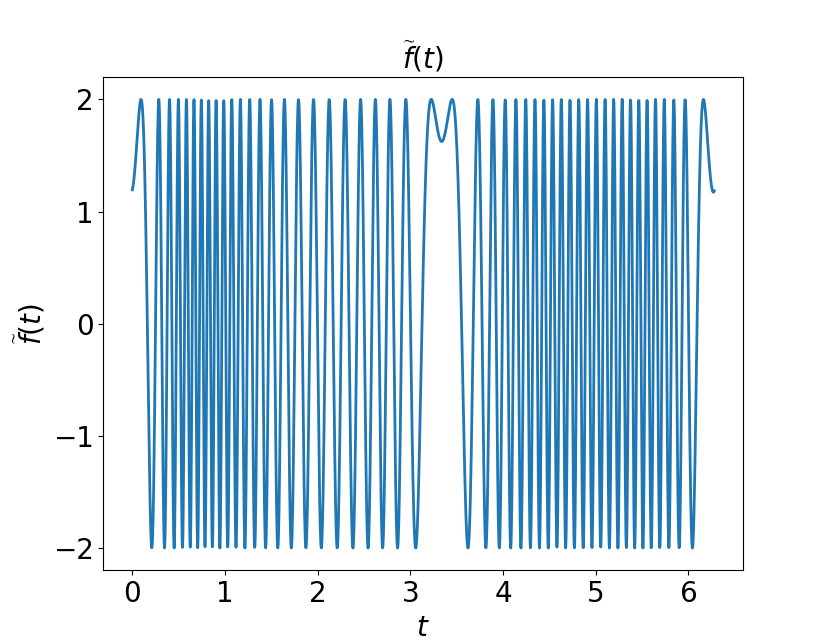}
		\subcaption{The real part of the right-hand side function $\widetilde{f}$ of the BIE.}
	\end{subfigure}
	\begin{subfigure}{.3\linewidth}
		\includegraphics[width=\textwidth]{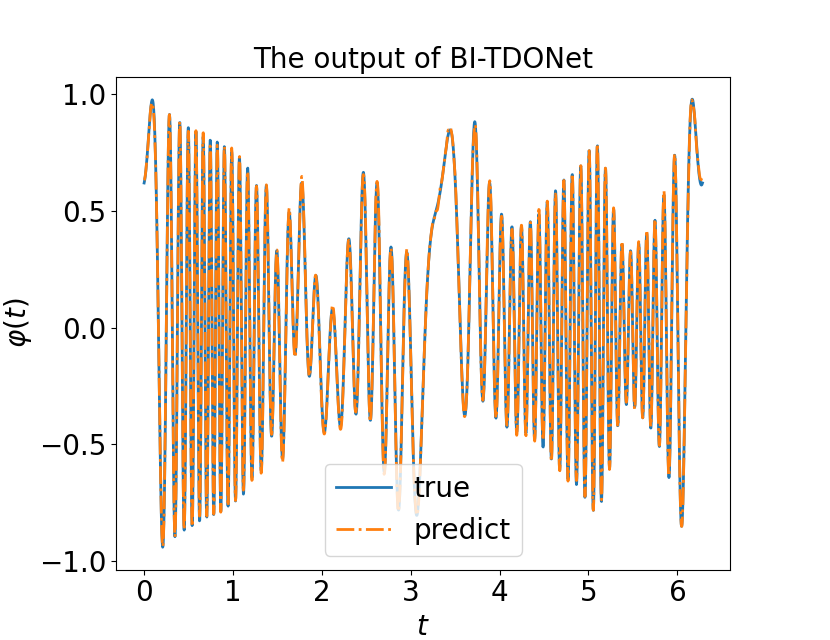}
		\subcaption{The real part of the predicted solution of BI-TDONet.}
	\end{subfigure}
	\begin{subfigure}{.3\linewidth}
		\includegraphics[width=\textwidth]{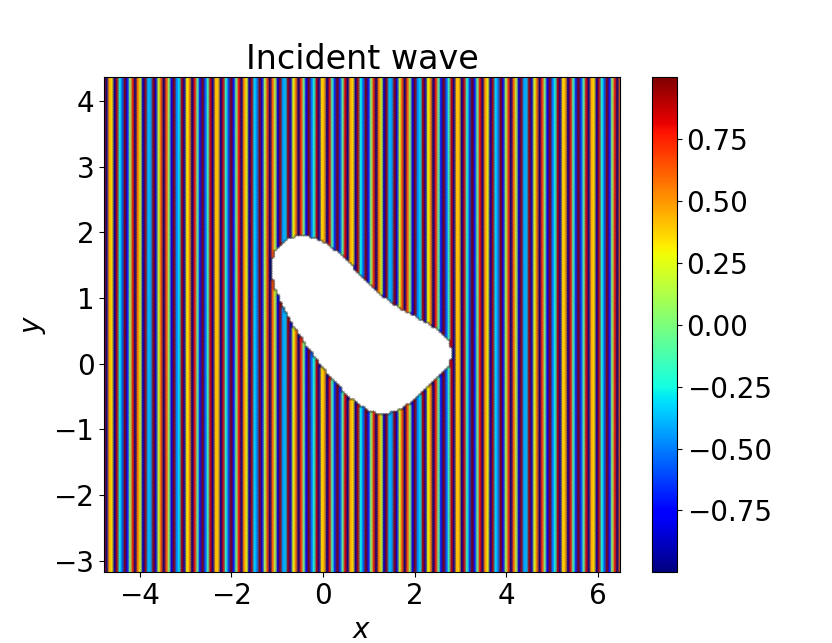}
		\subcaption{The real part of the plane incident wave in the direction $[1,0]$.}
	\end{subfigure}
	\begin{subfigure}{.3\linewidth}
		\includegraphics[width=\textwidth]{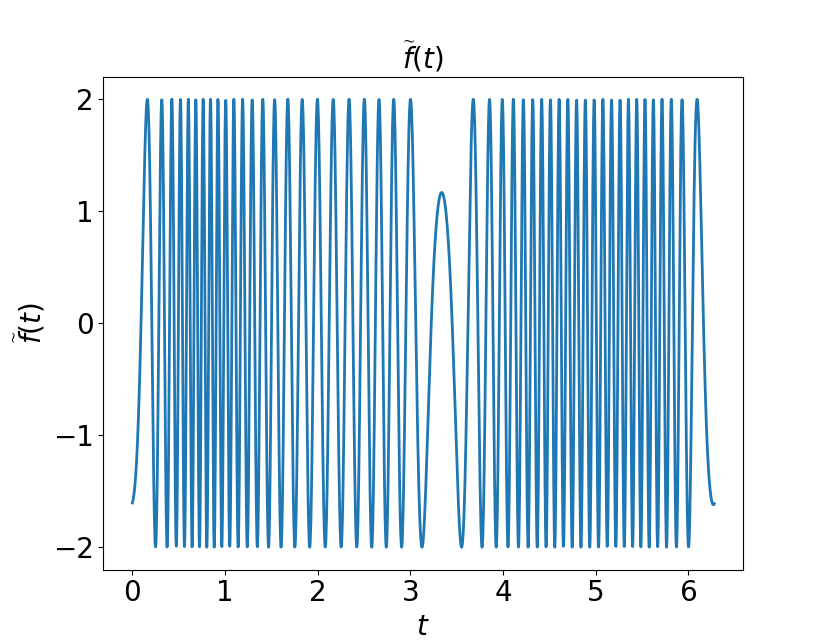}
		\subcaption{The imaginary part of the right-hand side function $\widetilde{f}$ of the BIE.}
	\end{subfigure}
	\begin{subfigure}{0.3\textwidth}
		\centering
		\includegraphics[width=\textwidth]{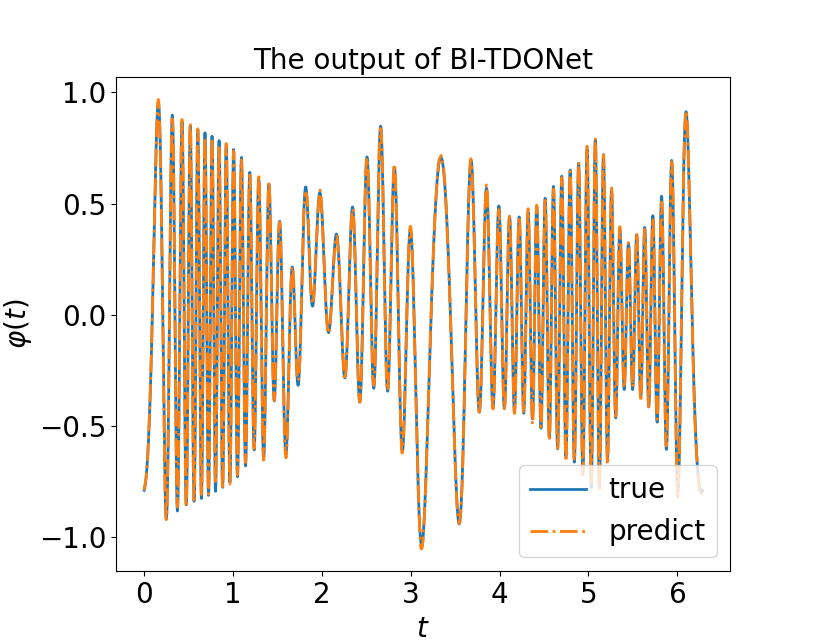}
		\subcaption{The imaginary part of the predicted solution of BI-TDONet.}
	\end{subfigure}
	\begin{subfigure}{0.3\textwidth}
		\centering
		\includegraphics[width=\textwidth]{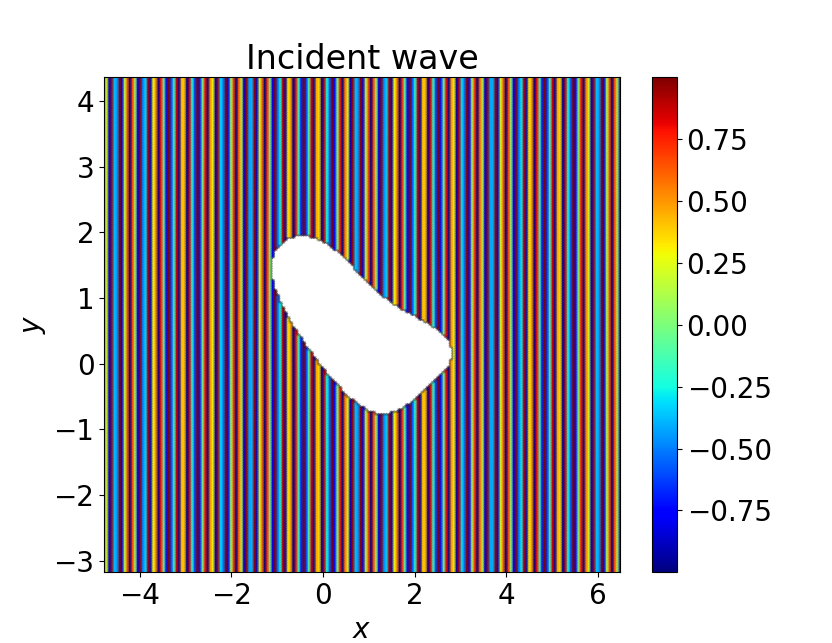}
		\subcaption{The imaginary part of the plane incident wave in the direction $[1,0]$.}
	\end{subfigure}
	\begin{subfigure}{0.3\textwidth}
		\centering
		\includegraphics[width=\textwidth]{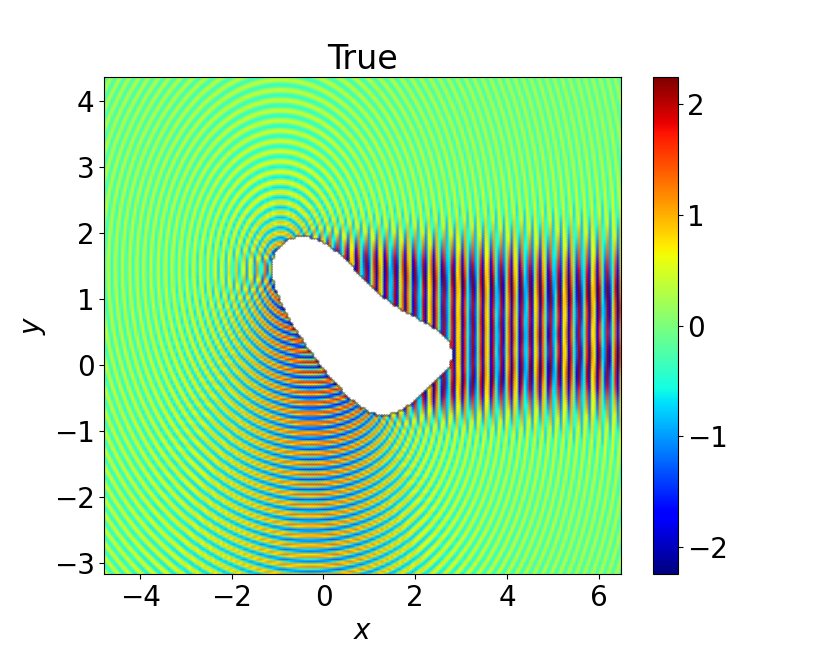}
		\subcaption{The real part of the true solution of the scattered field .}
	\end{subfigure}
	\begin{subfigure}{0.3\textwidth}
		\centering
		\includegraphics[width=\textwidth]{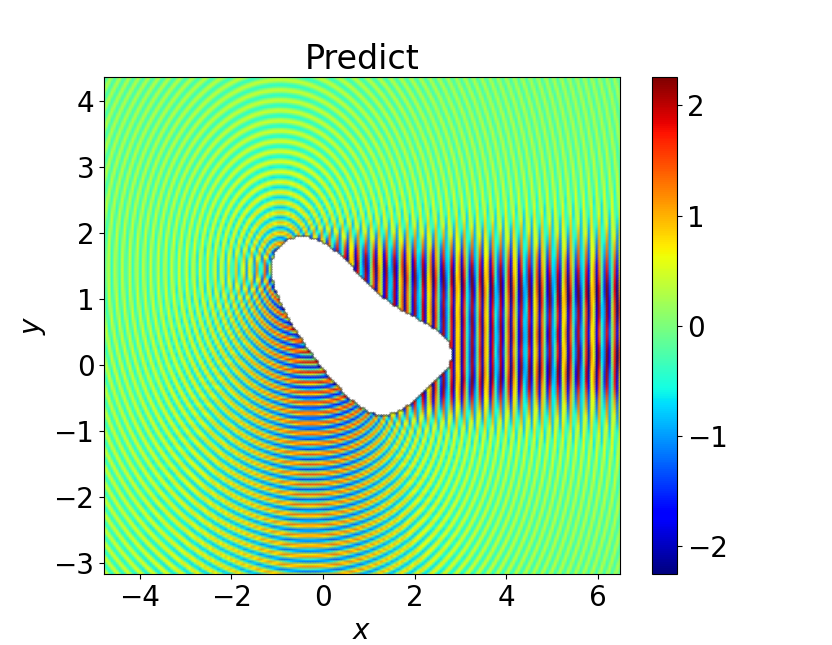}
		\subcaption{The real part of the predicted solution of the scattered field.}
	\end{subfigure}
	\begin{subfigure}{0.3\textwidth}
		\centering
		\includegraphics[width=\textwidth]{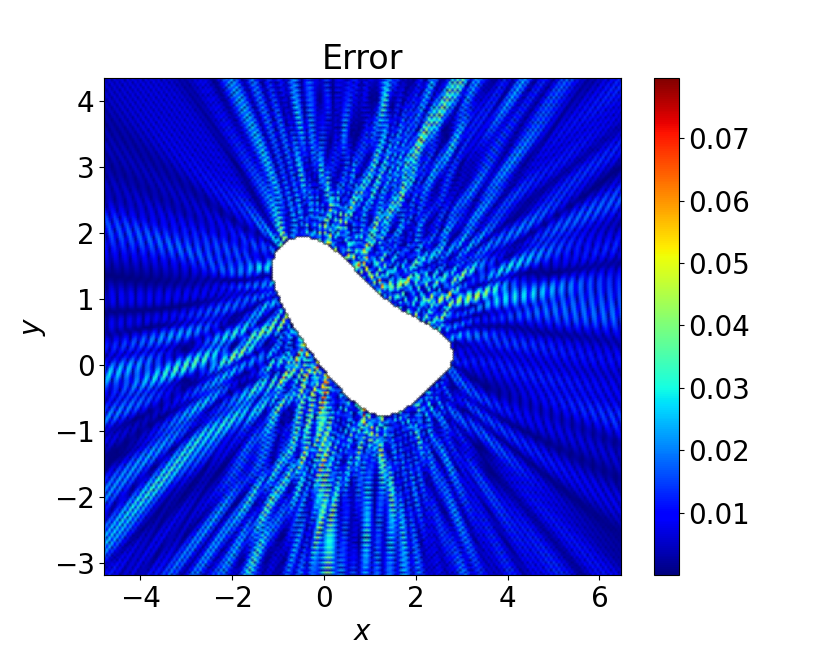}
		\subcaption{The real part of the absolute error between the true and predicted solutions.}
	\end{subfigure}
	\begin{subfigure}{.3\linewidth}
		\includegraphics[width=\textwidth]{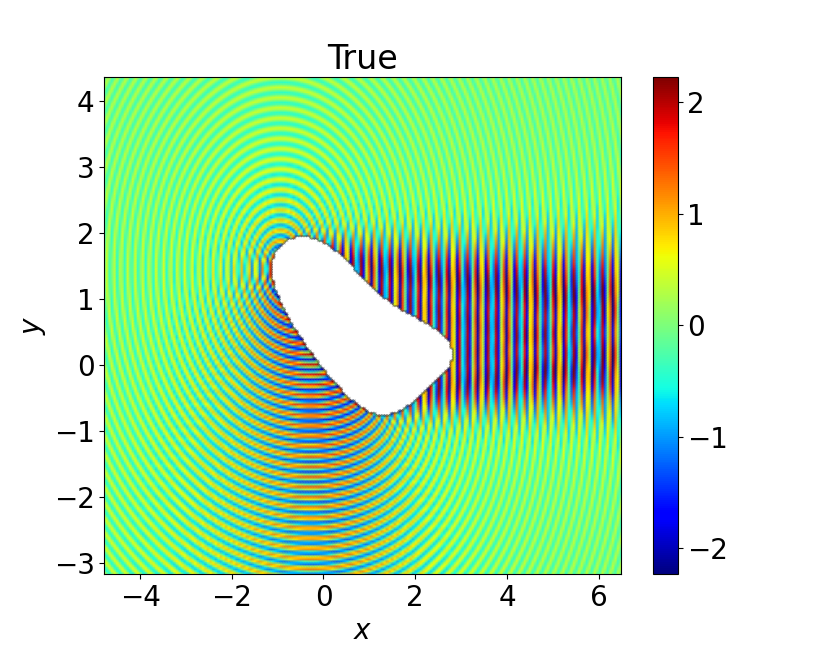}
		\subcaption{The imaginary part of the true solution of the scattered field .}
	\end{subfigure}
	\begin{subfigure}{0.3\textwidth}
		\centering
		\includegraphics[width=\textwidth]{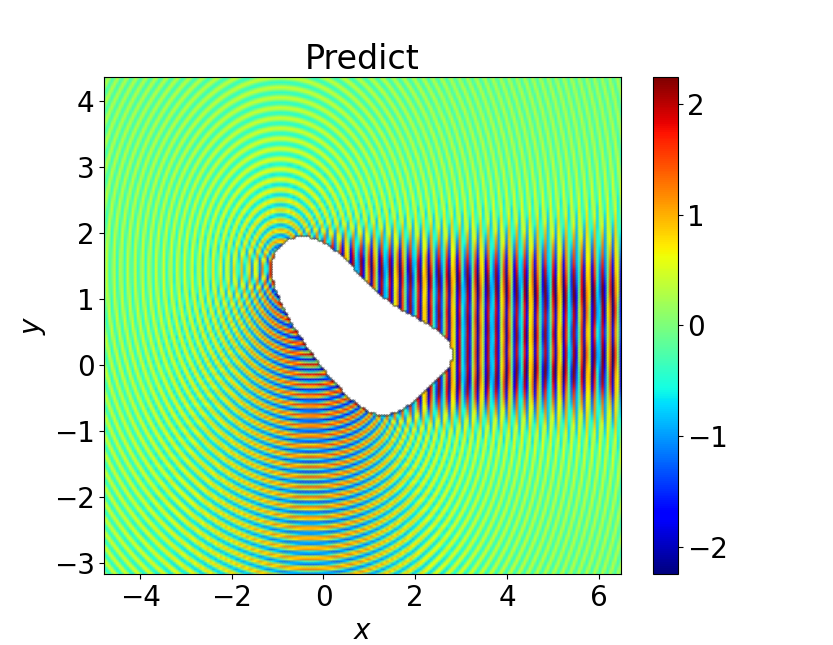}
		\subcaption{The imaginary part of the predicted solution of the scattered field.}
	\end{subfigure}
	\begin{subfigure}{0.3\textwidth}
		\centering
		\includegraphics[width=\textwidth]{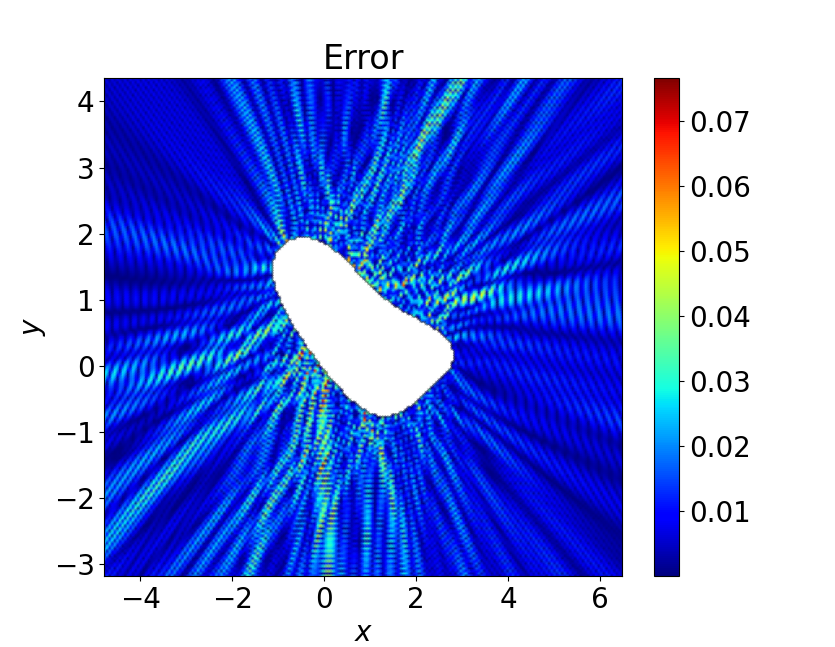}
		\subcaption{The imaginary part of the absolute error between the true and predicted solutions.}
	\end{subfigure}
	\caption{The performance of BI-TDONet in the first example of the acoustic wave obstacle scattering problem.}
	\label{helmholtz_result}
\end{figure}

\begin{figure}[htb]
	\vspace{-1.cm}
	\centering
	\begin{subfigure}{.3\linewidth}
		\includegraphics[width=\textwidth]{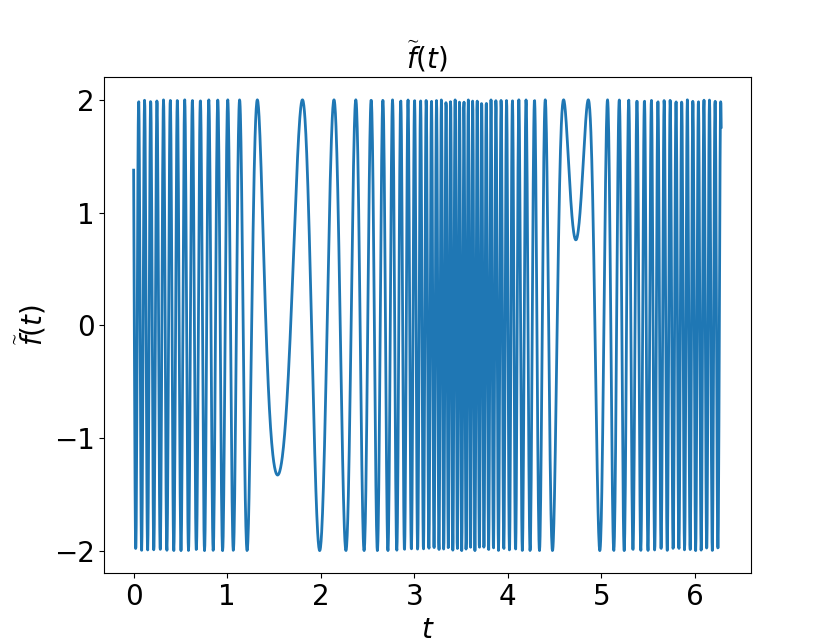}
		\subcaption{The real part of the right-hand side function $\widetilde{f}$ of the BIE.}
	\end{subfigure}
	\begin{subfigure}{.3\linewidth}
		\includegraphics[width=\textwidth]{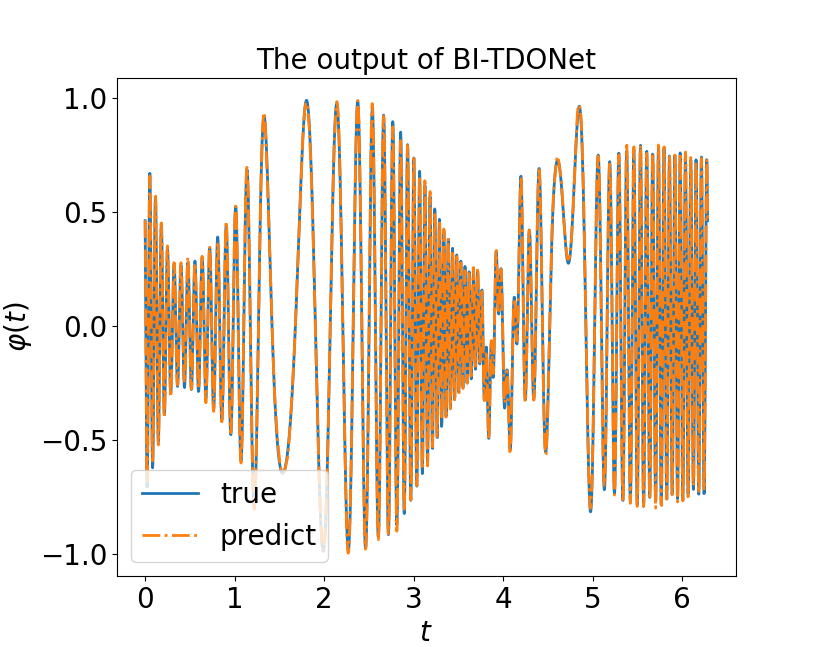}
		\subcaption{The real part of the predicted solution of BI-TDONet.}
	\end{subfigure}
	\begin{subfigure}{.3\linewidth}
		\includegraphics[width=\textwidth]{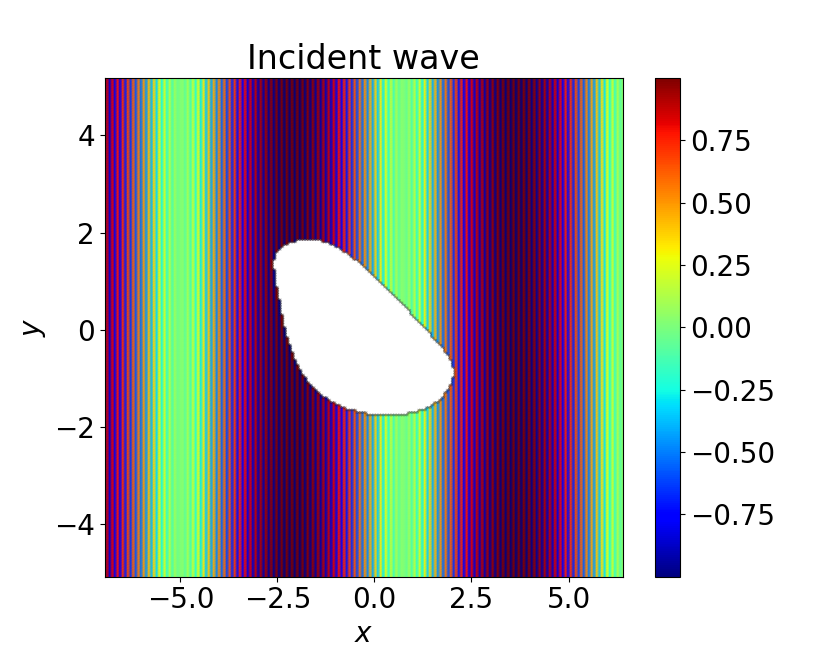}
		\subcaption{The real part of the plane incident wave in the direction $[1,0]$.}
	\end{subfigure}
	\begin{subfigure}{.3\linewidth}
		\includegraphics[width=\textwidth]{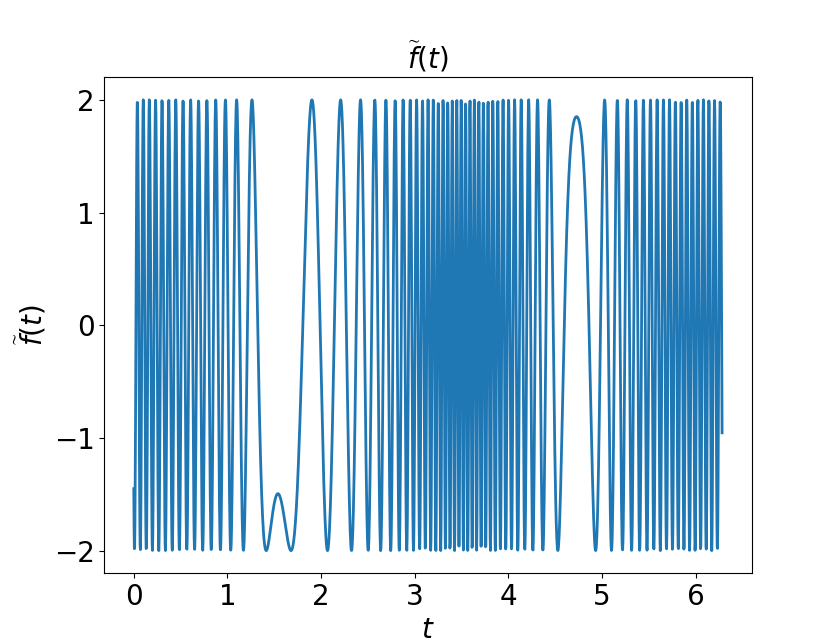}
		\subcaption{The imaginary part of the right-hand side function $\widetilde{f}$ of the BIE.}
	\end{subfigure}
	\begin{subfigure}{0.3\textwidth}
		\centering
		\includegraphics[width=\textwidth]{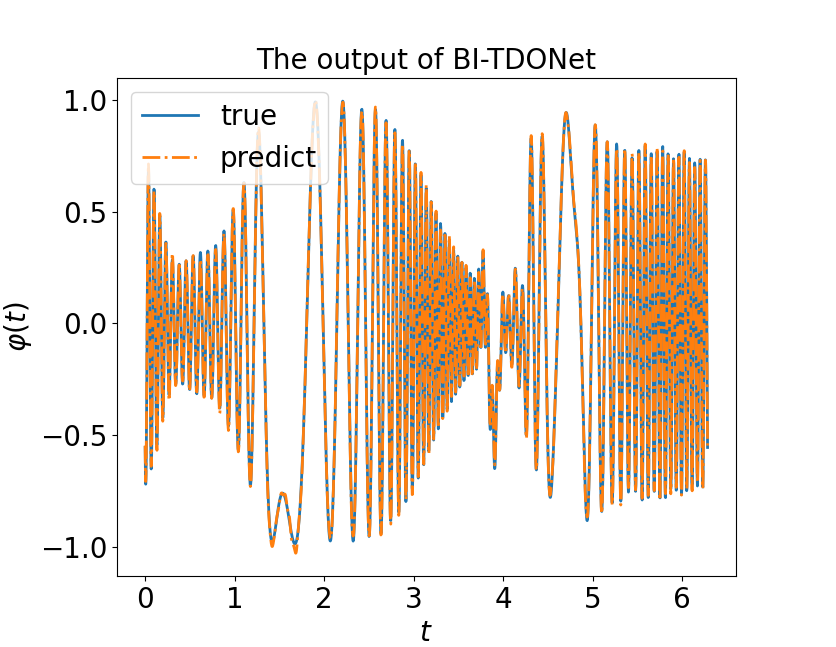}
		\subcaption{The imaginary part of the predicted solution of BI-TDONet.}
	\end{subfigure}
	\begin{subfigure}{0.3\textwidth}
		\centering
		\includegraphics[width=\textwidth]{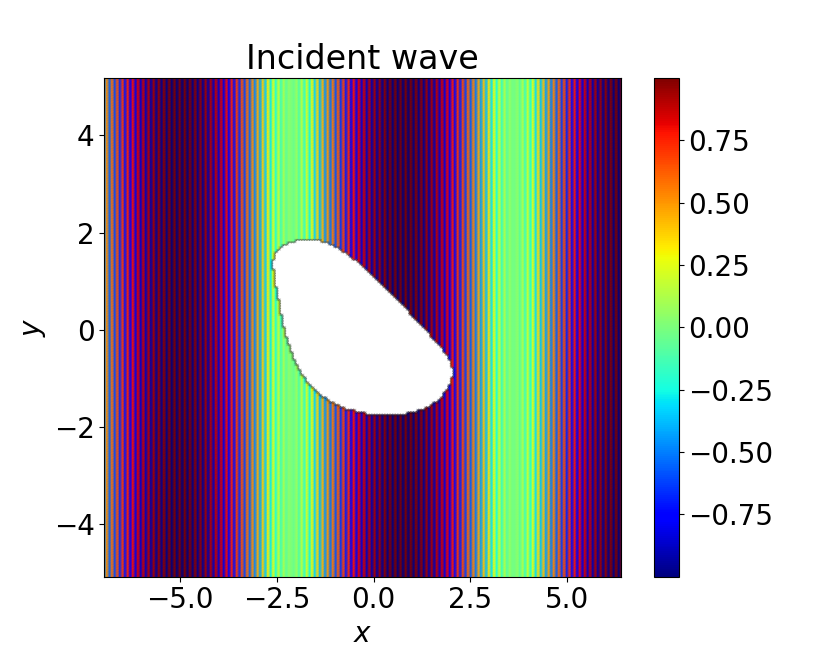}
		\subcaption{The imaginary part of the plane incident wave in the direction $[1,0]$.}
	\end{subfigure}
	\begin{subfigure}{0.3\textwidth}
		\centering
		\includegraphics[width=\textwidth]{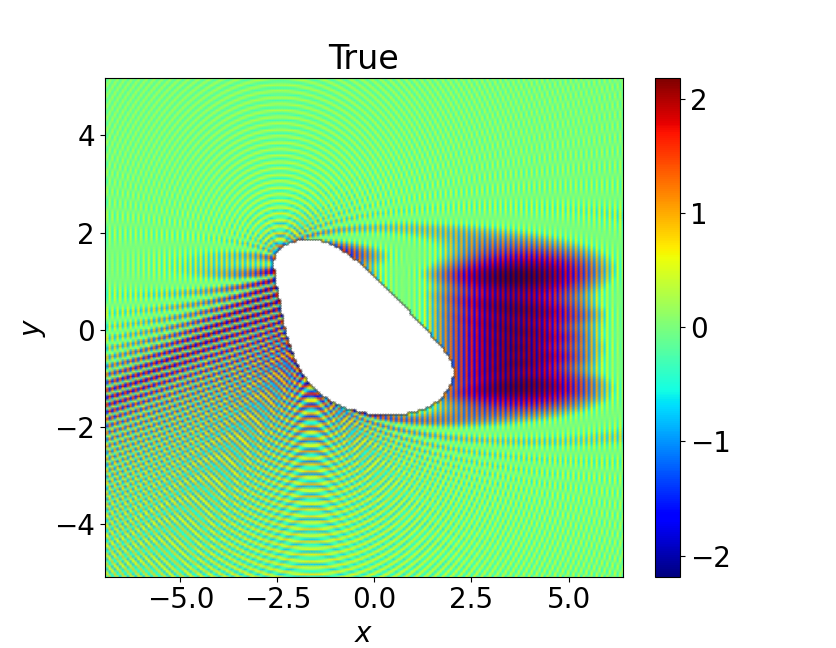}
		\subcaption{The real part of the true solution of the scattered field .}
	\end{subfigure}
	\begin{subfigure}{0.3\textwidth}
		\centering
		\includegraphics[width=\textwidth]{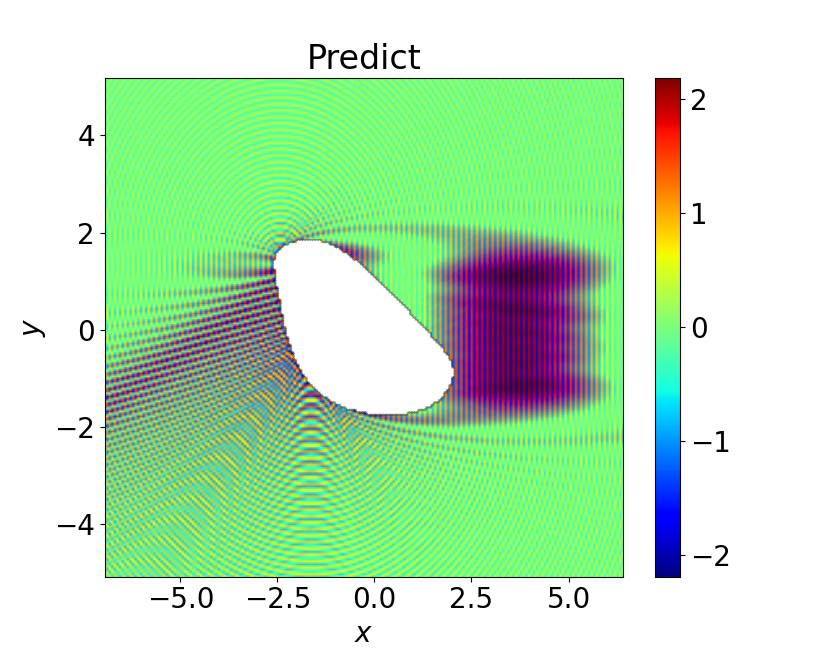}
		\subcaption{The real part of the predicted solution of the scattered field.}
	\end{subfigure}
	\begin{subfigure}{0.3\textwidth}
		\centering
		\includegraphics[width=\textwidth]{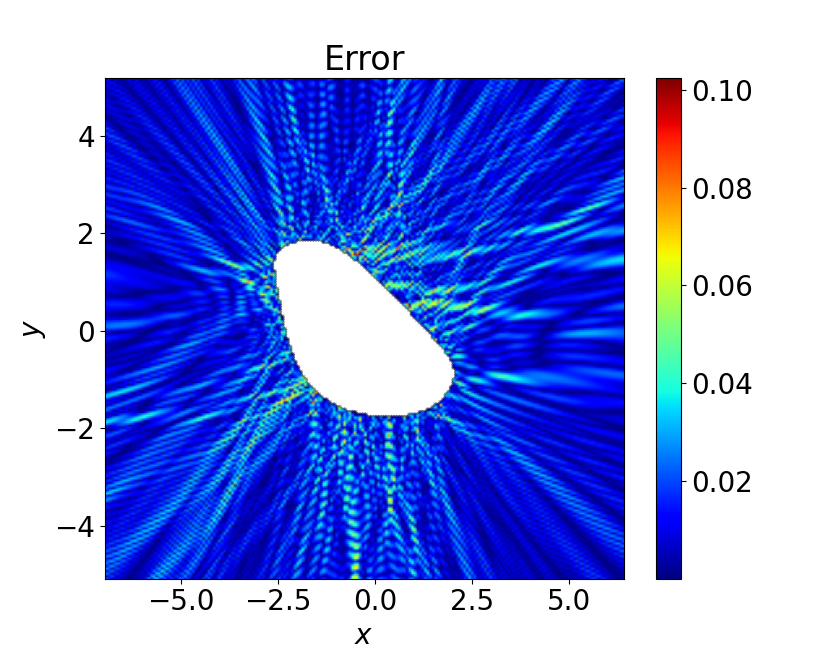}
		\subcaption{The real part of the absolute error between the true and predicted solutions.}
	\end{subfigure}
	\begin{subfigure}{.3\linewidth}
		\includegraphics[width=\textwidth]{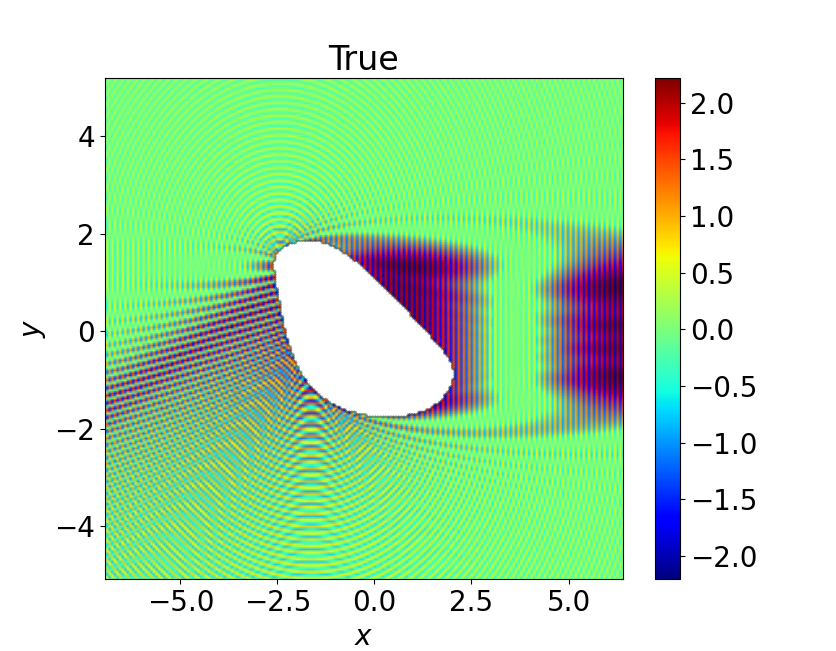}
		\subcaption{The imaginary part of the true solution of the scattered field .}
	\end{subfigure}
	\begin{subfigure}{0.3\textwidth}
		\centering
		\includegraphics[width=\textwidth]{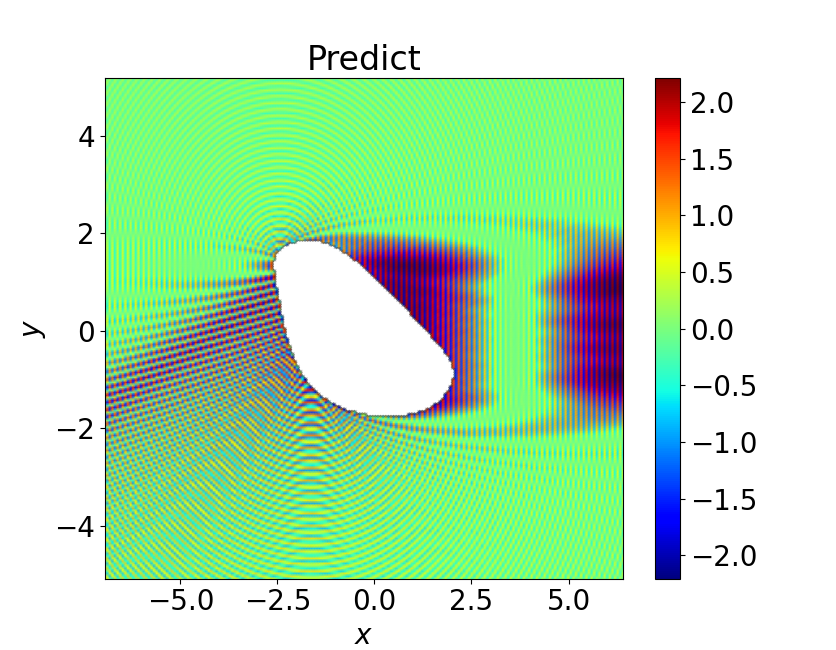}
		\subcaption{The imaginary part of the predicted solution of the scattered field.}
	\end{subfigure}
	\begin{subfigure}{0.3\textwidth}
		\centering
		\includegraphics[width=\textwidth]{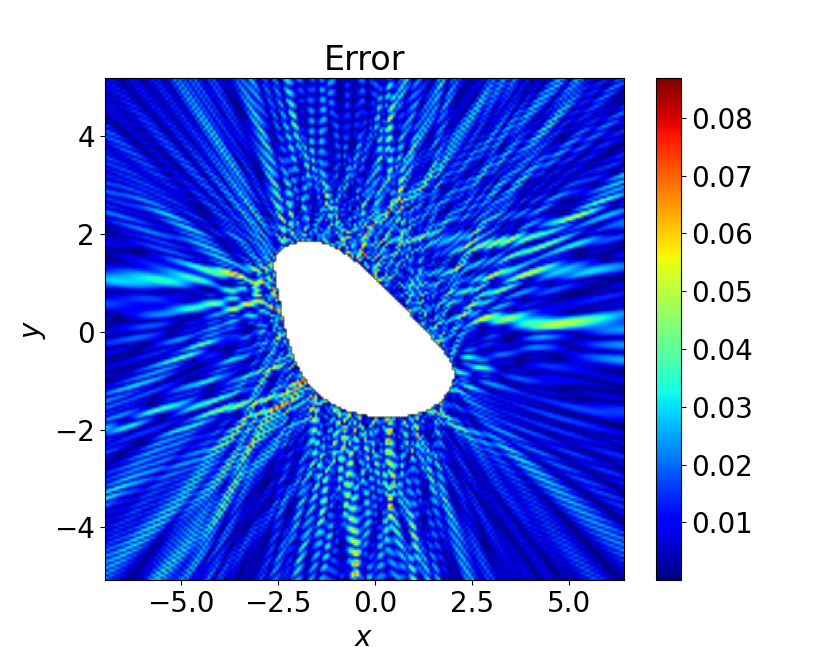}
		\subcaption{The imaginary part of the absolute error between the true and predicted solutions.}
	\end{subfigure}
	\caption{The performance of BI-TDONet in the second example of the acoustic wave obstacle scattering problem.}
	\label{helmholtz_result1}
\end{figure}

We randomly select two samples from the test set to demonstrate the effectiveness of BI-TDONet.  The visualization results for these two examples are displayed in Figures \ref{helmholtz_result} and \ref{helmholtz_result1}. Figures \ref{helmholtz_result} and \ref{helmholtz_result1} (a) and (d) show the real and imaginary parts of the right-hand side of the BIE. Figures \ref{helmholtz_result} and \ref{helmholtz_result1} (b) and (e) respectively illustrate the real and imaginary parts of both the true solution of the BIE and the predicted solution from BI-TDONet. Figures \ref{helmholtz_result} and \ref{helmholtz_result1} (c) and (f) display the real and imaginary parts of the plane incident wave. Figures \ref{helmholtz_result} and \ref{helmholtz_result1} (g) and (j) depict the real and imaginary parts of the true scattered field. Figures \ref{helmholtz_result} and \ref{helmholtz_result1} (h) and (k) show the real and imaginary parts of the predicted solution, computed using the potential integrals \eqref{scatter} from BI-TDONet's output. Finally, Figures \ref{helmholtz_result} and \ref{helmholtz_result1} (i) and (l) present the absolute errors between the predicted and true solutions of the scattered field.

From the visuals in (a), (b), (d), and (e) of Figures \ref{helmholtz_result} and \ref{helmholtz_result1}, it is clear that both the input and output of BI-TDONet exhibit high levels of oscillation. When the input and output possess such highly oscillatory features, frameworks like BI-DeepONet, which rely on direct function values as input and output, encounter significant challenges. In contrast, BI-TDONet, which uses trigonometric coefficients as input and output, naturally captures the frequency characteristics, thereby achieving superior predictive performance. These results underscore BI-TDONet's capability to effectively handle the acoustic sound-soft obstacle scattering problem for wavenumbers ranging from $40$ to $50$, demonstrating its robustness and efficiency.

\section{Conclusion}

Based on the boundary integral equation method, this paper proposes an operator learning framework capable of solving PDEs in various two-dimensional connected domains. This framework overcomes the drawback of existing deep operator learning methods that require retraining when dealing with PDEs defined on different geometrical regions. Utilizing this framework, we developed two neural operator learning models: BI-DeepONet and BI-TDONet. BI-DeepONet is based on the DeepONet architecture, while BI-TDONet is built on the singular value decomposition of bounded linear operators and can incorporate frequency domain characteristics.

Numerical experiment results show that BI-DeepONet and BI-TDONet exhibit outstanding performance in solving LBVPs, potential flow around obstacles, and elastostatics problems. In these experiments, the mean $l_2$ relative error (MRE) of BI-TDONet is approximately $10^{-2}$ to $10^{-1}$ of that of BI-DeepONet. In the acoustic wave sound-soft obstacle scattering problem, BI-TDONet achieves an error of $3.51 \times 10^{-2}$, whereas  BI-DeepONet fails to converge. We speculate this is due to the high-frequency information present in the input functions, which BI-DeepONet cannot effectively capture. Additionally, it is worth noting that our proposed operator learning model takes only $10^{-5}$ to $10^{-6}$ seconds on average to predict the solutions of the boundary integral equations, a speed significantly faster than the fast boundary integral equation algorithms developed over the past decade.

It should be noted that the BIE-based operator learning framework proposed in this paper cannot be directly used to solve general linear or nonlinear PDEs. For general linear PDEs, this is mainly because the Green's functions of such PDEs are difficult to obtain. However, the method proposed by Nicolas Boulle et al. \cite{Nicolas2022}, which uses neural networks to learn Green's functions, offers a potential solution to this problem. Furthermore, for certain nonlinear PDEs, we can transform them into a series of linear PDEs using implicit-explicit Euler methods or Rothe's method. Given the significant advantages of BIE-based neural operators, we are considering their application to nonlinear PDEs such as the Navier-Stokes equations and exploring their use in three-dimensional problems like $3$D LBVPs.

 \bibliographystyle{abbrv}
 \bibliography{cas-refs}

\end{document}